\documentclass[12pt,a4paper]{article}
\usepackage[OT2,T1]{fontenc}
\usepackage{amssymb}
\usepackage{amsmath}
\usepackage{stmaryrd}
\usepackage{mathrsfs}
\usepackage{mathtools}
\usepackage{bm}
\usepackage{graphicx}
\usepackage[leftcaption]{sidecap}
\usepackage{caption}
\usepackage{subcaption}
\usepackage{here}
\usepackage[margin=1in]{geometry}
\usepackage{cite}
\DeclareMathAlphabet{\mathscrbf}{OMS}{mdugm}{b}{n}
\usepackage{tikz}
\usetikzlibrary{decorations.pathmorphing}
\usepackage{parskip}
\usepackage{enumerate}

\let\OLDthebibliography\thebibliography
\renewcommand\thebibliography[1]{
  \OLDthebibliography{#1}
  \setlength{\parskip}{0pt}
  \setlength{\itemsep}{3pt plus 0.3ex}
}

\usepackage[colorlinks,allcolors=blue]{hyperref}

\usepackage{amsthm}

\newcommand{\be}{\begin{equation}}
\newcommand{\ee}{\end{equation}}

\newcommand{\bbgamma}{g}
\newcommand{\CL}{generalised conformal}
\newcommand{\cf}{\itt{cf.}\ }
\newcommand{\dd}{\text{d}}
\newcommand{\DD}{D}
\newcommand{\EE}{E}
\newcommand{\FF}{F}
\newcommand{\GG}{G}
\newcommand{\der}{\partial}

\newcommand{\eg}{\textit{e.g.}\ }
\newcommand{\eps}{\varepsilon}
\newcommand{\hs}{{\tt hs}}
\newcommand{\hxi}{\xi}
\newcommand{\ie}{\textit{i.e.}\ }
\newcommand{\itt}{\textit}
\newcommand{\mX}{\mathfrak{X}}
\newcommand{\nn}{\nonumber}
\newcommand{\phii}{\varphi}
\newcommand{\sproj}{super-projectable}
\newcommand{\tw}{\texttt{w}}
\newcommand{\UIR}{{\sc uir}}

\newcommand\bq{{\bf q}}
\newcommand\bx{{\bf x}}

\newcommand{\cD}{{\cal D}}

\newcommand{\cH}{{\cal H}}

\newcommand{\cL}{{\cal L}}

\newcommand{\cO}{{\cal O}}
\newcommand{\cR}{{\cal R}}

\newcommand{\cU}{\,{\cal U}}
\newcommand{\cV}{{\cal V}}

\newcommand{\sI}{\mathscr{I}}
\newcommand{\sM}{\mathscr{M}}
\newcommand{\sN}{\mathscr{N}}
\newcommand{\sU}{\mathscr{U}}

\newcommand{\CC}{\mathbb{C}}

\newcommand{\RR}{\mathbb{R}}

\begin{document}

\begin{center}
\makebox[\textwidth][c]{{\Large{\textbf{Massless Scalars and Higher-Spin BMS in Any Dimension}}}}\\[1.5em]
{\large{Xavier Bekaert$^a$ and Blagoje Oblak$^b$}}\\[1.5em]
$^a$ Institut Denis Poisson, Unit\'e Mixte de Recherche 7013,\\ Universit\'e de Tours, Universit\'e d’Orl\'eans, CNRS,\\
Parc de Grandmont, F-37200 Tours, France.\\[.2em]
{\tt xavier.bekaert@lmpt.univ-tours.fr}\\[1em]
$^b$ CPHT, CNRS, \'Ecole Polytechnique, IP Paris,\\ F-91128 Palaiseau, France.\\[.2em]
{\tt blagoje.oblak@polytechnique.edu}\\
\vspace{5em}
\begin{minipage}{.823\textwidth}
{\bf Abstract.} Starting from the asymptotic kinematics of massless scalar fields near null infinity in any spacetime dimension, we build two higher-spin extensions of the Carrollian definition of the BMS group and its generalisations. The first extension exhibits conformal properties reminiscent of the singleton in Anti-de Sitter space. The second acts on the space of radiative solutions of the d'Alembert equation, \ie on Sachs's representation of BMS, which we relate to the scalar massless Poincar\'e representation and extend to any Carrollian manifold. The corresponding enveloping algebra is a higher-spin extension of BMS that can be interpreted as the asymptotic symmetry of a putative exotic higher-spin gravity theory around Minkowski spacetime. Along the way, we provide a pedagogical introduction to Carrollian geometry and its relation to BMS.
\end{minipage}
\end{center}
\vspace{2em}

\newpage
\tableofcontents

\newpage
\section{Motivation and summary}
\label{semott}

Ever since the early seminal works on conserved quantities in general relativity \cite{Arnowitt,Bondi,Sachs1,Sachs2,Brown}, it has been well established that asymptotic symmetries are crucial for the quantisation of all gauge theories. This has become even more manifest in recent years, following a series of striking discoveries relating Bondi-Metzner-Sachs (BMS) symmetry \cite{Bondi,Sachs1,Sachs2} to flat space holography \cite{Barnich:2009se,Barnich:2010eb,Bagchi:2009my,Bagchi:2012yk} and scattering amplitudes \cite{He:2014laa,Strominger:2017zoo}, and their modern blend known as ``celestial holography'' \cite{Raclariu:2021zjz,Prema:2021sjp,Pasterski:2021raf}. As a result, the BMS group and its extensions \cite{Barnich:2009se,Barnich:2010eb,Campiglia:2014yka,Campiglia:2015yka} are now key ingredients in any putative holographic description of quantum gravity around flat backgrounds.

The present work originates from the interplay between these concepts and higher-spin gravity theories. Indeed, despite heavy constraints imposed by no-go theorems on interacting higher-spins in Minkowski spacetime (see \eg \cite{Bekaert:2010hw,Rahman:2015pzl,Ponomarev:2022vjb} and references therein), it is still of interest to study such systems and their symmetries: they are relevant for string theory in the tensionless limit \cite{Gross:1988ue,Sagnotti:2003qa}, and flat space physics more generally has key applications. One thus naturally wonders what are the higher-spin analogues of Minkowskian asymptotic symmetries, and in particular of the BMS group. Conversely, one may ask if there exist Minkowskian cousins of well-known actors in the higher-spin AdS/CFT correspondence (see \eg the reviews \cite{Bekaert:2012ux,Giombi:2016ejx,Sleight:2017krf}), such as the singleton \cite{Dirac:1963ta} that may be seen as the fundamental representation of the higher-spin algebra, or the Flato-Fronsdal theorem \cite{Flato:1978qz} that provides the decomposition of the tensor product of two singletons. The goal of this paper is to shed light on these questions and put forward possible answers.

This is not the first time such issues are raised in the literature. For instance, higher-spin asymptotic symmetries and their implications are well known in spacetime dimension three, in both AdS \cite{Henneaux:2010xg,Campoleoni:2010zq,Gaberdiel:2010ar,Gaberdiel:2010pz,Campoleoni:2011hg} and Minkowski backgrounds \cite{Afshar:2013vka,Gonzalez:2013oaa,Campoleoni:2015qrh,Campoleoni:2016vsh,Ammon:2020fxs}. The situation is much less clear-cut in dimensions $\geqslant4$: in AdS, the standard higher-spin holographic dictionary bears no obvious relation to enhanced asymptotic symmetries, while in flat space the only known construction of higher-spin asymptotic symmetries is that of \cite{Campoleoni:2017mbt,Campoleoni:2017qot,Campoleoni:2018uib,Campoleoni:2020ejn}, where it was shown that Weinberg's soft theorem for higher-spin gauge bosons can be rephrased as the Ward identity of an extension of standard gravitational BMS symmetry. (See also \cite{Campoleoni:2021blr}, whose motivations are  analogous to ours.)

Another key ingredient of this work is the mathematical language of curved Carrollian geometry \cite{Duval:2014uoa}, based on the Carroll symmetry group \cite{Levy1965,Gupta1966}. Its first appearance in the physics literature dates back to \cite{Henneaux:1979vn} in the ``zero signature limit'' of the Hamiltonian formulation of general relativity. More importantly for our purposes, it was first related to BMS in \cite{Duval:2014uva,Duval:2014lpa} as a modern take on Penrose's classical approach to asymptotic symmetries \cite{Penrose:1965am,Penrose:1972ea,Geroch1977,Ashtekar:1987tt}. Since then, Carrollian geometry has been ubiquitous in flat space physics, including \eg ultra-relativistic hydrodynamics at infinity \cite{Ciambelli:2018xat,Ciambelli:2018wre,Campoleoni:2018ltl,Petkou:2022bmz} and celestial holography \cite{Ciambelli:2019lap,Donnay:2022aba}.

Despite this large body of existing knowledge, the line of thought pursued here appears to be new. Namely, inspired by the key role of the singleton---also known as the ``Rac'' \cite{Flato:1978qz}---in higher-spin AdS/CFT, we study the asymptotic kinematics of a massless scalar field near null infinity \cite{Campiglia:2017dpg,Campiglia:2017xkp,Satishchandran:2019pyc} and use it to define higher-spin BMS algebras based on Carrollian geometry. Our paper is thus a step towards flat higher-spin holography, understood as an analogue of the holographic dictionary between a scalar bulk field with critical mass in AdS and the higher-spin symmetry of its boundary data. In practice, the analysis is carried out in arbitrary bulk spacetime dimension $d+2$, involving $d$-dimensional celestial spheres $S^d$, and yields two inequivalent Minkowskian analogues of the usual singleton:
\begin{itemize}
\item[1.] \textbf{The Wick-rotated Rac (WRac)} consists of time-independent field configurations at null infinity that stem from ``overleading'' solutions of the d'Alembert equation. It is essentially obtained upon replacing $\mathfrak{so}(d,2)$ by $\mathfrak{so}(d+1,1)$ in the group-theoretic definition of the Rac \cite{Flato:1978qz} and should therefore look familiar to higher-spin experts, as it is formally described by the same equations up to proper changes of signature. In fact, it shares several important qualitative features with the Rac---it has no bulk degrees of freedom and lives on the conformal boundary---save for one cardinal property: it is \itt{not} unitarisable. Another unsatisfactory feature of the WRac is that is \itt{not} a faithful representation of BMS (nor of Poincar\'e); only the Lorentz subgroup is represented faithfully.
\item[2.] \textbf{The Sachs module} \cite{Sachs2} consists of time-dependent configurations on null infinity, determined by radiative solutions of the d'Alembert equation. It seems to be the best candidate for an analogue of the Rac in flat spacetime, furnishing a representation of Poincar\'e and BMS that is both faithful and unitary \cite{Sachs2,Mccarthy:1972ry,McCarthy01,McCarthy00,McCarthy317,McCarthy301,McCarthy489}. Its main weakness is its generality: as we shall see, it furnishes a unitary representation not only of BMS, but of the immensely larger group of all Carrollian bundle automorphisms.
\end{itemize}
The peculiarities of these two setups suggest that flat higher-spin holography must be quite different from its AdS cousin. We shall proceed nonetheless and define new Carrollian higher-spin symmetries by considering algebras of differential operators on null infinity that preserve these structures. The WRac will thus yield an algebra isomorphic to the standard one of bosonic higher-spin gravity \cite{Eastwood:2002su,Vasiliev:2003ev} on a de Sitter spacetime in one less dimension, signalling that it is ``too small'' to provide a suitable starting point for higher-spin gravity in Minkowski spacetime since all supertranslations are quotiented out. On the other hand, the symmetry algebra of the Sachs module will define a novel candidate higher-spin extension of BMS, whose spectrum of generators turns out to contain the BMS Killing tensors obtained in \cite{Campoleoni:2017qot} from the asymptotic symmetries of free massless higher-spin fields. Since the space of symmetries encountered in \cite{Campoleoni:2017mbt,Campoleoni:2017qot,Campoleoni:2018uib,Campoleoni:2020ejn} only had a vector structure, our present work may be seen as a way to endow it with a Lie bracket. A corollary of this proposal is also a sharp distinction between symmetry algebras stemming from distinct choices of fall-off conditions; choosing the ``correct'' higher-spin BMS algebra for a putative higher-spin theory around Minkowski spacetime requires the selection of certain preferred fall-offs.

The paper is organised as follows. We start in section \ref{seSCAL} by writing the asymptotic solution of the d'Alembert equation near null infinity. Its leading piece then yields either the WRac, or the Sachs module of the BMS group. In particular, we extend the Sachs Hermitian form to any dimension and relate it to the standard Poincar\'e-invariant one for scalar massless unitary representations. Section \ref{seCAR} is devoted to a detailed, self-contained review of Carrollian geometry and its relation to (generalised) BMS symmetry. This also allows us to write the Sachs inner product in a coordinate-independent way such that invariance under bundle automorphisms becomes manifest. We stress that sections \ref{seSCAL}--\ref{seCAR} are devoid of higher spins and may be of general interest to anyone working on asymptotic symmetries and flat holography. Finally, section \ref{seHS} is devoted to the definition of higher-spin extensions of BMS based on Carrollian geometry and the two aforementioned representations (WRac and Sachs). We briefly conclude with a discussion on the fall-offs and spectra of bulk gauge fields needed for a putative higher-spin theory to admit these algebras as asymptotic symmetries.

\section{BMS action on massless scalars}
\label{seSCAL}

In this section, we solve d'Alembert's equation $\square\Phi=0$ as an asymptotic series in the inverse distance away from the origin in Bondi coordinates, and find that the space of solutions supports a linear action of the BMS algebra. We then briefly recall certain definitions pertaining to the transformations of densities on manifolds under diffeomorphisms, before identifying the densities that occur in the asymptotic expansion of scalars. Thus, an ``overleading'' choice of fall-offs naturally leads to the WRac solution mentioned in the \hyperref[semott]{introduction}, while radiative fall-offs are such that the field's leading component defines the Sachs module. Related considerations have been put forward in \cite{Campiglia:2017dpg,Campiglia:2017xkp,Satishchandran:2019pyc,Nguyen:2022nnx} in the context of  asymptotic symmetries. The Carrollian perspective on these results is relegated to section \ref{seCAR}, while their higher-spin implications will be treated in section \ref{seHS}.

\subsection{Asymptotic scalar dynamics}
\label{sescad}

Here we briefly recall the expression of Poincar\'e and BMS generators in Bondi coordinates, then study the asymptotic expansion of solutions of the d'Alembert equation near null infinity. This will motivate the existence of two distinguished kinds of boundary data: the first, time-dependent but unconstrained, will eventually lead to the Sachs module. The other, time-independent but singular, will yield a WRac.

\begin{figure}[t]
\centering
\begin{tikzpicture}
    \node at (0,0) {\includegraphics[width=.3\textwidth]{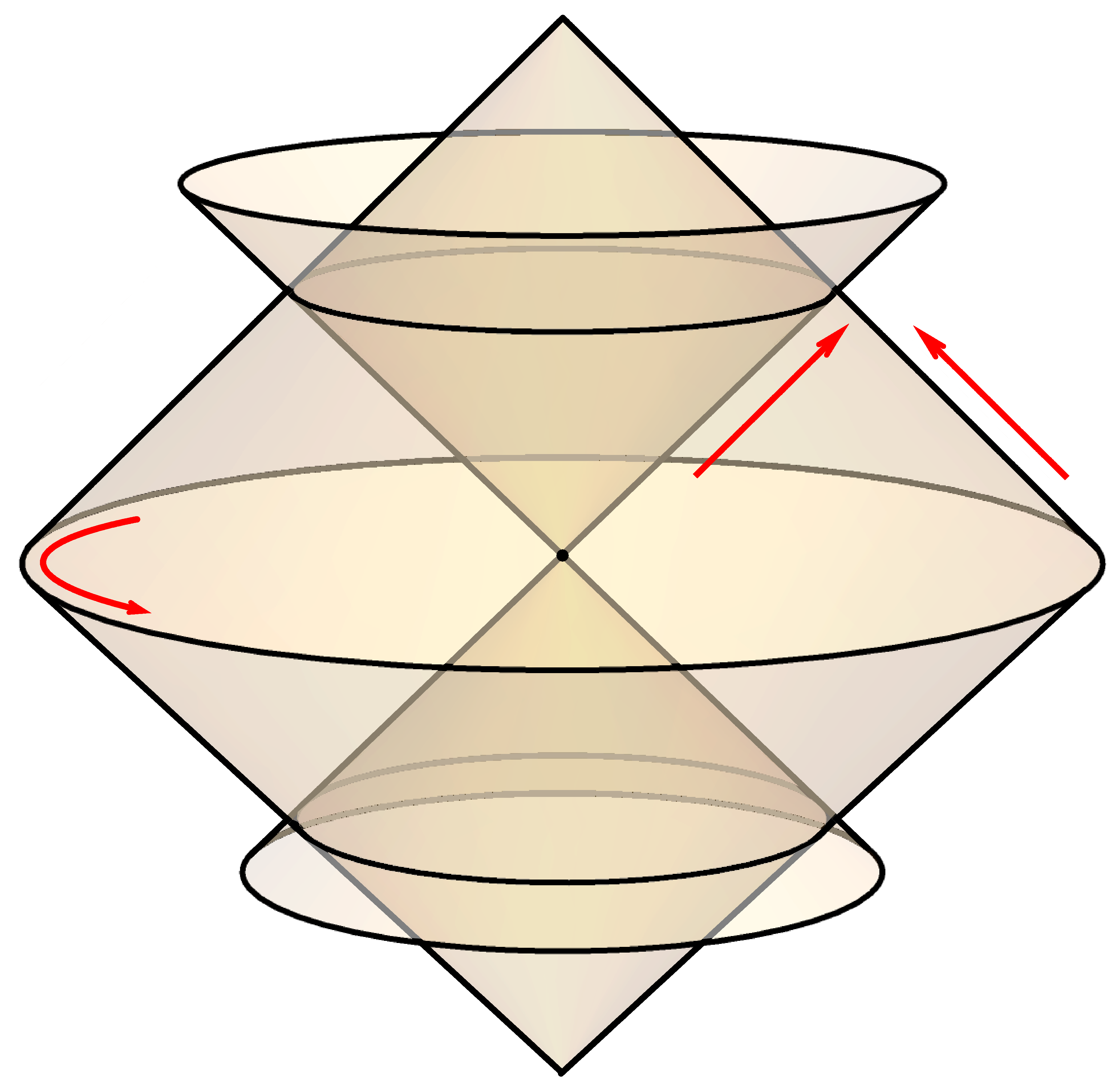}};
    \node[red] at (2,.7) {$u$};
    \node[red] at (1.1,.5) {$r$};
    \node[red] at (-1.6,0) {$x^a$};
    \draw[red,decorate,decoration={snake,segment length=.5em,amplitude=.1em}] (-2.35,0.1)--(-0.15,2.3);
    \node[red] at (-2,.9) {$\sI^+$};
    \draw[red,decorate,decoration={snake,segment length=.5em,amplitude=.1em}] (2.4,-0.2)--(0.1,-2.35);
    \node[red] at (2,-1) {$\sI^-$};
    \node[red] at (0.02,2.25) {\textbullet};
    \node[red,above] at (0.02,2.25) {$i^+$};
    \node[red] at (0.015,-2.25) {\textbullet};
    \node[red,below] at (0.015,-2.25) {$i^-$};
\end{tikzpicture}
~~~~~~
\begin{tikzpicture}
\draw[-,thick] (-2.25,0)--(0,2.25)--(2.25,0)--(0,-2.25)--cycle;
\draw[-] (-1.5,-1.5)--(1.5,1.5);
\draw[-] (-1.5,1.5)--(1.5,-1.5);
\draw[-,red,line width=.2em] (-1.05-.05,-1.05+.05) -- (-.07-.05,-.07+.05);
\draw[-,red,line width=.2em] (1.05+.05-2.25,-1.05+.05) -- (.07+.05-2.25,-.07+.05);
\draw[-,blue,line width=.2em] (1.05+.05,-1.05+.05) -- (.07+.05,-.07+.05);
\draw[-,blue,line width=.2em] (-1.05-.05+2.25,-1.05+.05) -- (-.07-.05+2.25,-.07+.05);
\draw[-,green,line width=.2em] (-1.05-.05,1.05-.05) -- (-.07-.05,.07-.05);
\draw[-,green,line width=.2em] (1.05+.05-2.25,1.05-.05) -- (.07+.05-2.25,.07-.05);
\draw[-,yellow,line width=.2em] (1.05+.05,1.05-.05) -- (.07+.05,.07-.05);
\draw[-,yellow,line width=.2em] (-1.05-.05+2.25,1.05-.05) -- (-.07-.05+2.25,.07-.05);
\draw[-,black,line width=.2em] (-1.05+.05,-1.05-.05) -- (-.07+.05,-.07-.05);
\draw[-,black,line width=.2em] (1.05-.05,-1.05-.05+2.25) -- (.07-.05,-.07-.05+2.25);
\draw[-,gray,line width=.2em] (1.05-.05,-1.05-.05) -- (.07-.05,-.07-.05);
\draw[-,gray,line width=.2em] (-1.05+.05,-1.05-.05+2.25) -- (-.07+.05,-.07-.05+2.25);
\draw[-,purple,line width=.2em] (1.05-.05,1.05+.05-2.25) -- (.07-.05,.07+.05-2.25);
\draw[-,purple,line width=.2em] (-1.05+.05,1.05+.05) -- (-.07+.05,.07+.05);
\draw[-,orange,line width=.2em] (-1.05+.05,1.05+.05-2.25) -- (-.07+.05,.07+.05-2.25);
\draw[-,orange,line width=.2em] (1.05-.05,1.05+.05) -- (.07-.05,.07+.05);
\draw[-,dashed] (-1.125,1.125)--(-1.125,-1.125);
\draw[-,dashed] (1.125,1.125)--(1.125,-1.125);
\node[red,above] at (0.02,2.25) {\color{white}{$i^+$}};
\node[red,below] at (0.015,-2.25) {\color{white}{$i^-$}};
\end{tikzpicture}
\caption{\itt{Left:} A Penrose diagram of Minkowski spacetime, displaying Bondi coordinates $(r,u,x^a)$ with $a=1,\cdots\!,d$, along with the past and future light-cones of the origin. Each circle is really a sphere $S^d$. The conformal boundary of the diagram consists of future and past null infinities, respectively denoted $\sI^+$ and $\sI^-$. \itt{Right:} The past and future light-cones of the origin are related to null infinities by an inversion $x^{\mu}\to x^{\mu}/x^2$, written here in inertial coordinates. Any two segments of the same colour are mapped on one another under this map, and the two dashed lines indicate its fixed points. See footnote \ref{footman} for a more detailed statement.}
\label{fiPen}
\end{figure}
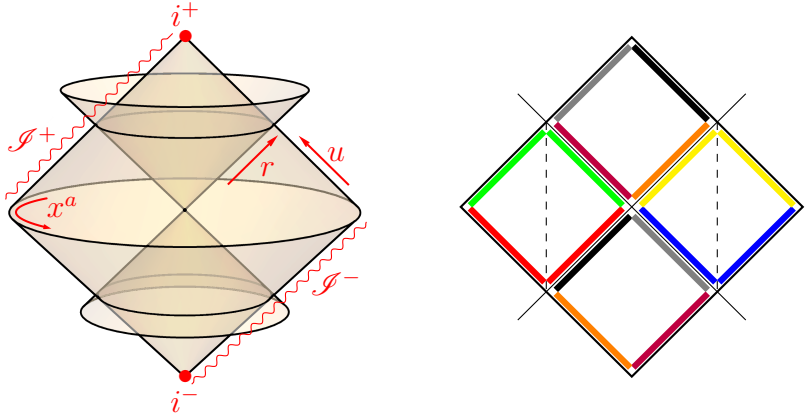

\paragraph{Bondi coordinates and BMS generators.} Consider retarded Eddington-Finkelstein (Bondi) coordinates $(r,u,x^a)$ on $(d+2)$-dimensional flat spacetime $\RR^{d+1,1}$, where $u$ is  retarded time, $r$ is the distance away from some arbitrary spatial origin, and $x^a$ ($a=1,\ldots\,,d$) are angles on a celestial sphere $S^d$ (see fig.\ \ref{fiPen}). In these terms, the Minkowski metric reads 
\be
\label{mmetric}
\dd s^2
=
-\dd u^2-2\,\dd u\,\dd r+r^2\bbgamma_{ab}(x)\,\dd x^a\,\dd x^b\,,
\ee
so $\der_r$ is tangent to radial outgoing null rays. The isometries of \eqref{mmetric} span the Poincar\'e group, whose generators are vector fields $\xi=\xi^{\mu}\der_{\mu}$ with components (see \cite[eqs.\ (4.7)--(4.8)]{Barnich:2010eb} or \cite[eq.\ (2.13)]{Campoleoni:2017mbt})
\begin{align}
\label{xiir}
\xi^r
&=
-\frac{r+u}{d}\,\nabla_aX^a(x)+\frac{1}{d}\,\nabla^2\alpha(x)\,,\\
\label{xiiu}
\xi^u
&=
\alpha(x)+\frac{u}{d}\,\nabla_aX^a(x)\,,\\
\label{xiia}
\xi^a
&=
X^a(x)-\frac{1}{r}\,g^{ab}(x)\big[\der_b\alpha(x)+\frac{u}{d}\,\der_b\nabla_cX^c(x)\big]\,,
\end{align}
where $X^a\der_a$ is a conformal Killing vector field on $S^d$, $\nabla$ is the Levi-Civita connection on $S^d$ with Laplacian $\nabla^2$, and $\alpha(x)$ is any function on $S^d$ such that
\be
\tfrac{1}{d}\,\nabla^2\alpha+\alpha
=
\alpha_0\,,
\label{acons}
\ee
with $\alpha_0$ the zero-mode of $\alpha(x)$.\footnote{Eq.\ \eqref{acons} can equivalently be written as $\der_a(\nabla^2+d)\alpha=0$, without mentioning the zero mode. It is also equivalent to the ``good-cut equation in Bondi frame'' $\nabla_a\nabla_b\alpha\propto g_{ab}$ (see \eg \cite[eq.\ (2.6)]{Ashtekar:1987tt}).} The elements of the pair $(X,\alpha)$ respectively generate Lorentz transformations and spacetime translations. In particular, Lorentz generators correspond to conformal maps on the celestial sphere and induce angle-dependent rescalings of the radial coordinate (since they leave the term $r^2\dd\Omega^2$ invariant in the metric \eqref{mmetric}). The Lie bracket of such pairs reads
\be
\label{liba}
\big[(X,\alpha),(Y,\beta)\big]
=
\Big([X,Y],X^a\der_a\beta-\tfrac{1}{d}\,\beta\,\nabla_aX^a-Y^a\der_a\alpha+\tfrac{1}{d}\,\alpha\,\nabla_aY^a\Big)\,,
\ee
exhibiting the standard semi-direct sum structure of the Poincar\'e algebra $\mathfrak{iso}(d+1,1)=\mathfrak{so}(d+1,1)\inplus\RR^{d+2}$. The BMS generalisation consists in relaxing the restriction \eqref{acons} and allowing the function $\alpha(x)$ to be arbitrary (as opposed to having only modes of angular momentum $\ell\leqslant1$); it is then known as a \itt{supertranslation} \cite{Bondi,Sachs1,Sachs2}. We return to this in much greater detail below, starting in section \ref{sebemin}. 

\paragraph{Asymptotic d'Alembert equation.} Now let $\Phi$ be a complex massless scalar field in $\RR^{d+1,1}$ with Lagrangian density $\cL=\sqrt{g}\,\nabla_{\mu}\Phi^*\nabla^{\mu}\Phi$, whose equation of motion is the d'Alembert equation
\be
\square\Phi
=
\der_r^2\Phi
-2\,\der_r\dot\Phi
+\tfrac{d}{r}\,\der_r\Phi
-\tfrac{d}{r}\,\dot\Phi
+\tfrac{1}{r^2}\nabla^2\Phi
=
0\,,
\label{daleq}
\ee
where $\dot\Phi:=\der_u\Phi$. Our goal is to expand this equation near future null infinity (where $r\to\infty$ with fixed $u$ and $x$) and relate the various degrees of the expansion to BMS representations. Indeed, in the spirit of ``asymptotic quantisation'' \cite{Ashtekar:1981bq,Ashtekar:1987tt}, the symplectic form expressed in terms of data $\phi$ at null infinity reads
\be
\label{symplectic}
\Omega[\delta\phi]
=
\lim_{r\to\infty}
\Big[
\,\,i\,r^d\int\dd u\,\dd^dx\,
\sqrt{g(x)}\,\delta\Phi^*\wedge\delta\dot{\Phi}\,\Big]\,,
\ee
where the wedge product is ``vertical'' and involves one-forms in field space. The corresponding energy functional is
\be
E[\phi]
=\frac{1}{2}
\lim_{r\to\infty}\Big[\,r^d\int\dd u\,\dd^dx\,\sqrt{g(x)}\,|\dot\Phi|^2\,\Big],
\label{s5}
\ee
so it is clear that fall-offs at (null) infinity play a key role for the scalar field's phase space. Accordingly, write the solution of \eqref{daleq} as an asymptotic series\footnote{The analogue of \eqref{expa} for a field with mass $m>0$ involves exponentials $e^{-mr}$. The resulting radial expansion is different from that of the massless case; we do not consider it here.}
\be
\Phi(r,u,x)
\sim
\frac1{r^{\Delta}}\sum_{n=0}^{\infty}\frac{\phi_n(u,x)}{r^n}
\qquad\text{as }r\to\infty\,,
\label{expa}
\ee
where $\phi_0\neq0$ by definition and $\Delta$ is some number that we leave free for now. Plugging this ansatz in the d'Alembert equation \eqref{daleq} yields \cite{Satishchandran:2019pyc}
\begin{align}
\label{first}
\cO(r^{-(\Delta+1)}): & & (d-2\Delta)\dot\phi_0&= 0\,,\\
\label{second}
\cO(r^{-(\Delta+2)}): & & (d-2-2\Delta)\dot\phi_1+\Delta(d-1-\Delta)\phi_0-\nabla^2\phi_0&= 0\,,\\
\label{third}
\cO(r^{-(\Delta+3)}): & & (d-4-2\Delta)\dot\phi_2+(\Delta+1)(d-2-\Delta)\phi_1-\nabla^2\phi_1&= 0\,.
\end{align}
The structure of this sequence is reminiscent of the Fefferman-Graham expansion of scalar dynamics in AdS (see \eg \cite[sec.\ 5]{Skenderis:2002wp}), but it is quite distinct from it in several respects. Most crucially, it does not uniquely fix the value of $\Delta$ since the leading-order equation \eqref{first} yields
\be
\label{timeindep}
\Delta=d/2
\qquad\text{or}\qquad
\dot\phi_0=0\,.
\ee
Here the first option imposes no constraint on $\phi_0$ but fixes $\Delta$ in a way that allows the energy \eqref{s5} to be finite. It is indeed this first possibility that one typically considers in the context of asymptotic symmetries \cite{Campiglia:2017dpg,Campiglia:2017xkp,Satishchandran:2019pyc}, where the corresponding solutions of the d'Alembert equation are said to be ``radiative''. By contrast, the second option in \eqref{timeindep} leaves $\Delta$ arbitrary, and turns out to be fairly natural from a higher-spin perspective. Accordingly, we now investigate some immediate features of these two families of solutions. More elaborate considerations regarding their conformal properties and their transformation laws under BMS will be presented in sections \ref{serac}--\ref{sebemin}.

\paragraph{Recursive solution and GJMS operators.} Let us first analyse the branch of solutions obtained by choosing $\Delta=d/2$ in \eqref{timeindep}. Then eq.\ \eqref{first} is automatically satisfied, with $\phi_0$ to be thought of as an arbitrary initial condition on null infinity; we shall typically assume that $\dot\phi_0$ goes to zero in the far past and future in order for the energy \eqref{s5} to be finite, but this is not essential here. The subsequent hierarchy of equations \eqref{second}, \eqref{third}, etc.\ is then solved in terms of time-independent ``integration functions'' $F_1(x)$, $F_2(x)$, etc.\ and time integrals of $\phi_0$. For instance, eqs.\ \eqref{second}--\eqref{third} yield
\begin{align}
\label{s6b}
\phi_1(u,x)
&=
F_1(x)+\tfrac{1}{2}\int_0^u\dd v\,\big(\nabla^2-\tfrac{d}{2}(\tfrac{d}{2}-1)\big)\phi_0(v,x)\,,\\
\phi_2(u,x)
&=
F_2(x)-\tfrac{u}{4}\big(\nabla^2-(\tfrac{d}{2}+1)(\tfrac{d}{2}-2)\big)F_1(x)\nonumber\\
\label{ss5t}
&\quad+\tfrac{1}{8}\int_0^u\dd v\int_0^v\dd w\,\prod_{j=0}^1\big(\nabla^2-(\tfrac{d}{2}+j)(\tfrac{d}{2}-j-1)\big)\phi_0(w,x)\,,
\end{align}
where $F_1(x)$ and $F_2(x)$ are smooth but otherwise arbitrary. Similar expressions hold for all $\phi_n$'s, $n\geqslant1$: each starts with an integration function $F_n(x)$, then involves a polynomial in $u$ where the coefficient of $u^k$ is linear in $F_{n-k}(x)$ ($k=1,\cdots\!,n-1$), and ends with an $n$-fold time integral of $\phi_0(u,x)$, acted upon by the differential operator
\be
\label{s5t}
P_{2n}
:=
\prod_{j=0}^{n-1}\Big(\nabla^2-(\tfrac{d}{2}+j)(\tfrac{d}{2}-j-1)\Big)
\ee
of order $2n$. For $n=1$, this is the \itt{Yamabe operator} on $S^d$, \ie the conformally covariant completion of the Laplacian. (See \eg \cite[sec.\ 3]{Eastwood:2002su} for a concise review.) More generally, \eqref{s5t} is  the so-called \itt{GJMS operator} of order $2n$ on $S^d$ \cite{GJMS,Branson}. It is conformally covariant when acting on a scalar primary field of scaling dimension $\tfrac{d}2-n$ on $S^d$ (see section \ref{senote} for the terminology). More generally, it is Weyl-covariant when it acts on conformal densities of weight $n-\tfrac{d}2$ on an arbitrary curved manifold.

\paragraph{Truncated hierarchy.} We now turn to the branch of solutions of eq.\ \eqref{daleq} obtained by choosing the second option in \eqref{timeindep}: $\phi_0(u,x)=\phi_0(x)$. Then the power $\Delta$ is arbitrary and the hierarchy of equations starting with \eqref{second}--\eqref{third} can again be integrated iteratively in terms of successive time integrals of $\phi_0$ and time-independent integration functions $F_n(x)$, similarly to eqs.\ \eqref{s6b}--\eqref{ss5t}. The key difference with respect to the choice $\Delta=d/2$ is that (i) the energy \eqref{s5} is now generally either infinite or zero,\footnote{Since $\dot\phi_0=0$, one might be tempted to conclude that \eqref{s5} vanishes; the reason this is not the case is because subleading terms in the expansion \eqref{expa} are time-dependent, and contribute to \eqref{s5}.} and (ii) the differential operators involved in the hierarchy of solutions are no longer conformally covariant, since they do not take the GJMS form \eqref{s5t}. This second issue can in fact be cured in one specific class of choices of $\Delta$: if
\be
\label{t6t}
\Delta
=
\frac{d}{2}-N
\ee
for some positive integer $N$, then the expansion \eqref{expa} can be split as
\be
\label{s6t}
\Phi(r,u,x)
\sim
\frac{1}{r^{d/2-N}}\sum_{n=0}^{N-1}\frac{\phi_n(u,x)}{r^n}
+\frac{1}{r^{d/2}}\sum_{n=0}^{\infty}\frac{\phi_{N+n}(u,x)}{r^n}
\ee
where $\phi_N(u,x)$ is again completely unconstrained while $\phi_0,\phi_1,\ldots\,,\phi_{N-1}$ satisfy a closed system of equations that can be solved explicitly. To see this, note that \eqref{t6t} truncates the first $N$ equations of the hierarchy \eqref{first}--\eqref{second}--\eqref{third}--$\,\cdots$ so that
\begin{align}
\dot\phi_0
&=0\,,\\
2(N-1)\dot\phi_1+(\tfrac{d}{2}-N)(\tfrac{d}{2}+N-1)\phi_0-\nabla^2\phi_0
&=0\,,\\
2(N-2)\dot\phi_2
+(\tfrac{d}{2}-N+1)(\tfrac{d}{2}+N-2)\phi_1-\nabla^2\phi_1
&=0\,,\\
&\;\,\,\vdots\nn\\
\label{s5q}
(\tfrac{d}{2}-N+N)(\tfrac{d}{2}+N-N-1)\phi_{N-1}-\nabla^2\phi_{N-1}
&=0\,.
\end{align}
The solution can again be found by writing $\phi_1,\ldots\,,\phi_{N-1}$ as described around eq.\ \eqref{ss5t}, with the difference that time integrals can now be carried out explicitly. Furthermore, the last equation \eqref{s5q} involves no time derivatives, so it boils down to a pure constraint that needs to hold at all times, eventually yielding the set of conditions
\be
\label{s6q}
\prod_{j=0}^{N-1-k}\Big(\nabla^2-(\tfrac{d}{2}+j)(\tfrac{d}{2}-j-1)\Big)F_k(x)
=
0
\qquad\forall\,k=0,1,\cdots\!,N-1,
\ee
where $F_0:=\phi_0(x)$. Each integration function $F_k$ must therefore lie in the kernel of a GJMS operator \eqref{s5t} of order $2(N-k)$. In general, such kernel conditions have no smooth solution, so most $F_k$'s are either singular, or vanish. As a result, the ``overleading'' branch of solutions of \eqref{daleq} built in this way is similarly either singular or trivial; requiring that solutions be \itt{smooth} then sets $\phi_0=\phi_1=\cdots=\phi_{N-1}=0$ in the expansion \eqref{s6t}, reducing it to what one would have obtained by choosing $\Delta=d/2$ in the first place. This is yet another sense in which the choice $\Delta=d/2$ is ``canonical'' in \eqref{timeindep}, supplementing the energy argument above.

An exception occurs for $d$ even and $N\geqslant d/2\geqslant 1$: in that case, the GJMS kernel condition \eqref{s6q} involves factors of the form $\nabla^2+\ell(\ell+d-1)$, where $\ell=j-\tfrac{d}{2}+1\geqslant0$ in terms of the index $j$ used in \eqref{s6q}. Each such factor \itt{does} have a non-trivial smooth kernel that consists of spherical harmonics with angular momentum $\ell$, where ``angular momentum'' is meant in the sense of the index $\ell$ in the value $\ell(\ell+d-1)$ of the quadratic Casimir of SO($d+1$). In terms of the field $\Phi$ solving the d'Alembert equation \eqref{daleq}, each such spherical harmonic is multiplied by some power of $r$ and some positive power of $u$. It is thus perfectly possible---at least in even spacetime dimensions---to build smooth solutions of d'Alembert that do \itt{not} satisfy the canonical boundary condition $\Phi=\cO(r^{-d/2})$. One should keep in mind, however, that their energy \eqref{s5} is typically infinite and that their spacetime dependence is heavily constrained, since the set of such smooth solutions is finite-dimensional.

\paragraph{Combining the two branches.} To summarise, assuming that the only powers of $r$ in the expansion \eqref{expa} are integers modulo $d/2$, the space of solutions of the d'Alembert equation \eqref{daleq} splits in two branches, roughly in the spirit of AdS/CFT:
\be
\label{s6qq}
\Phi(r,u,x)
:=
R(r,u,x)+S(r,u,x)\qquad\text{(on-shell)}.
\ee
Here the second term has a radial expansion dictated by the canonical choice $\Delta=d/2$ in \eqref{timeindep}, namely
\be
\label{e52}
S(r,u,x)
\sim
\frac{1}{r^{d/2}}\sum_{n=0}^{\infty}\frac{\phi_n(u,x)}{r^n}
\qquad\text{with}\qquad
\phi_0(u,x):=\lim_{r\to\infty}[\,r^{\frac{d}{2}}\,S(r,u,x)\,]\,,
\ee
where $\phi_0(u,x)$ is unconstrained while the subleading terms $\phi_{n\geqslant1}$ are fixed by $\phi_0$ and integration functions $F_n(x)$, in a hierarchy that begins with eqs.\ \eqref{s6b}--\eqref{ss5t}. The energy \eqref{s5} of $S$ is solely determined by the leading term $\phi_0$, and it is finite provided $\dot\phi_0$ decays to zero sufficiently fast at early and late times. We shall refer to this canonical branch of solutions as the \itt{Sachs module} of BMS, in reference to the work \cite{Sachs2} where it was first studied from the perspective of asymptotic symmetries; this is also why we write this branch as ``$S$'' in eq.\ \eqref{s6qq}. By contrast, the first term on the right-hand side of \eqref{s6qq} typically has an ``overleading'' and finite radial expansion
\be
\label{rexpa}
R(r,u,x)
=
\sum_{n=1}^N\frac{\psi_n(u,x)}{r^{d/2-n}}\,,
\ee
where the $\psi_n$'s are again determined by integration functions $F_k$, but these are now constrained by GJMS kernel conditions of the form \eqref{s6q}. The corresponding solutions are generally singular and have infinite energy. Non-singular, non-trivial solutions only occur in special circumstances ($N\geqslant d/2\in\mathbb{N}$), then spanning a finite-dimensional set of field configurations given by spherical harmonics. While this overleading branch of solutions is somewhat awkward from the viewpoint of asymptotic symmetries, it is quite natural from a higher-spin perspective that likens it to the singleton, or \itt{Rac}, in AdS \cite{Flato:1978qz,Angelopoulos:1980wg}, whose boundary behaviour is indeed constrained by a d'Alembertian condition. In the flat case, this condition becomes a set of \itt{Euclidean} Laplacian constraints \eqref{s6q} on celestial spheres, so we shall refer to this branch of solutions as the \itt{Wick-rotated Rac (WRac)} of Minkowski space; hence the notation ``$R$'' in eq.\ \eqref{s6qq}.

The remainder of this section is devoted to a detailed group-theoretic analysis of these leading and subleading components: following a brief aside (section \ref{senote}) on densities and weights, the WRac is discussed in section \ref{serac} while the BMS transformation law of the various terms of the expansion \eqref{expa}, and in particular the unitary Sachs module, are studied in section \ref{sebemin}.

\subsection{An interlude on densities and weights}
\label{senote}

This work is concerned with transformation laws of various (mostly scalar) fields on celestial spheres under the BMS group, and in particular under celestial diffeomorphisms or conformal maps, so it is crucial to fix our conventions in that respect. What follows therefore spells out our definitions for Radon densities, volumic densities, conformal densities, conformal primaries, and the accompanying terminology.

A word of caution may be in order here. The abstract content of this subsection may seem off-putting to physicists, especially as we start with some measure-theoretic concepts that are not normally covered in the physics literature. However, to the extent that our first objective is to understand BMS transformations of the $\phi_n$'s in the expansion \eqref{expa}, it is essential to define an action of diffeomorphisms on densities when the underlying metric (\eg that of a celestial sphere) is not flat. This problem turns out to require a bit of mathematical machinery that, to our knowledge, has been overlooked in the literature on BMS; we attempt to fill that gap here, as it will even play a Carrollian role in section \ref{se34}. (For an introduction to measure theory, see \eg \cite{Royden}.)

\paragraph{Radon densities.} Consider an orientable $d$-dimensional manifold $\sM$ and endow it with a measure $\mu$.\footnote{For any measurable set $\sN\subseteq\sM$, one writes $\mu(\sN):=\int_\sN\dd\mu$.} The latter is equivalent to a volume form $\eps$, \ie a nowhere-vanishing top form; given local coordinates $x^a=(x^1,\cdots,x^d)$, any such form reads\footnote{We abuse notation slightly by using the same letter $\eps$ to denote both the volume form $\eps$ and its ``density function'' $\eps(x)$. We also write $\dd\mu$ for the volume form corresponding to the measure $\mu$. This notation is  common but rather abusive since this differential form is \textit{not} exact in general. }
\be
\label{murho}
\dd\mu
=
\eps
=
\eps(x)\,\dd^dx
\ee
where $\eps(x)$ is some strictly positive function on $\sM$ and $\dd^dx:=\dd x^1\wedge\cdots\wedge\dd x^d$. This applies in particular when $\sM$ carries a metric $g_{ab}$, in which case $\eps(x)=\sqrt{g(x)}:=\sqrt{\det g_{ab}(x)}$.

Now suppose some group acts on $\sM$ by orientation-preserving diffeomorphisms $x\to x'=F(x)$. In general, such transformations deform the measure in the sense that $F^*\eps\neq\eps$\,: the pullback by $F$ need not preserve the volume form. However, it is certainly true that $F^*\eps=\rho_F\,\eps$  (or, equivalently, in coordinates $\dd\mu\big(F(x)\big)=\rho_F(x)\,\dd\mu(x)$\,) for some positive function $\rho_F(x)$ known as the \itt{Radon-Nikodym derivative} of $\mu$ \cite[sec.\ 18.4]{Barut}, proportional to the Jacobian of $F$. It is indeed straightforward to compute $\rho_F$ for any diffeomorphism acting on a volume form: writing the diffeomorphism as $x\to F(x)$, eq.\ \eqref{murho} yields
\be
\label{RNN}
\rho_F(x)
=
\frac{\dd\mu\big(F(x)\big)}{\dd\mu(x)}
=
\frac{\eps\big(F(x)\big)}{\eps(x)}\,\bigg|\frac{\der F}{\der x}\bigg|\,,
\ee
where $|\der F/\der x|:=|\det(\der x^{\prime a}/\der x^b)|$ is the Jacobian of the map $x\to x'=F(x)$ in local coordinates. It is then straightforward to verify the cocycle property $\rho_{F_1\circ F_2}=F_2^*(\rho_{F_1})\,\rho_{F_2}$ \cite[sec.\ 3.2.3]{Oblak:2016eij}, either from the pullback definition $F^*\eps=\rho_F\,\eps$, or from the coordinate expression \eqref{RNN}. Again, all this applies to measures induced by metrics, in which case
\be
\rho_F(x)
=
\sqrt{\frac{g\big(F(x)\big)}{g(x)}}\,\,\bigg|\frac{\der F}{\der x}\bigg|\,.
\label{s12t}
\ee
We stress that the presence of the same metric $g$ in both the numerator and the denominator of \eqref{s12t} is not a typo: the metric was \itt{not} transformed with the usual transformation law under the diffeomorphism $x\to x'$ because the  measure density $\eps(x)=\sqrt{g(x)}$ is understood as being fixed in the present setting. In fact, had the metric been transformed following the usual formula
\be
\label{standardtransfo}
g'_{ab}(x')
=
\frac{\der x^c}{\der x'^a}\frac{\der x^d}{\der  x'^b}\,g_{cd}(x)\,,
\ee
the na\"ive candidate $\sqrt{g'(x')/g(x)}\,|\der x'/\der x|$ would have equalled one. By contrast, the actual Radon-Nikodym derivative \eqref{s12t} is typically a highly non-trivial function on $\sM$.

Having fixed the terminology, we are now ready to state the following definition: a \itt{Radon density} with weight $\tw$ on $\sM$ is a function $\phi(x)$ that transforms under any diffeomorphism $F:x\to x'$ as $\phi\to F\cdot\phi:=(\rho_{F^{-1}})^{\tw}\,\phi\circ F^{-1}$. More explicitly, writing $F\cdot\phi:=\phi'$ and using $\rho_{F^{-1}}\circ F=\rho_F^{-1}$, the transformation law reads
\be
\label{trabi}
\phi'(x')
=
\rho_F(x)^{-\tw}\,\phi(x)
\stackrel{\text{\eqref{s12t}}}{=}
\left(\frac{g(x)}{g(x')}\right)^{\tw/2}
\bigg|\,\det\left(\frac{\der x^{\prime a}}{\der x^b}\right)\,\bigg|^{-{\tw}}
\,\phi(x)\,.
\ee
Here we chose to write the Radon-Nikodym derivative \eqref{s12t} only for the special case where the measure on $\sM$ is inherited from a metric; this is by far the most relevant situation we will encounter below. In particular, it is essential for later reference to work out the action of vector fields on densities from the infinitesimal version of eq.\ \eqref{trabi}: letting $x'^a=x^a+\epsilon X^a(x)$ and defining $\delta\phi:=-\lim_{\epsilon\to0}\tfrac{1}{\epsilon}(\phi'-\phi)$, the expansion of \eqref{trabi} yields
\be
\label{deltabi}
\delta\phi
=
\cL_X\phi+\tw\,\nabla_aX^a\,\phi
\ee
where $\cL_X\phi:=X^a\der_a\phi$ and $\nabla_aX^a$ is the covariant divergence of the vector field $X$. More generally, for an arbitrary volume form $\eps$ on $\sM$, the factor $\nabla_aX^a$ would be replaced by the divergence $\text{div}(X)$ defined thanks to the Lie derivative
\be
\cL_X\eps
=
\text{div}(X)\,\eps\,.
\label{didef}
\ee
In coordinates, it follows from \eqref{murho} that $\text{div}(X)=\eps(x)^{-1}\partial_a\big(\eps(x)X^a(x)\big)=\partial_a X^a(x)+X^a(x)\,\partial_a\big(\ln\eps(x)\big)$. For a scalar field ($\tw=0$), eq.\ \eqref{deltabi} reduces to $\delta\phi=X^a\der_a\phi$. An example of scalar density with non-zero weight is provided by (super)translations, whose weight $\tw=-1/d$ is visible in the BMS bracket \eqref{liba}. More generally, we shall see in section \ref{sebemin} that eqs.\ \eqref{trabi}--\eqref{deltabi} provide the basic form of the BMS transformation law of the $\phi_n$'s in the expansion \eqref{expa}.

One readily verifies that the definition \eqref{trabi} furnishes a representation of the group of orientation-preserving diffeomorphisms of $\sM$, in the sense that $F_1\cdot(F_2\cdot\phi)=(F_1\circ F_2)\cdot\phi$. Equivalently, the commutator of two transformations of the form \eqref{deltabi} generated by $X,Y$ is itself of the form \eqref{deltabi} generated by the Lie bracket $[X,Y]$. The vector space carrying this representation is the set $C^{\infty}(\sM)$ of all smooth functions on $\sM$. In what follows we denote this space by $C^{\infty}_{\tw}(\sM)$ to stress the weight $\tw$; in particular, $C^{\infty}_0(\sM)=C^{\infty}(\sM)$ is the usual space of scalar fields on $\sM$. Note that the generalisation to tensor-valued densities is straightforward: simply add the appropriate Jacobian matrices on the right-hand side of \eqref{trabi}. In terms of infinitesimal transformations, eq.\ \eqref{deltabi} then remains valid with $\cL_X\phi$ the usual Lie derivative of a tensor field.

We stress that the transformation laws \eqref{trabi}--\eqref{deltabi} are the most important ones for our purposes, and for BMS generally: all densities in sections \ref{seCAR}--\ref{seHS} will be Radon densities in the sense defined here, unless explicitly stated otherwise. Accordingly, the hasty reader may skip the next few paragraphs and go straight to section \ref{serac}. For now, we present a few related notions of densities, both for completeness and for comparison with other patches of the literature.

\paragraph{Volumic densities.} By contrast with Radon densities, a \itt{(volumic) density} with weight $\tw$ is a scalar field $\varphi(x)$ whose transformation law reads\footnote{\label{remvol}The adjective ``volumic'' is very much non-standard. We use it only in this subsection, for the sole purpose of clearly distinguishing inequivalent concepts of densities: Radon, volumic, and conformal.}
\be
\label{trab}
\phii'(x')
=
\bigg|\,\det\left(\frac{\der x^{\prime a}}{\der x^b}\right)\,\bigg|^{-{\tw}}
\,\phii(x)\,.
\ee
More generally, a tensor-valued (volumic) density of weight $\tw$ is a tensor field whose usual transformation law under reparametrisations involves an extra Jacobian factor to the power $\tw$, exactly as in the scalar definition \eqref{trab}. The corresponding infinitesimal transformation law is given by \eqref{deltabi} except that the divergence of $X$ on the right-hand side now involves standard derivatives instead of covariant ones:
\be
\label{deltab}
\delta\phii
=\cL_X\phii+\tw\,\der_aX^a\,\phii\,,
\ee
where $\cL_X$ is the Lie derivative along $X$ acting on the tensor field $\phii$; for a scalar field ($\tw=0$) it reduces to $\delta\phii=X^a\der_a\phii$. Scalar volumic densities of weight $\tw=1$ on a manifold $\sM$ (not necessarily orientable) are the objects that can be integrated over $\sM$ in a coordinate-independent way.

It is clear from these definitions that volumic densities and Radon densities are closely related notions: their transformations \eqref{trabi} and \eqref{trab} are identical save for a slightly different convention in handling the metric-induced measure. This is why we denote their weights by the same symbol $\tw$. Note that the transformation law \eqref{trab} can be defined on \itt{any} manifold, with or without metric or measure. This is perhaps the reason why volumic densities---normally just called \itt{densities}$^{\text{\ref{remvol}}}$---are more familiar in the literature.

\paragraph{Conformal densities.} Independently of a density's behaviour under diffeomorphisms, one can also define a notion of weight under Weyl transformations. The transformation law of a (scalar or tensor) \itt{conformal density} $\psi$ of conformal weight $w$ under Weyl transformations is given by
\be
\label{traw}
g_{ab}(x)\to g'_{ab}(x)
=
\Omega^2(x)\,g_{ab}(x)\,,
\qquad
\psi'(x)\,
=
\,\Omega(x)^w\,\psi(x)\,.
\ee
This is the notion of weight mentioned below the GJMS operator \eqref{s5t}.

Note that a field may well be a volumic density and a conformal density simultaneously. For instance, the metric $g_{ab}$ is a tensor density of volumic weight zero (since it is a covariant tensor field whose transformation law under diffeomorphisms is \eqref{standardtransfo}) and conformal weight two (since its transformation law under Weyl transformations is \eqref{traw}). Similarly, the volume density $\sqrt{g}$ on a manifold of dimension $d$ is a scalar volumic density with weight $\tw=1$ in the sense of eqs.\ \eqref{standardtransfo} and \eqref{trab}, and conformal weight $w=d$ in the sense of \eqref{traw}.

\paragraph{Conformal primaries.} Finally, one may compose a Weyl transformation $g_{ab}\to\Omega^2 g_{ab}$ and a conformal map $x\to x'$ such that the metric remains unchanged: $g'_{ab}(x')=g_{ab}(x)$, thereby fixing the Weyl parameter to $\Omega(x)=|\det(\der x'/\der x)|^{1/d}$. As a result, the corresponding conformal transformation law of a scalar density $\chi$ with volumic weight $\tw$ (in the sense of \eqref{trab}) and conformal weight $w$ (in the sense of \eqref{traw}) reads
\be
x\to x',
\qquad
\chi(x)\to\chi'(x')=\bigg|\,\det\left(\frac{\der x^{\prime a}}{\der x^b}\right)\,\bigg|^{-\tw+w/d}\chi(x).
\ee
The same would hold for a tensor density up to additional Jacobian matrices. This is by definition the transformation law of a \itt{conformal primary} under a conformal transformation. In particular, on a conformally flat manifold, we say that $\chi$ has \itt{scaling dimension} $\Delta$ if it transforms under dilations as
\be
\label{transfolawdensityty}
x\to x'=\lambda x
\,,
\qquad
\chi'(x')\,=\,\lambda^{-\Delta}\,\chi(x)\,,
\ee
with $\lambda>0$. Note that the scaling dimension of a tensor field is not independent of its ranks and weights: if $\chi$ is $r$ times contravariant and $s$ times covariant, a volumic density of weight $\tw$, and a conformal density of weight $w$, then its scaling dimension is $\Delta=d\,\tw-w+s-r$.

For instance, the GJMS operator \eqref{s5t} is conformally covariant when it acts on a conformal density of weight $w=n-\tfrac{d}2$; in particular, on the conformally flat sphere $S^d$, such a density is a conformal primary of scaling dimension $\Delta=\tfrac{d}{2}-n$. Another example is provided by the fluctuation $h_{ab}$ of the metric around a conformally flat background, which has rank $s=2$ and is a primary field of scaling dimension zero (since $w=s=2$). 
On the celestial sphere, both the metric and its determinant are conformal primary fields of scaling dimension $\Delta=0$. We will soon encounter similar transformation laws for the terms of the scalar expansion \eqref{expa}; indeed, our notation in \eqref{transfolawdensityty} is consistent with that in \eqref{expa}, as
$\phi_0$ will turn out to have scaling dimension $\Delta$.

\subsection{The WRac and its avatars}
\label{serac}

Having reviewed the terminology of densities and weights, let us now return to the massless scalar field of section \ref{sescad}. Consider the simplest singular case $\Delta=\frac{d}{2}-1$ among the choices \eqref{t6t}, whereupon the boundary field
\be
\label{e55}
\phi_0(x)
:=
\lim_{r\to\infty}\big[\,r^{\frac{d}{2}-1}\Phi(r,u,x)\,\big]
\ee
is time-independent by virtue of \eqref{timeindep} and the perturbative form of d'Alembert's equation truncates immediately. There is then a single GJMS kernel condition \eqref{s6q} that applies to $\phi_0$ itself, and it states that $\phi_0$ lies in the kernel of the aforementioned Yamabe operator (\ie \eqref{s5t} with $n=1$):
\be
\label{yamyam}
\left(%
\nabla^2-\frac{d-2}{4(d-1)}\,{\cR}
\right)\phi_0=0 \,,
\ee
where $\cR=d(d-1)$ is the scalar curvature of the unit sphere $S^d$. This condition generally has no non-zero smooth solutions, so $\phi_0$ is singular at best. A virtue of eq.\ \eqref{yamyam}, however, is that it is invariant under Weyl transformations of the metric  provided $\phi_0$ is a conformal density with weight $w=-\Delta=1-d/2$. This will indeed turn out to be the case: see eq.\ \eqref{deph} below.

We shall return to the Yamabe equation \eqref{yamyam} in great detail in section \ref{symsrac}, but for now it is worth stressing some properties of the corresponding bulk field $\Phi$. Since the expansion \eqref{rexpa} truncates to a single term $\Phi(r,u,x)=r^{1-d/2}\phi_0(u,x)$, one may equally well investigate it at large $r$, finite $r$, or even small $r$. Actually, in Cartesian coordinates $x^{\mu}$, such singular configurations form a class of solutions of \eqref{daleq} that are homogeneous of degree $1-\frac{d}{2}$: they solve both $\Box\Phi=0$ and $(x^{\mu}\der_{\mu} +\tfrac{d}{2}-1)\Phi=0$.\footnote{Note that homogeneity means $(u\der_u+r\der_r +\tfrac{d}{2}-1)\Phi=0$ in Bondi coordinates, so the ansatz $\Phi(r,u,x)=r^{1-d/2}\phi_0(u,x)$ implies $\der_u\phi_0=0$ at $u\neq 0$, which is indeed satisfied by the WRac.} This agrees with the fact that the solutions $\phi_0$ of the Yamabe equation on $S^d$ admit an ``ambient'' description as massless scalars in Minkowski spacetime with a suitable homogeneity degree. (This goes back to Dirac \cite{Dirac:1936fq}; see also \cite[sec.\ 3]{Eastwood:2002su} or \cite[sec.\ 3.5]{Bekaert:2011js} for reviews.) In such cases, the celestial sphere $S^d$ is usually realised as a projectivised light-cone through the origin, \ie as the set of past- or future-oriented  null directions, seen as the base space of a fibre bundle
\be
\RR^{d+1}\backslash\{0\}\twoheadrightarrow S^d:\boldsymbol{x}\mapsto[\boldsymbol{x}]=\{\lambda\,\boldsymbol{x}\in\RR^{d+1}\,|\,\lambda>0\}\,.
\label{projcone}
\ee
It is geometrically appealing that one can equivalently see the celestial sphere as projectivised null infinity; this is really no surprise, as the past (resp.) future light-cone through the origin and future (resp. past) null infinity are related by an inversion $x^{\mu}\to x^{\mu}/x^2$ provided one restricts attention to the interior $x^2\leqslant 0$ (see fig.\ \ref{fiPen}).\footnote{\label{footman}To be precise, the inversion is actually discontinuous on the light-cone: time-like points near the past/future light-cone are mapped on the time-like part of future/past null infinities, whereas space-like points near the past/future light-cone are mapped on the space-like part of past/future null infinities.}

Note that this ambient interpretation is standard in higher-spin AdS/CFT, where the scalar singleton \cite{Dirac:1963ta}, usually called ``Rac'' \cite{Flato:1978qz}, is instrumental: it is the minimal unitary irreducible representation (\UIR) of $\mathfrak{so}(d,2)$, described from the CFT$_d$ perspective as a primary scalar field solving d'Alembert's equation on the boundary of AdS$_{d+1}$, with a scaling dimension that saturates the unitarity bound. (See \cite{Kobayashi:2001nq} for a review of the relevant representations, and \cite{Bekaert:2011js} for a review on singletons in AdS/CFT.) The ambient version of this construction consists in seeing the singleton as the space of homogeneous solutions of the Laplace equation on the embedding space $\RR^{d,2}$ with a ``two-time'' signature. The analogy with the singular field $\phi_0$ appearing in \eqref{yamyam} is immediate: the  field $\Phi$ defines a space of homogeneous solutions of d'Alembert's equation in $\RR^{d+1,1}$, so one is tempted to think of $\phi_0$ as the ``singleton'' of a higher-spin dS$_{d+1}$/ECFT$_d$ correspondence \cite{Anninos:2011ui} involving de Sitter space and a Euclidean CFT defined on the celestial sphere $S^d$. This reiterates our motivation expressed above to refer to $\phi_0$ as a ``Wick-rotated Rac''.

As explained in section \ref{sescad}, these remarks extend to singular solutions with $\Delta=\frac{d}{2}-N$ for integers $N>1$. The Yamabe condition \eqref{yamyam} is then replaced by GJMS conditions \eqref{s6q}, and the boundary fields $\phi_0,\phi_1,\cdots\!,\phi_{N-1}$ may be seen as Wick-rotated, now time-dependent, higher-order analogues of the Rac,\footnote{See \cite{Bekaert:2013zya} on the holographic link between higher-order Rac and partially-massless higher-spin gravity.} with respective weights ranging from $\tw=\Delta/d=\frac{1}{2}-N/d$ to $\tw=\frac{1}{2}-1/d$.

\subsection{The unitary Sachs module}
\label{sebemin}

In its strictest sense, the BMS group is the semi-direct product between the Lorentz group and the vector group of supertranslations. Lorentz transformations are thus seen as (globally well-defined) conformal transformations of celestial spheres \cite{Oblak:2015qia}, while supertranslations are scalar densities with volumic weight $\tw=-1/d$, in the sense of eqs.\ \eqref{trabi}--\eqref{deltabi}, acting on retarded time as angle-dependent shifts \cite{Sachs2}. We now show how this group of transformations affects the components of the asymptotic expansion \eqref{expa} of a massless scalar field satisfying the d'Alembert equation near null infinity. Superrotations \cite{Barnich:2009se} are initially discarded for simplicity, but we will eventually see that diffeomorphisms of celestial spheres may be included at no cost, showing that Sachs's module \cite{Sachs2} is really a unitary representation of the generalised BMS group of \cite{Campiglia:2014yka,Campiglia:2015yka}. Further extensions to ``\sproj'' transformations and beyond will be addressed in sections  \ref{seCAR}--\ref{seHS}.

\paragraph{BMS generators again.} BMS transformations are diffeomorphisms of spacetime: they are generated by vector fields that depend on an infinity of parameters, generalising the finite-dimensional set of Poincar\'e-generating vector fields \cite{Barnich:2010eb}. In Minkowski spacetime and in terms of Bondi coordinates, any such BMS vector field $\xi=\xi^{\mu}\der_{\mu}$ has components \eqref{xiir}--\eqref{xiia} with a completely arbitrary supertranslation function $\alpha(x)$. The Poincar\'e subalgebra consists of the same vector fields with the added constraint \eqref{acons}, whereupon supertranslations reduce to standard spacetime translations. The weight $\tw=-1/d$ of supertranslations under Lorentz transformations---and diffeomorphisms more generally---can then be read off from the Lie bracket \eqref{liba}, whose right-hand side involves transformations \eqref{deltabi} for $\alpha,\beta$.

Now act with a vector field of the form \eqref{xiir}--\eqref{xiia} on the radial expansion \eqref{expa}, with an arbitrary parameter $\Delta$, to read off the BMS transformation laws
\begin{align}
\label{deph}
\delta\phi_0
&=
\big(\alpha+\tfrac{u}{d}\,\nabla_a X^a\big)\dot\phi_0
+X^a\der_a\phi_0
+\tfrac{\Delta}{d}\,\nabla_a X^a\,\phi_0\,,\\
\delta\phi_n
&=
\big(\alpha+\tfrac{u}{d}\,\nabla_a X^a\big)\dot\phi_n
+X^a\der_a\phi_n
+\tfrac{\Delta+n}{d}\,\nabla_a X^a\,\phi_n\nn\\
\label{dephin}
&\quad
+
\big(%
\tfrac{u}{d}\,\nabla_aX^a+\tfrac{1}{d}\,\nabla^2\alpha
\big)
(\Delta+n-1)
\phi_{n-1}
-
g^{ab}
\big(\der_b\alpha+\tfrac{u}{d}\,\der_b\nabla_a X^a\big)
\der_a\phi_{n-1}\,,
\end{align}
where $n\geqslant1$. Recall that the fields $\phi_n$ ($n\geqslant0$) are in general functions of both $u$ and $x$. Eq.\ \eqref{deph} means for instance that $\phi_0$ has weight $\tw=\Delta/d$ under conformal maps, reducing to $\tw=1/2$ for $\Delta=d/2$: this will play a role below.

It is important, for future reference, to write down the finite coordinate transformations at null infinity generated by BMS vector fields \eqref{xiir}--\eqref{xiia}. Accordingly, let $F$ be a diffeomorphism of $S^d$ obtained from the flow of $X^a\der_a$ in \eqref{xiia}, and let $\alpha$ be a finite supertranslation function. Then the corresponding BMS transformation of $\sI$ is $(u,x)\to(u',x')$ with
\be
\label{coortaf}
u'=\rho_F(x)^{1/d}u+\alpha(x')
\qquad\text{and}\qquad
x'^{a}=F^a(x)\,,
\ee
where $\rho_F(x)$ is the Radon-Nikodym derivative \eqref{s12t} of the measure on $S^d$ under $F$. As we shall explain in sections \ref{se33}--\ref{se34}, this is actually a conformal transformation of null infinity as a whole when $F$ is a conformal map, and it is a generalised conformal transformation when $F$ is an arbitrary diffeomorphism. The field transformation $\phi_0\to\cU_{(F,\alpha)}\phi_0$ that accompanies this map can then be found by integrating eq.\ \eqref{deph} to obtain
\be
\label{bact}
\Big(\cU_{(F,\alpha)}\phi_0\Big)
\big(u',x'\big)
=
\rho_F(x)^{-\Delta/d}\,\phi_0(u,x)\,,
\ee
where the notation anticipates that we wish to think of $\cU_{(F,\alpha)}$ as a unitary operator. This equation is nothing but the transformation law \eqref{trabi} of a density with weight $\tw=\Delta/d$, up to the presence of an extra transformation of retarded time. We shall see shortly that unitarity occurs for $\tw=1/2$, \ie for the radiative fall-off condition \eqref{e52}. More generally, it is clear at this point that the radial expansion \eqref{expa} furnishes a representation of BMS, organised as a hierarchy of the form \eqref{deph}--\eqref{dephin}.\footnote{It would be amusing to find a $w_{1+\infty}$-like structure in the sequence \eqref{deph}--\eqref{dephin} when $d=2$, as has been achieved in gravity \cite{Strominger:2021mtt,Himwich:2021dau,Freidel:2021ytz}. We will not attempt to do that here.\label{winfty}}

\paragraph{Sachs module.} Let us now show that the group action \eqref{bact} with $\Delta=d/2$ furnishes a massless irreducible unitary representation of the BMS group. To achieve this we take a somewhat unexpected route: we start by recalling the construction of (massless) Poincar\'e representations in momentum space, then perform a Fourier transform along retarded time to recover eq.\ \eqref{bact} and argue that any massless Poincar\'e representation lifts to a representation of BMS, and even to a representation of the generalised BMS group where superrotations are allowed to be generic diffeomorphisms of celestial spheres \cite{Campiglia:2014yka,Campiglia:2015yka}. The argument is somewhat similar to that of \cite{Longhi:1997zt,Gomis:2015ata,Batlle:2017llu,Batlle:2022hwf}.

Consider a massless scalar \UIR\ of the Poincar\'e group in spacetime dimension $d+2$. The orbit (mass shell) is the future light-cone \eqref{projcone} of null momenta with positive energy: $\cO=\{(|\bq|,\bq)\in\RR^+\times\RR^{d+1}\}\cong\RR^+\times S^d$. The Hilbert space $L^2(\cO,\mu)$ consists of square-integrable wavefunctions with a Hermitian form\footnote{We write wavefunctions on $\cO$ as $\Phi$, $\Psi$, etc. Despite the similar notation, this has nothing to do with the bulk field $\Phi$ that solved the d'Alembert equation \eqref{daleq} in section \ref{sescad}.}
\be
\label{s1}
\langle\Phi|\Psi\rangle
:=
\int_{\RR^{d+1}}\frac{\dd^{d+1}\bq}{|\bq|}\,\Phi^*(\bq)\Psi(\bq)
\ee
where we chose the Lorentz-invariant measure for later convenience. The Poincar\'e group is represented unitarily on this Hilbert space, according to
\be
\label{t1}
\big(\,\cU_{(F,\alpha)}\Phi\big)\big(F(\bq)\big)
:=
e^{\,i\,\langle \,F(q)\,,\,\alpha\,\rangle}\,\Phi(\bq)\,,
\ee
where $F$ is a Lorentz transformation, $\alpha$ is a translation and $\langle q,\alpha\rangle:= q_{\mu}\,\alpha^{\mu}$ is the bilinear pairing (scalar product) of spacetime momenta and translations. Our goal is to rephrase eqs.\ \eqref{s1}--\eqref{t1} in terms of Bondi-like coordinates on the orbit $\cO$.

Accordingly, let $E=|\bq|>0$ be the energy of the momentum vector $\bq$ and let $\bx\in S^d\subset\RR^{d+1}$ be a unit vector such that $\bq=E\,\bx$, so that $q=(E,E\bx)$. Then the Hermitian form \eqref{s1} can be recast as
\be
\label{pp}
\langle\Phi|\Psi\rangle
=
\int_0^{\infty}\dd E\,E^{d-1}\int_{S^d}\dd^d\bx\,\sqrt{g(\bx)}\,\Phi^*(E,\bx)\Psi(E,\bx)
\ee
where $\dd^d\bx\,\sqrt{g(\bx)}$ is the standard volume form on the unit sphere $S^d$. Note that the wavefunctions $\Phi$ and $\Psi$ vanish for $E\leqslant 0$ since their support is the positive-energy zero-mass shell $\cO$. Now introduce a notion of ``retarded time'' $u$ thanks to the Fourier transform
\be
\label{def}
\phi(u,\bx)
:=
\int\limits_{-\infty}^{+\infty}
\tfrac{\dd E}{\sqrt{2\pi}}\,E^{\frac{d}{2}-1}\,e^{-iEu}\,\Phi(E,\bx)\,,
\qquad
\Phi(E,\bx)
=
\frac{1}{E^{\frac{d}{2}-1}}\int\limits_{-\infty}^{+\infty}\tfrac{\dd u}{\sqrt{2\pi}}\,e^{iEu}\,\phi(u,\bx)\,,
\ee
where the power $E^{\frac{d}{2}-1}$ will be justified shortly from an asymptotic argument. Setting this aside for now, eq.\ \eqref{def} allows us to rewrite the inner product \eqref{s1}--\eqref{pp} in a way that exactly coincides with Sachs's Hermitian form introduced in the early days of BMS symmetry \cite{Sachs2} (see also \cite[thm 12.3]{Carmeli:2000af}), namely
\be
\label{sachspo}
\langle\phi|\psi\rangle
:=
\langle\Phi|\Psi\rangle
=
i\int\limits_{-\infty}^{+\infty}\dd u\int\limits_{S^d}\dd^d\bx\,\sqrt{g(\bx)}\,
\,\phi^*(u,\bx)\,\der_u\psi(u,\bx)\,.
\ee
We stress that we have thus matched an inner product of quantum wavefunctions, in a Poincar\'e \UIR, with what was understood in \cite{Sachs2} as an inner product of radiative solutions \eqref{e52} of the d'Alembert equation. The matching goes further: the quantum expectation value of the Poincar\'e Hamiltonian in a state $|\phi\rangle$ is nothing but the energy \eqref{s5} of the corresponding classical field configuration \eqref{e52}.

The justification of the weighted Fourier transform \eqref{def} stems from the asymptotic behaviour of massless fields in Bondi coordinates---a property that was instrumental in the seminal works relating asymptotic symmetries to soft theorems \cite{He:2014laa} (see also \cite[Sec.\ 9, Ex.\ 4]{Strominger:2017zoo}). Indeed, consider the on-shell bulk field
\begin{align}
\label{defourier}
\Phi(r,u,\bx)
&=
\frac{1}{(2\pi)^{(d+1)/2}}
\int_{\RR^{d+1}}\frac{\dd^{d+1}\bq}{|\bq|}\,
\,e^{-i(u+r)\,|\bq|\,+\,ir\,\bx\cdot\bq}\,\Phi(\bq)
\\
\label{e555}
&=
\int\limits_{0}^{+\infty}\frac{\dd E}{\sqrt{2\pi}}\,E^{d-1}\,e^{-iEu}
\int_{S^d}\frac{\dd^d\hat\bq\,\sqrt{g(\hat\bq)}}{(2\pi)^{d/2}}
\,e^{iEr\,(\cos\theta-1)}\,\Phi(E,\hat\bq)
\,,
\end{align}
where $E=|\bq|>0$ is the norm of an integrated momentum and $\hat\bq:=\bq/|\bq|$ so that $\dd^{d+1}\bq=E^d\,\dd E\,\dd^d\hat\bq\,\sqrt{g(\hat\bq)}$, while $\theta\in[0,\pi]$ denotes the angle between $\bq$ and the spatial direction $\bx$. We stress that $\bq$ and $\bx$ are \textit{unrelated} at this stage. However, near null infinity $r\to\infty$, the stationary phase approximation of \eqref{defourier} results in the collinearity $\theta\approx0$, \ie $\bq\approx E\,\bx$ for non-zero $E$. The integral over the celestial sphere then satisfies the asymptotic behaviour
\begin{align}
\int_{S^d}\dd^d\hat\bq\,\sqrt{g(\hat\bq)}
\,e^{iEr\,(\cos\theta-1)}\,\Phi(E,\hat\bq)
&=
\int_0^\pi \dd\theta\, (\,\sin\theta\,)^{d-1}\int_{S^{d-1}}
\dd\Omega\,e^{iEr\,(\cos\theta-1)}
\,\Phi(E,\hat\bq)\nonumber\\
&\sim
\int_0^\pi\dd\theta\,
\theta^{d-1}\int_{S^{d-1}}
\dd\Omega\,e^{-iEr\theta^2/2}
\,\Phi(E,\bx)
%\,\Big(1+{\cal O}(1/r)\Big)
\nonumber\\
&\propto
\frac{1}{(Er)^{d/2}}
\,\Phi(E,\bx)
%\,\Big(1+{\cal O}(1/r)\Big)
\end{align}
where we factorised the measure on $S^d$ as $\dd^d\hat\bq\,\sqrt{g(\hat\bq)}=(\sin\theta)^{d-1}\dd\theta\,\dd\Omega$, with $\dd\Omega$ the standard measure on $S^{d-1}$. In the last line we also neglect irrelevant constant factors (such as the volume of $S^{d-1}$) that can be eliminated by a redefinition of $\Phi$. Thus eq.\ \eqref{e555} yields
\be
\label{deffinal}
\Phi(r,u,\bx)
\,\stackrel{r\to\infty}{\sim}\,
r^{-\frac{d}2}\int\limits_{-\infty}^{+\infty}\tfrac{\dd E}{\sqrt{2\pi}}\,
E^{\tfrac{d}2-1}
\,e^{-iEu}\,\Phi(E,\bx)
=
r^{-d/2}\phi(u,x)\,,
\ee
which is to say that the asymptotic behaviour of the field $\Phi(r,u,\bx)$ in position space identifies with the weighted Fourier transform of the wavefunction $\Phi(E,\bx)$ in momentum space. Up to the factor $r^{-d/2}$, this is exactly the Fourier transform introduced in \eqref{def}. (As before, we let $\Phi(E,\bx)$ be supported in the region $E>0$.)

Let us now return to wavefunctions with the Hermitian form \eqref{s1}. Since all Poincar\'e transformations act unitarily on their Hilbert space, let us rewrite their action in terms of coordinates $(u,x)$, and identify the transformation of a wavefunction with that given by \eqref{bact} for a field at infinity. We start from the unitary operator \eqref{t1}, now applied to the wavefunction \eqref{def}, which yields
\be
\big(\cU_{(F,\alpha)}\,\phi\big)(u,\bx)
=
\frac{1}{\sqrt{2\pi}}\int_0^{\infty}\dd E\,E^{\frac{d}{2}-1}\,e^{-iEu}\,
e^{i\,\langle\,(E,E\bx)\,,\,(\alpha^0,\boldsymbol\alpha)\,\rangle}\,\,\Phi\big(F^{-1}(E,\bx)\big)\,.
\label{ouf}
\ee
Here the argument of $\Phi$ on the right-hand side involves the action of a Lorentz transformation $F$ on a point $(E,\bx)$ on the light-cone. Specifically, $F^{-1}(E,\bx)=\big(E\,f(\bx),F^{-1}(\bx)\big)$, where $\bx\to F^{-1}(\bx)$ is a conformal transformation of $S^d$ and $f(\bx)$ is a positive function on $S^d$ that can be found thanks to the fact that the measure $E^{d-1}\dd E\,\sqrt{g(\bx)}\,\dd^d\bx$ is Lorentz-invariant. Indeed, one has $f(x)=\big(\rho_{F^{-1}}(x)\big)^{-1/d}$ in terms of the Radon-Nikodym derivative \eqref{s12t}. Furthermore, the exponent in the integrand of \eqref{ouf} is $\langle (E,E\bx),(\alpha^0,\boldsymbol\alpha)\rangle=E\alpha^0+E\,\bx\cdot\boldsymbol{\alpha}:= E\,\alpha(\bx)$, where $\alpha(\bx)$ is the supertranslation function on the celestial sphere implementing the spacetime translation $\alpha$. Using all this in \eqref{ouf} gives
\be
\label{eq33}
\big(\cU_{(F,\alpha)}\,
\phi\big)(u,\bx)
=
\frac{1}{\sqrt{2\pi}}
\int_0^{\infty}\dd E\,E^{\frac{d}{2}-1}\,e^{-iEu}\,
e^{iE\alpha(\bx)}
\,\Phi\big(E\,f(\bx),F^{-1}(\bx)\big)\,.
\ee
The coordinates $\bx$ are mere spectators here, so one can change the integration variable into $\tilde E:= E\,f(\bx)$ to find
\begin{align}
\label{e53}
\big(\cU_{(F,\alpha)}\,\phi\big)(u,\bx)
&=
\big(\rho_{F^{-1}}(\bx)\big)^{1/2}
\phi\Big(\big(\rho_{F^{-1}}(\bx)\big)^{1/d}\big(u-\alpha(\bx)\big),F^{-1}(\bx)\Big)\,.
\end{align}
This is it: upon evaluating the left-hand side at $(u',x')$ given by \eqref{coortaf} instead of $(u,x)$, eq.\ \eqref{e53} coincides with the scalar field's transformation law \eqref{bact} with $\Delta=d/2$. In other words, the radiative branch \eqref{e52} of solutions the d'Alembert equation, endowed with the Sachs form \eqref{sachspo} and acted upon by BMS transformations according to eq.\ \eqref{bact} with $\Delta=d/2$, is equivalent to a scalar massless \UIR\ of the Poincar\'e group. Conversely, any such \UIR\ lifts to a corresponding \UIR\ of BMS, since eqs.\ \eqref{bact}--\eqref{e53} make sense for any supertranslation $\alpha$ and define a representation that leaves the Sachs form invariant.\footnote{More generally, \itt{all} faithful \UIR{}s of the Poincar\'e group ISO(3,1) lift to \UIR{}s of BMS$_4$. Equivalently, all \UIR{}s of BMS$_4$ induced from BMS little groups corresponding to Poincar\'e little groups of faithful \UIR{}s of ISO(3,1) remain irreducible upon restriction. The proof is a corollary of \cite[Th.\ 7]{McCarthy489}, which states that the branching rules for the restriction from BMS to Poincar\'e are obtained from the branching rules of the corresponding little groups. Note that this result may seem to contradict earlier statements in \cite{Mccarthy:1972ry,McCarthy01,McCarthy00,McCarthy317}, but the latter rely on overly restrictive choices of topology for the functional space of supertranslations, whereas the ``nuclear'' topology of \cite{McCarthy489} is better suited for physical applications.} (See also \cite{Longhi:1997zt,Gomis:2015ata,Batlle:2017llu} for massive scalars.) Perhaps more strikingly, the lift holds regardless of the restriction to conformal transformations: it even applies when Lorentz transformations are enhanced to arbitrary diffeomorphisms of $S^d$. Indeed, eq.\ \eqref{bact} always defines a representation of the extended group $\text{Diff}(S^d)\ltimes C_{\tw}^{\infty}(S^d)$, where the first factor consists of ``superrotations'' in the sense of \cite{Campiglia:2014yka,Campiglia:2015yka,Colferai:2020rte} and $\tw=-1/d$ for supertranslations. Unitarity with respect to the Sachs form \eqref{sachspo} then holds provided $\Delta=d/2$. We shall return to this example in great detail in sections \ref{se34} and \ref{symsSachs}.

\paragraph{The non-unitary WRac.} We conclude this section by briefly returning to the WRac of section \ref{serac}, given by \eqref{e55} in terms of the bulk field. It is then straightforward to verify that the space of such (singular) fields also carries a representation of the BMS group, simply by exponentiating the transformation law \eqref{deph} to obtain \eqref{bact} with $\Delta=\frac{d}{2}-1$. This representation, however, is not faithful, since $\phi_0(x)$ is now time-independent so that all (super)translations act trivially. Furthermore, the lack of smoothness and the weight $\tw=\Delta/d=\frac{1}{2}-1/d$ prevent the existence of a natural Hermitian form on such fields that would be BMS-invariant (or even Poincar\'e-invariant, for that matter). We will nevertheless return to this singular construction in section \ref{symsrac}, as it is a natural cousin of the singleton in AdS/CFT.

As a final comment, note that the saddle-point analysis of eqs.\ \eqref{defourier}--\eqref{deffinal} na\"ively suggests that all massless scalars satisfy the fall-offs $\Phi=\cO(r^{-d/2})$, apparently ruling out the overleading fall-offs of the WRac. The way out of this paradox can be found by going back to eq.\ \eqref{e555}, which shows that saddle point of the Fourier transform in momentum space occurs \textit{either} for collinear momenta, \textit{or} at zero energy. It so happens that the WRac realizes this second possibility, in accordance with the time-independence ensured by the equation of motion $\dot\phi_0=0$ in \eqref{timeindep}. (By the way, this is true not only of the WRac, but of any massless field that satisfies $\Phi=\cO(r^{-\Delta})$ with $\Delta\neq d/2$.)

\section{Carrollian geometry and BMS groups}
\label{seCAR}

Our approach has been pragmatic so far: starting from the massless field equation \eqref{daleq}, we found that its space of radiative solutions carries a unitary representation of the BMS group. In order to relate this structure to higher-spin transformations, we now adopt a more intrinsic perspective based on Carrollian geometry. The latter may be seen as an ultra-relativistic limit of Lorentzian geometry when the speed of light vanishes and the  metric becomes degenerate due to the appearance of one vanishing eigenvalue. This motivates the definition of automorphisms analogous to ``isometries'' or ``conformal maps'', except that they must preserve this null direction. Accordingly, we now review the description of BMS transformations as Carrollian conformal maps, following the intrinsic (\ie purely from the boundary) and geometric (\ie global and coordinate-free) approach of Penrose \cite{Penrose:1965am}, translated in Carrollian language in \cite{Duval:2014uva}. We start with generalities on principal $\RR$-bundles, their automorphisms and connections; we then introduce Carrollian spacetimes, isometries and conformal maps; finally, we define Carrollian volume forms and their symmetries to revisit the Sachs module \eqref{sachspo} from a Carrollian viewpoint.

We stress that the language reviewed here will be crucial in section \ref{seHS}, as it simplifies computations whose expression in coordinates is otherwise unwieldy. Still, the presentation will unavoidably be quite mathematical, so we have attempted to frame it in a pedagogical manner accessible to theoretical physicists. It is self-contained, save for basic notions of differential geometry and fibre bundles that are not reviewed in any depth; we refer \eg to \cite{Steenrod,Lee,Nakahara} for the necessary background. A pedagogical introduction to Carrollian conformal geometry is also provided \eg in \cite{Ciambelli:2019lap,Herfray:2021qmp}. Our terminology and notation will mostly follow \cite[appendix A]{Bekaert:2015xua}.

\subsection{Principal \texorpdfstring{$\boldsymbol\RR$}{RR}-bundles and their automorphisms}
\label{sebundle}

Here we review elementary concepts, needed later, on principal $\RR$-bundles. We start from the notion of fundamental vector field to define projectable, \sproj, and invariant vector fields as generators of suitable automorphisms. (These families of vector fields will provide the basic building blocks of higher-spin differential operators in section \ref{seHS}.) We conclude by defining basic and invariant tensor fields, crucial for the later definition of Carrollian clocks (section \ref{se32}), metrics (section \ref{se33}) and volumes (section \ref{se34}).

\begin{figure}[t]
\centering
\includegraphics[width=.33\textwidth]{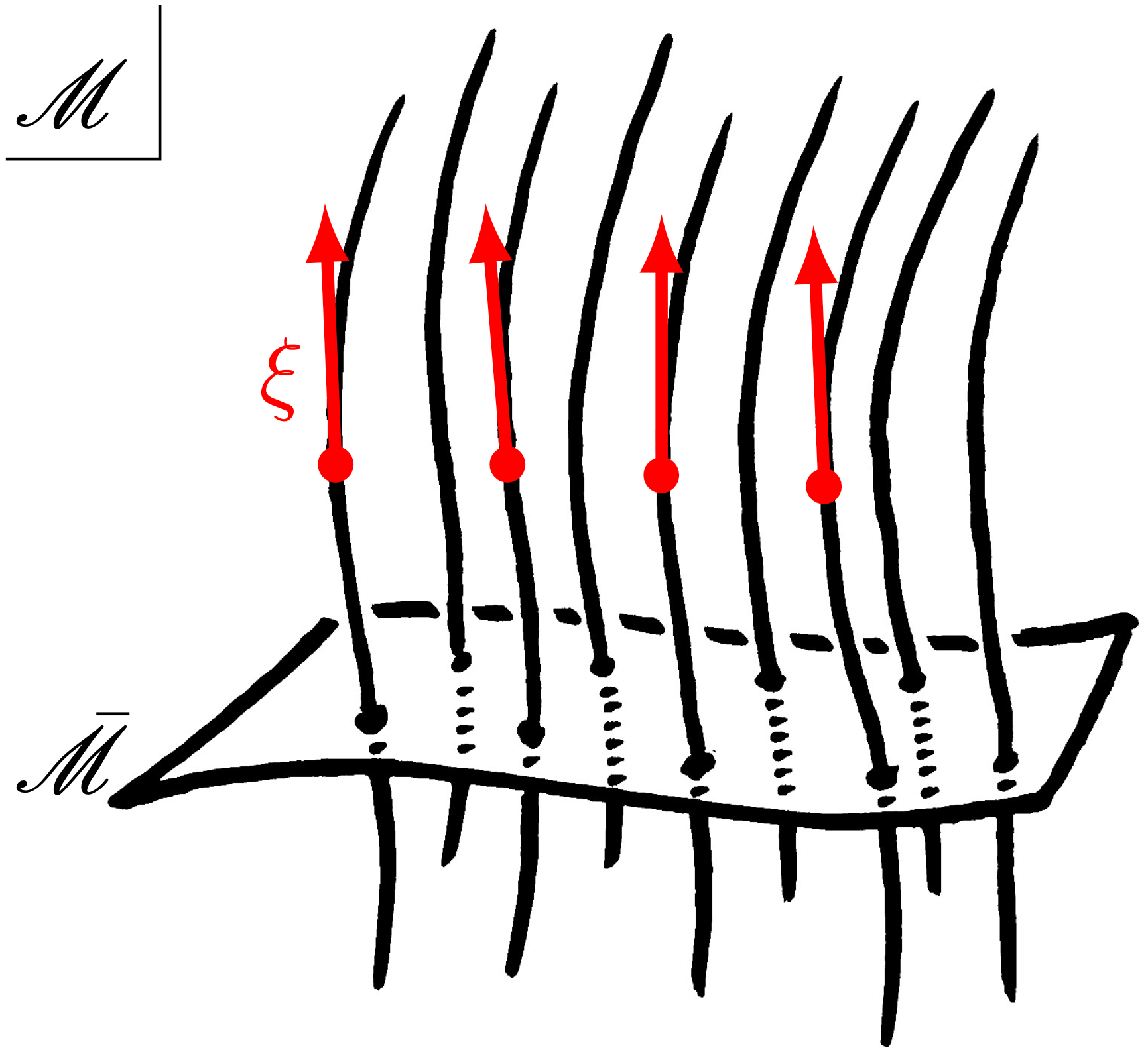}
\caption{A principal $\RR$-bundle $(\sM\twoheadrightarrow\bar\sM,\xi)$. Each curve is a fibre, and all points of that fibre are mapped by the projection \eqref{s14} on a single point on the base $\bar\sM$. The bundle $\sM$ is locally isomorphic to a product $\bar\sM\times\RR$, but in the case at hand this is even true globally, thanks to the global existence of a smooth $(\RR,+)$ group action.}
\label{finew}
\end{figure}

\paragraph{Ray bundle and fundamental vector field.} The structure underlying any Carrollian geometry is a principal $\RR$-bundle---a \itt{ray bundle}---whose $(d+1)$-dimensional total space $\sM_{d+1}$ is a Carrollian ``spacetime'' while the $d$-dimensional base manifold $\bar\sM_d:=\sM_{d+1}/\RR$ is Carrollian ``space''. We let
\be
\pi\,:\,\sM_{d+1}\twoheadrightarrow\bar\sM_d
\label{s14}
\ee
be the projection from the bundle to its base, such that the preimage $\pi^{-1}(x)\cong \RR$ of any $x\in\bar\sM$ is a fibre in $\sM$. The additive group $\RR$ acts freely (and properly\footnote{Requiring the $\RR$-action to be ``proper'' ensures that the quotient $\sM/\RR$ is a manifold. This rules out pathological cases such as a torus acted upon by $\RR$ via translations along an irrational slope.}) on $\sM$ by \itt{Carrollian time translations} whose orbits are fibres; each fibre may thus be seen as the worldline of a test mass (see fig.\ \ref{finew}). Note that our convention for dimensions matches that of section \ref{seSCAL}: the $d$-dimensional celestial sphere $\bar\sM_d=S^d$ is the base manifold of $(d+1)$-dimensional null infinity $\sM_{d+1}=\sI_{d+1}^{\pm}\cong S^d\times\RR$, which is indeed Carrollian. In what follows we will lighten notation by omitting the dimension subscript in $\sM_{d+1}$ and $\bar\sM_d$.

Given the principal $\RR$-bundle $\sM\twoheadrightarrow\bar\sM$, its \itt{fundamental vector field} is the vector field $\xi\in\mX(\sM)$ whose flow consists of fibrewise translations; its integral curves are orbits of the $\RR$-action on $\sM$, \ie Carrollian worldlines. In particular, $\xi$ vanishes nowhere, and it is vertical in the sense that $\pi_*(\xi)=0$ in terms of the pushforward of the projection \eqref{s14}. Conversely, choosing a nowhere-vanishing vertical vector field $\xi$ on a fibre bundle $\pi:\sM\twoheadrightarrow\bar\sM$ with fibres $\pi^{-1}(x)\cong\RR$ is equivalent to choosing a free action of $\RR$ on $\sM$, and automatically endows $\sM$ with the structure of a principal $\RR$-bundle. The data defining a principal $\RR$-bundle will therefore be denoted as a pair $(\sM,\xi)$ from now on.

Locally, one can always find adapted coordinates $(u,x^a)$ on $\sM$ with $u\in\RR$ and $x^a$ ($a=1,\ldots\!,d$) coordinates on (a patch in) $\bar\sM$, such that Carrollian time translations read $u\to u+\text{cst}$. The corresponding fundamental vector field is $\xi=\der_u$, with integral curves labelled by their initial point $x^a=x^a_0$ and parametrised by null time $u$. The projection \eqref{s14} then simply reads 
\be
\label{fibration}
\pi
\,:\,
\sM\twoheadrightarrow\bar\sM
\,:\,
(u,x^a)\mapsto x^a\,.
\ee
Note that any principal $\RR$-bundle is trivial, \ie it is globally true that $\sM\cong\bar\sM\times\RR$ as fibre bundles, and the fibre coordinate $u$ is always globally well-defined.\footnote{This crucially relies on the \itt{principal} bundle structure, requiring the existence of a globally well-defined action of $\RR$ on $\sM$. It would not be true for arbitrary bundles with fibres homeomorphic to $\RR$ (think \eg of a M\"obius band, which is crucially \itt{not} a principal $\RR$-bundle).} In this sense, it is often simpler to just work in a chart that makes the splitting between $\bar\sM$ and $\RR$ manifest, as in the Bondi coordinates of section \ref{seSCAL}. Still, sticking to intrinsic geometry (as opposed to local charts) reveals the bundle structure underlying Carrollian manifolds. This will simplify some tedious computations involving differential operators in section \ref{seHS}. In what follows we therefore go back and forth between local coordinates and global statements.

\paragraph{Examples: null infinity and light-cone.} Most importantly for our purposes, future and past null infinities $\sI^\pm\cong S^d\times\RR$, on the conformal boundary of compactified Minkowski spacetime $\RR^{d+1,1}$, are typically described in terms of a retarded/advanced time coordinate $u\in\RR$ along with angular coordinates $x^a$ on a celestial sphere $S^d$ (see fig.\ \ref{fiPen} above). Both are thus trivial $\RR$-bundles over $S^d$ with a projection \eqref{fibration} in terms of Bondi coordinates. The fundamental vector field $\xi=\der_u$ generates time translations $u\to u+\text{cst}$, so its congruence of integral curves spans a cone with its tip at $i^{\pm}$ removed.

Another aforementioned example of Carrollian structure is provided by light-cones (say at the origin) in Minkowski spacetime, which may be seen as images of null infinity by the inversion $x^{\mu}\to x^{\mu}/x^2$. Each light-cone $\sN\cong S^d\times\RR$ is thus a principal bundle \eqref{projcone} whose trivial structure is obvious. Furthermore, it can be used as the celebrated M\"obius model of conformally flat geometry, \ie the projective null cone identified with the celestial sphere, seen as a conformally flat manifold. (See \eg \cite[sec.\ 3]{Eastwood:2002su} or \cite[sec.\ 2.2]{Bekaert:2011js} for reviews.) This makes conformal symmetry manifest since Lorentz transformations of $\RR^{d+1,1}$ induce conformal transformations of $S^d$, as we used below eq.\ \eqref{ouf}. We shall return to conformal (BMS) maps on Carrollian manifolds in section \ref{se33}, but this first requires that we define families of vector fields generating suitable symmetries of $\RR$-bundles.

\paragraph{(Super-)projectable and invariant vector fields.} The bundle structure \eqref{s14} and the fundamental vector field $\xi$ readily motivate the definition of several subspaces of vector fields on $\sM$; these will be crucial in section \ref{seHS} to build differential operators spanning higher-spin symmetries. Accordingly, a vector field $X\in\mX(\sM)$ is \itt{projectable} if $\cL_{\xi}X$ is vertical, \ie if
\be
\label{procond}
\cL_{\xi}X=f\,\xi
\ee
for some $f\in C^{\infty}(\sM)$. It is \itt{\sproj}\ if $\cL_{\xi}f=0$, and it is \itt{invariant} if $f=0$. All three such subsets of vector fields span Lie algebras, respectively denoted $\mX_{\text{pro}}(\sM)$, $\mX_{\text{spro}}(\sM)$ and $\mX_{\text{inv}}(\sM)$, with the obvious inclusions
\be
\mX_{\text{inv}}(\sM)\subset\mX_{\text{spro}}(\sM)\subset\mX_{\text{pro}}(\sM)\subset\mX(\sM)\,.
\label{s19b}
\ee
Now for the geometric interpretation. First, the projectability condition \eqref{procond} ensures that the flow of $X$ maps fibres on fibres, \ie that $X$ generates automorphisms of $\sM\twoheadrightarrow\bar\sM$, seen as a fibre bundle without any extra structure. In particular, any projectable vector field $X$ has a well-defined projection $\bar X:=\pi_*(X)$ on $\bar\sM$. The super-projectability condition $\cL_\xi f=0$ is somewhat tricky to justify for now, but we will return to it shortly in local coordinates; we will also see in sections \ref{se33}--\ref{se34} that it specifies the class of vector fields that includes all conformal/BMS transformations and their generalisations. Finally, the invariance condition $\cL_\xi X=0$ states that $X$ generates automorphisms of $\sM\twoheadrightarrow\bar\sM$, seen as a \itt{principal} $\RR$-bundle.

In coordinates $(u,x^a)$ such that $\xi=\der_u$, any projectable vector field reads $X=X^a(x)\der_a+g(u,x)\der_u$ where $g(u,x)=\alpha(x)+\int_0^u\dd v\,f(v,x)$ and the functions $X^a(x)$, $\alpha(x)$, $f(u,x)$ are arbitrary. It is \sproj\ if $f(u,x)$ is independent of $u$, \ie if
\be
X=X^a(x)\,\der_a+\big(\alpha(x)+uf(x)\big)\der_u\,,
\label{sprovec}
\ee
and it is invariant if all its components are time-independent (\ie $f=0$). Note the similarity between eq.\ \eqref{sprovec} and the $(u,a)$ components \eqref{xiiu}--\eqref{xiia} of the BMS vector field, whose $r\to\infty$ limit is clearly \sproj\ on null infinity: this was actually the reason for introducing \sproj\ vector fields in the first place.

\begin{figure}[t]
\centering
\begin{subfigure}{0.25\textwidth}
  \centering
  \includegraphics[width=\linewidth]{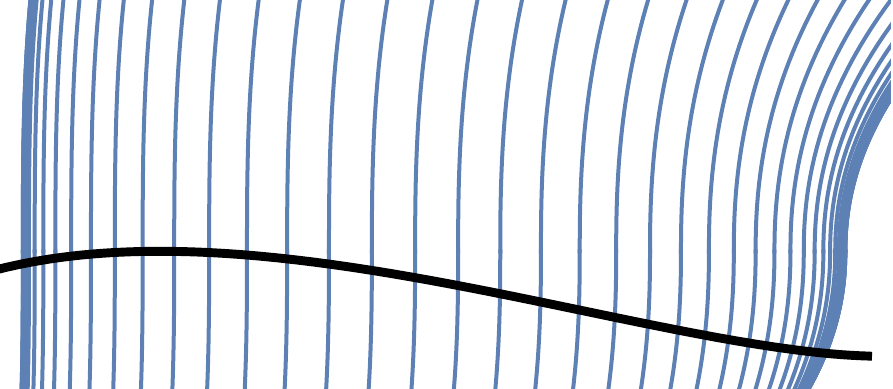}
\end{subfigure}%
$~\xleftarrow{~~~~~~}~$%
\begin{subfigure}{0.25\textwidth}
  \centering
  \includegraphics[width=\linewidth]{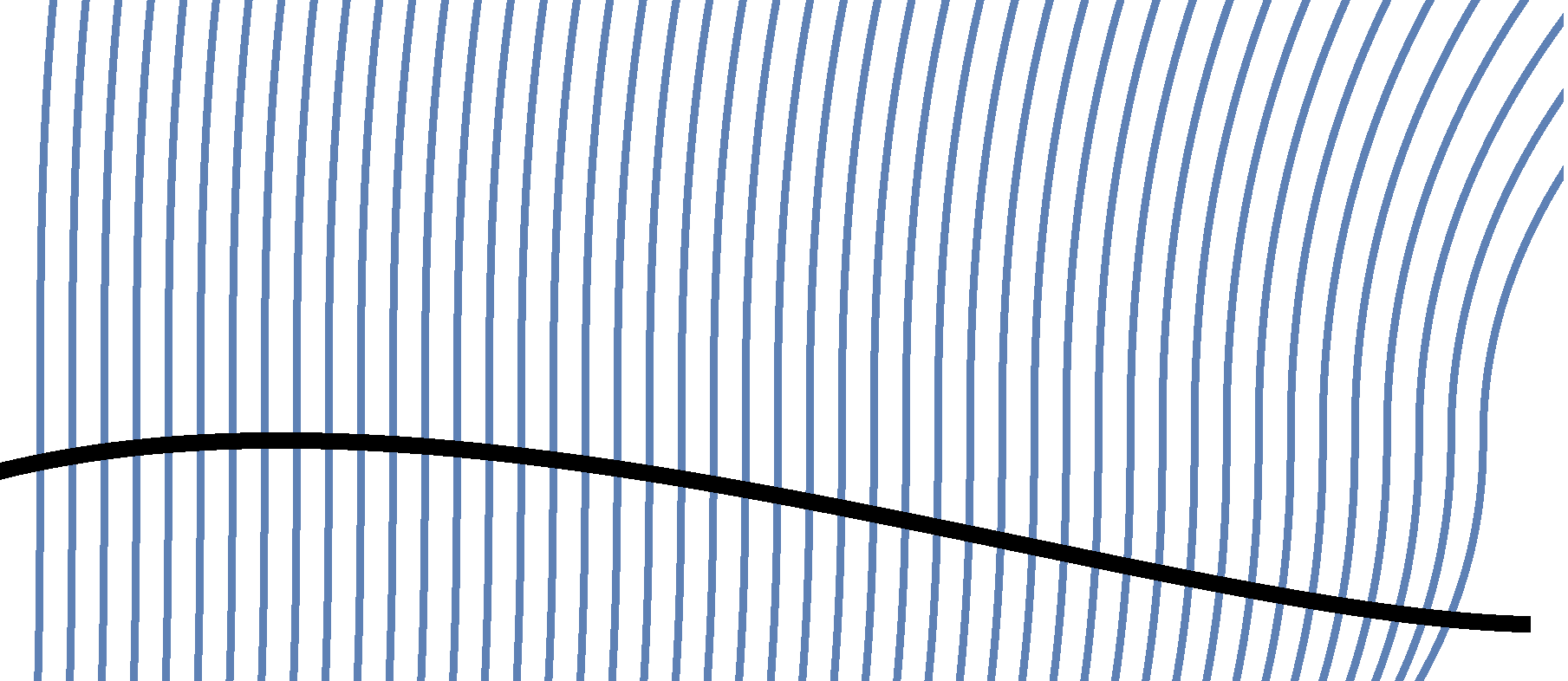}
\end{subfigure}
$~\xrightarrow{~~~~~~}~$%
\begin{subfigure}{0.25\textwidth}
  \centering
  \includegraphics[width=\linewidth]{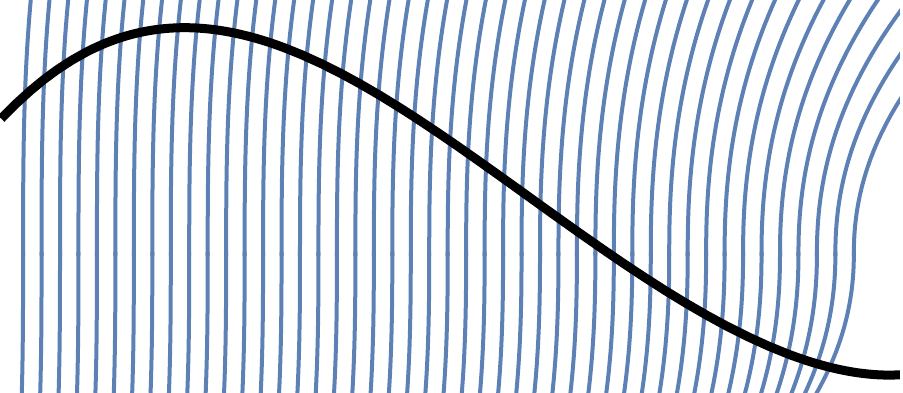}
\end{subfigure}
\caption{Given the bundle $\sM\twoheadrightarrow\bar\sM$, any bundle automorphism maps $\bar\sM$ to itself smoothly, but generally shuffles its points so that different fibres are mapped on one another (left). Furthermore, automorphisms locally change the origin of the Carrollian time coordinate (right). They also affect the local normalisation of the fundamental vector field (not shown here). In the BMS context, local translations along fibres are supertranslations, while superrotations are automorphisms projecting onto conformal maps of the base space: see sections \ref{se33}--\ref{se34}.}
\label{FiCarroll}
\end{figure}

Note that we could have \itt{started} from the definition of bundle automorphisms as maps that preserve the fibre bundle structure $\pi:\sM\twoheadrightarrow\bar\sM$; each such map is a diffeomorphism $F:\sM\to\sM$ such that there exists a base diffeomorphism $\bar F:\bar\sM\to\bar\sM$ for which
\be
\bar F\circ\pi=\pi\circ F\,,
\label{fipi}
\ee
ensuring that $F$ maps fibres on fibres (see \eg \cite{Steenrod}). In adapted coordinates $(u,x^a)$, it takes the form of a smooth transformation
\be
\label{e68}
(u,x)\to F(u,x)=\big(u'(u,x),x'(x))
\ee
where $x\to x'(x)$ is any diffeomorphism of the base space $\bar\sM$ and $u'$ is any function of $(u,x)$; see the examples in fig.\ \ref{FiCarroll}. Vector fields whose flow consists of such maps are precisely the projectable ones defined above, while \sproj\ vector fields generate diffeomorphisms \eqref{e68} for which $u'=f(x)u+\alpha(x)$ with $f(x)>0$. Similarly, we could have defined \itt{principal bundle} automorphisms as bundle automorphisms $F:\sM\to\sM$ that are equivariant with respect to the action of $\RR$ on $\sM$ and thus leave the fundamental vector field invariant in the sense that $F_*(\xi)=\xi$. In the notation \eqref{e68}, one then has $u'=u+\alpha(x)$. The corresponding vector fields are invariant, as defined above.

The reason we avoided this route (from finite transformations to their infinitesimal counterparts) is due to the standard recipe for higher-spin extensions of symmetries, which consists in first considering infinitesimal spacetime symmetries realised as first-order differential operators, then allowing for higher-order operators by turning to the enveloping algebra. It is indeed this approach that will be pursued in section \ref{seHS}, and it exclusively involves infinitesimal generators in Lie algebras as opposed to finite transformations in Lie groups.

\paragraph{Basic and invariant tensor fields.} The notion of invariance, and other similar constraints, is readily extended to covariant tensor fields (\eg differential forms) instead of vectors. Accordingly, a covariant tensor field $T_{\mu\nu\cdots}$ on $\sM$ is called \itt{invariant} if $\cL_{\xi}T=0$, \itt{horizontal} if any of its contractions with $\xi$ vanishes, and \itt{basic} if it is both invariant and horizontal. In this last case, $T$ is the pullback by the projection \eqref{s14} of a tensor $\bar T$ on the base manifold $\bar\sM$; in other words, $T$ is basic iff $T=\pi^*\bar T$.

For example, a function $f$ on $\sM$ is invariant if $\cL_{\xi}f=0$; in local coordinates $(u,x^a)$, it is any time-independent function $f(x)$. Similarly, a one-form $A$ on $\sM$ is invariant if its components in adapted coordinates are $u$-independent (\ie if $A=f(x)\,\dd u+A_a(x)\,\dd x^a$); it is horizontal if its $u$ component vanishes, so basic one-forms read $A=A_a(x)\dd x^a$ in local coordinates. Other examples of basic tensor fields will be provided by Carrollian metrics in section \ref{se33}.

\subsection{Ehresmann connections as Carrollian clocks}
\label{se32}

The concepts presented so far in section \ref{seCAR} apply to any principal $\RR$-bundle, without any Carrollian spacetime interpretation save for time translations. We now push this intuition much further by defining Carrollian clocks and horizontal surfaces, all based on a choice of Ehresmann connection on $\sM$. This eventually leads to the notion of Euler vector fields, and it will allow us to write the algebras \eqref{s19b} as semi-direct sums that start to resemble BMS. The spacetime picture will later be further completed by the addition of a Carrollian metric in section \ref{se33}. We refer \eg to \cite{Bekaert:2015xua,Ciambelli:2019lap,Petkou:2022bmz} for more on the natural appearance of Ehresmann connections in Carrollian geometry.

\paragraph{Carrollian clocks.} Let $\sM\twoheadrightarrow\bar\sM$ be a principal $\RR$-bundle with fundamental vector field $\xi$, and let $p\in\sM$ be a point. Then a non-zero vector $v$ tangent to $\sM$ at $p$ is \itt{time-like} if it is vertical, \ie if $v\propto\xi_p$; otherwise $v$ is \itt{space-like}. Thus the space of time-like vectors has measure zero within the set of all tangent vectors on a Carrollian manifold, which should be contrasted with Lorentzian manifolds where time-like vectors are as ``generic'' as space-like ones.

Now, one typically wishes to have a global prescription splitting any tangent vector, anywhere in $\sM$, as the sum of a time-like piece and a space-like ``horizontal'' one. This can be achieved with an Ehresmann connection on $(\sM,\xi)$, \ie an invariant one-form $A\in\Omega^1(\sM)$ such that $i_{\xi}A=1$ everywhere. In the Carrollian context, such a connection may be called a \itt{Carrollian clock}, since the normalization $i_\xi A=1$ suggests the interpretation of $A$ as a field of clocks ``dual'' to the fundamental vector field $\xi$ that generates time translations. In adapted coordinates $(u,x^a)$ such that $\xi=\der_u$, any clock reads $A=\dd u+A_a(x)\dd x^a$ with arbitrary time-independent components $A_a(x)$.

The standard terminology of connections and their curvature carries over to the Carrollian realm: a clock is \itt{locally synchronised} if the connection $A$ is flat ($\dd A=0$), and it is \itt{globally synchronised} if $A=\dd f$ is exact. Note that the curvature $\dd A$ is a basic two-form.\footnote{$i_\xi(\dd A)=\cL_\xi A=0$ and $\cL_\xi(\dd A)=\dd i_\xi\dd A=0$ thanks to Cartan's magic formula $\cL_\xi=\dd\circ i_{\xi}+i_{\xi}\circ\dd$.} This is actually manifest in adapted coordinates, where $\dd A=\der_{[a}A_{b]}(x)\,\dd x^a\wedge\dd x^b$ only involves the base form $A_a(x)\dd x^a$. In this sense, the only obstruction that may prevent a locally synchronised clock from being globally synchronisable occurs if the first de Rham class of $\bar\sM$ (\ie its first Betti number) happens to be non-trivial. The total space $\sM$ plays no role for this issue, in accordance with our comment above that all principal $\RR$-bundles are trivial.

\paragraph{Horizontal vectors and simultaneity.} Given a principal $\RR$-bundle $(\sM,\xi)$ with an Ehresmann connection $A$, a vector field $X\in\mX(\sM)$ is \itt{horizontal} if $i_XA=0$. The condition $i_\xi A=1$ then ensures that any horizontal vector is space-like; in adapted coordinates, any horizontal vector reads $X=X^a(u,x)\big(\der_a-A_a(x)\der_u\big)$ with arbitrary components $X^a(u,x)$. The notion of horizontality also allows us to define a linear lift $\ell$ from vector fields on the base $\bar\sM$ to invariant horizontal vector fields on $\sM$:
\be
\ell:
\mX(\bar\sM)\hookrightarrow\mX_{\text{inv}}(\sM):
\bar X\mapsto\ell(\bar X)\,,
\label{s22t}
\ee
where $X=\ell(\bar X)$ is the unique invariant and horizontal vector field such that $\pi_*(X)=\bar X$. In adapted coordinates, this amounts to the ``minimal coupling'' prescription $\ell:X^a(x)\,\der_a\mapsto X^a(x)\big(\der_a-A_a(x)\der_u\big)$. Note that the lift \eqref{s22t} is a morphism of Lie algebras iff the connection is flat; this is most easily verified in adapted coordinates.

Suppose now that one picks a point in $\sM$. The set of horizontal vectors at that point is a $d$-dimensional subspace of the $(d+1)$-dimensional tangent space at that point, so the Carrollian clock $A$ actually defines a distribution of $d$-planes in $\sM$. Each such plane is a local choice of ``absolute space''. When the clock is locally synchronised ($\dd A=0$), the distribution turns into a foliation of $\sM$ thanks to the Frobenius theorem (see \eg \cite[Chap.\ 11]{Lee} for details). If, in addition, the clock is globally synchronised ($A=\dd f$), then horizontal vector fields are those that solve $\cL_Xf=0$. In adapted coordinates, one has $f(u,x)=u+g(x)$ and horizontal vectors read $X=X^a(u,x)(\der_a-\der_ag(x)\der_u)$, so the leaves of the foliation are submanifolds specified by an equation of the form $u=\text{cst}-g(x)$. This defines a global \itt{Carrollian time} $u':=f(u,x)=u+g(x)$.

To conclude, suppose one has chosen \itt{two} clocks $A$ and $A'$. Then their difference $A'-A$ is necessarily basic. Accordingly, any change of clock mapping $A\to A'=A+\pi^*\bar B$ with $\bar B\in\Omega^1(\bar\sM)$ is called a local \itt{Carrollian boost}. In terms of the distribution of space-like $d$-planes specified by a clock, the boost's effect is to locally tilt the planes, \ie to locally change the notion of simultaneity. We will use such boosts in section \ref{se34} to prove that invariant volume forms on $\sM$ induced by a volume form on $\bar\sM$ are independent of one's choice of Carrollian clock.

\paragraph{Euler vector fields.} The notion of horizontality defined by a clock will soon allow us to unveil the semi-direct structure of projectable and invariant vector fields introduced in section \ref{sebundle}. In order to also include \sproj\ vector fields in this analysis, an aside is now needed on Euler vector fields. Namely, let $A=\dd f$ be a globally synchronised clock and let $\Sigma\subset\sM$ be the corresponding leaf at Carrollian time $f=0$. Then an \itt{Euler vector field} is a \sproj\ vector field $\eta$ on $\sM$ such that
\be
\label{e93}
\cL_\eta \xi=-\xi\,,
\qquad
\cL_\eta A=+A\,
,\qquad
(i_\eta A)|_\Sigma=0\,.
\ee
Any such vector field generates dilations of Carrollian time. Indeed, in adapted coordinates, it reads $\eta=u\der_u+X^a(x)\der_a$ with arbitrary components $X^a(x)$. Euler vector fields are thus not unique; in fact the difference between two Euler vector fields $\eta,\eta'$ is an invariant horizontal vector field $Y:=\eta'-\eta$ since $\cL_\xi Y=\cL_\xi\eta'-\cL_\xi\eta=0$ and $i_Y A\big|_{\Sigma}= i_{\eta'} A\big|_{\Sigma}-i_\eta A\big|_{\Sigma}=0$ (the latter equality extends to $\sM$ since $A$ and $Y$ are both invariant). It is natural to fix this ambiguity by calling \itt{canonical Euler vector field} the unique Euler vector field $\eta$ such that $\cL_\eta X=0$ for any invariant horizontal vector field $X\in\ell\big(\mX(\bar\sM)\big)$, where $\ell$ is the horizontal lift \eqref{s22t}. The canonical Euler vector field simply reads $\eta=u\,\der_u$ in adapted coordinates.

Together, the fundamental vector field $\xi$ and any Euler vector field $\eta$ span the affine algebra $\mathfrak{igl}(1)$ of the real line,  where $\xi$ and $\eta$ respectively generate fibrewise translations and dilations. In fact, a principal $\RR$-bundle endowed with a global section $\Sigma$ can be identified with a line bundle (\ie a vector bundle with one-dimensional fibre) with zero section $\Sigma$, in which case there is a canonical action of the affine algebra $\mathfrak{igl}(1)$ on each fibre. This affine structure will now allow us to write the algebra of \sproj\ vector fields as a semi-direct sum.

\paragraph{Algebras of (super-)projectable and invariant vector fields.} The arsenal of clocks and horizontal vectors allows us to gain some insight in the structure of the algebras \eqref{s19b} defined above; as it happens, this will also eventually provide the semi-direct structure of BMS transformations. Accordingly, note first that all three algebras $\mX_{\text{pro}}(\sM)$, $\mX_{\text{spro}}(\sM)$ and $\mX_{\text{inv}}(\sM)$ of section \ref{sebundle} admit an ideal consisting of vertical vector fields. This ideal is isomorphic to $C^{\infty}(\sM)$ in the projectable case, since any vertical vector field reads as $f\,\xi$ in terms of some function $f$ on $\sM$; it is isomorphic to $\{f\in C^{\infty}(\sM)\,|\,\cL^2_\xi f=0\}\cong C^{\infty}(\bar\sM)\otimes\mathfrak{igl}(1)$ in the \sproj\ case, where $\mathfrak{igl}(1)$ is spanned by the fundamental vector field $\xi$ and the canonical Euler vector field $\eta$; and it is isomorphic to $C_{\text{inv}}^\infty(\sM)=\{f\in C^{\infty}(\sM)\,|\,\cL_\xi f=0\}\cong C^{\infty}(\bar\sM)$ in the invariant case. It follows that the quotients
\be
\frac{\mX_{\text{pro}}(\sM)}{C^{\infty}(\sM)}
\cong
\frac{\mX_{\text{spro}}(\sM)}{C^{\infty}(\bar\sM)\otimes\mathfrak{igl}(1)}
\cong
\frac{\mX_{\text{inv}}(\sM)}{C^{\infty}(\bar\sM)}
\cong
\mX(\bar\sM)
\ee
are Lie algebras, all three isomorphic to the algebra $\mX(\bar{\sM})$ of vector fields on the base. It is tempting to deduce from this the semi-direct sum expressions
\begin{align}
\mX_{\text{pro}}(\sM)
&\cong
\mX(\bar\sM)
\inplus
C^{\infty}_0(\sM)\,,
\label{al1}\\
\label{s22q}
\mX_{\text{spro}}(\sM)
&\cong
\mX(\bar\sM)
\inplus
\big(C^{\infty}_0(\bar\sM)\otimes\mathfrak{igl}(1)\big)\,,\\
\mX_{\text{inv}}(\sM)
&\cong
\mX(\bar\sM)
\inplus
C^{\infty}_0(\bar\sM)\,,
\label{al3}
\end{align}
where the subscript in $C^{\infty}_0$ indicates that $\mX(\bar{\sM})$ acts on functions via the Lie derivative, \ie according to eqs.\ \eqref{deltabi}--\eqref{deltab} with $\tw=0$. Indeed, the decompositions \eqref{al1}--\eqref{al3} are correct but \itt{not} canonical, as they require a choice of flat connection $A$. To see this, consider projectable vector fields for definiteness: in order to write any such vector field $X$ as a pair $(\bar X,f)\in\mX(\bar\sM)\inplus C^{\infty}(\sM)$, one needs both the projection $\bar X=\pi_*(X)$ \itt{and} a horizontal lift \eqref{s22t} that defines the function $f$ by $f\,\xi:=X-\ell(\pi_*(X))$. Conversely, the same data is needed in order to associate with any pair $(\bar X,f)\in\mX(\bar\sM)\inplus C^{\infty}(\sM)$ a vector field on $\sM$ via $X=\ell(\bar X)+f\xi$. This lift only exists once a clock has been chosen, and it is a Lie algebra morphism only if the clock is synchronisable. This is why we could not yet write the isomorphisms \eqref{al1}--\eqref{al3} in section \ref{sebundle}.

The expressions \eqref{al1}--\eqref{al3} also provide information on the groups of automorphisms generated by projectable, \sproj\ and invariant vector fields. These groups are semi-direct products, and they respectively read
\be
\label{s12}
\text{Diff}(\bar\sM)\ltimes C_0^{\infty}(\sM),
\qquad
\text{Diff}(\bar\sM)\ltimes\big\{\text{Maps }\bar\sM\to\text{IGL(1)}\big\},
\qquad
\text{Diff}(\bar\sM)\ltimes C_0^{\infty}(\bar\sM).
\ee
The corresponding transformations in coordinates $(u,x^a)$ were written around \eqref{e68}. These semi-direct products manifestly resemble the BMS group, but we are not quite there yet. First, BMS vector fields in \eqref{xiir}--\eqref{xiia} are \sproj\ but not invariant, so they generate automorphisms of null infinity seen as a fibre bundle, but not as a principal $\RR$-bundle. In this sense, the first two groups in \eqref{s12} are too large, while the third is overly restrictive. Second, the Abelian normal subgroups in \eqref{s12} consist of functions on $\sM$ or $\bar\sM$ (as in the ``warped Virasoro group'' where $\bar\sM=S^1$ \cite{Detournay:2012pc,Afshar:2015wjm}), while the Abelian subgroup needed for BMS consists of densities with a suitable weight: recall the bracket \eqref{liba} or the weight $-1/d$ appearing in the BMS action in \eqref{coortaf}. Defining BMS thus requires an extra, conformal, structure, to which we now turn.

\subsection{BMS as conformal Carroll}
\label{se33}

Having reviewed the basics of principal $\RR$-bundles, we now consider genuine Carrollian physics by endowing these bundles with degenerate metrics. This will lead to Carrollian isometries whose natural generalisation \cite{Duval:2014uva,Duval:2014lpa} will allow us to introduce the Carrollian cousins of conformal groups, and in particular BMS algebras. The groups obtained in this way generally have no infinite-dimensional non-Abelian factor, save for exceptional circumstances (one- or two-dimensional celestial spheres) where conformal generators admit an infinite-dimensional enhancement. The generalisation to non-conformal diffeomorphisms of the base and ``generalised'' BMS \cite{Campiglia:2014yka,Campiglia:2015yka} will be treated in section \ref{se34}.

\paragraph{Carrollian spacetimes.} An (invariant) \itt{Carrollian metric} is a symmetric, covariant, rank-two tensor field $\gamma_{\mu\nu}$ on $\sM$ which is basic ($\cL_{\xi}\gamma=0$ and $\gamma(\xi,\cdot)=0$), positive semi-definite ($\gamma(X,X)\geqslant 0$) and whose radical is spanned by $\xi$ (\ie $\gamma(X,\cdot)=0$ iff $X$ is vertical). Since it is basic, it can be written as the pullback $\gamma=\pi^*\bbgamma$ of a Riemannian (non-degenerate, positive-definite) metric $\bbgamma$ on the base $\bar\sM$. Thus, in adapted coordinates $(u,x^a)$ with $\xi=\der_u$, a Carrollian metric reads
\be
\gamma
=
\bbgamma_{ab}(x)\,\dd x^a\,\dd x^b
\label{s15b}
\ee
where $\bbgamma_{ab}$ can be any Riemannian metric on the base. A triplet $(\sM,\xi,\gamma)$ consisting of a principal $\RR$-bundle $\sM$ with fundamental vector field $\xi$ and a Carrollian metric $\gamma$ will be called a \itt{Carrollian spacetime}. A typical example of Carrollian metric is again provided by null infinity, whose metric can be read off from the limit $r\to\infty$ of the Minkowski metric \eqref{mmetric} restricted to a surface with constant $r$. The result is nothing but eq.\ \eqref{s15b}, \ie the pullback by the projection \eqref{fibration} of the metric on a celestial sphere, up to a divergent normalisation. 

Note that the invariance condition on the Carrollian metric may be relaxed in full generality \cite{Duval:2014lpa}, which is crucial in certain cases. For instance, the induced metric on a light-cone in Minkowski spacetime is not invariant, although it obeys all the other axioms above. Moreover, in Penrose's approach the metric structure at null infinity is only defined up to a conformal factor, so a more general definition may seem in order (see \eg \cite[sec.\ 2]{Penrose:1972ea}). In practice the restriction to \itt{invariant} metrics at null infinity adopted here entails no loss of generality (see \eg \cite[sec.\ II.B.2]{Ashtekar:1987tt} or \cite[sec.\ II]{Ashtekar:2014zsa}), so we stick to it.

A final comment: in contrast to Riemannian geometry, there is no unique torsionless affine connection compatible with a given Carrollian structure $(\xi,\gamma)$---the reason essentially lies in the fact that the metric is not invertible. This is yet another instance where the usefulness of Ehresmann connections becomes manifest in Carrollian geometry: a unique notion of parallel transport on a Carrollian spacetime $(\sM,\xi,\gamma)$ becomes privileged only once it is further endowed with a clock $A$, \cf the ``special'' connection in \cite[app.\ A]{Bekaert:2015xua}. This is not crucial for our purposes, so we will not dwell on that point.

\paragraph{Carrollian isometries.} Given a Carrollian spacetime $(\sM,\xi,\gamma)$, its \itt{Carrollian isometries} are principal bundle automorphisms that preserve the Carrollian metric. Any such automorphism projects on an isometry of the base $(\bar\sM,\bbgamma)$. Thus, a vector field $X\in\mX(\sM)$ generates Carrollian isometries iff it is invariant and satisfies in addition the Killing equation on $\bar\sM$, that is,
\be
\label{s23b}
\cL_X\xi=0
\qquad\text{and}\qquad
\cL_{\bar X}\bbgamma=0\,.
\ee
In adapted coordinates $(u,x^a)$, any such vector field reads $X=X^a(x)\,\der_a+\alpha(x)\,\der_u$, where $\bar X=X^a(x)\,\der_a$ and $\alpha(x)$ are, respectively, a Killing vector field and an arbitrary function on $\bar\sM$ (recall eq.\ \eqref{sprovec}). The algebra of Carrollian isometry generators thus has a structure of semi-direct sum
\be
\mathfrak{isom}(\sM)
\cong
\mathfrak{isom}(\bar\sM)
\inplus
C^{\infty}_{\tw}(\bar\sM)
\label{t23b}
\ee
provided a synchronised clock has been chosen to define a horizontal lift \eqref{s22t}. Note that the weight $\tw$ of fibrewise translations under diffeomorphisms of $\bar\sM$ is arbitrary in \eqref{t23b}, since the Killing equation in \eqref{s23b} implies $\nabla_aX^a=0$ so that the coefficient of $\tw$ in the transformation laws \eqref{deltabi}--\eqref{deltab} vanishes. The simplest choice is $\tw=0$, but we will soon see that conformal invariance yields instead $\tw=-1/d$, as in the BMS bracket \eqref{liba}.

The Carrollian isometry group with Lie algebra \eqref{t23b} has a structure of semi-direct product $\text{Isom}(\bar\sM)\ltimes C^{\infty}_{\tw}(\bar\sM)$. The BMS group to be introduced now will extend it, in the same way that conformal maps extend standard isometries.

\paragraph{Carrollian conformal structure.} An equivalence class $(\sM,[\xi,\gamma])$ of Carrollian spacetimes, with respect to the equivalence relation
\be
\label{confequiv}
(\xi,\gamma)\sim(\Omega^{-1}\,\xi\,,\Omega^2\,\gamma)
%\xi\sim\Omega^{-1}\,\xi\,,\qquad \gamma\sim \Omega^2\,\gamma
\ee
where $\Omega=\pi^*\bar\Omega$ is any positive invariant function, is called a \itt{Carrollian conformal structure}.\footnote{It is also called a \itt{universal structure} in Penrose's approach to asymptotically flat spacetimes, since it is the natural kinematical structure at null infinity; see \eg  \cite[sec.\ II.B.2]{Ashtekar:1987tt} or \cite[sec.\ II]{Ashtekar:2014zsa}.} Its projection on the base defines a conformal structure $(\bar{\sM},[\bbgamma])$ for the class of Riemannian metrics $\bbgamma'\sim\overline{\Omega}^2\,\bbgamma$. The key point of \eqref{confequiv} is to link the conformal weight of $\gamma$ to that of $\xi$ in such a way that $\gamma\otimes\xi\otimes\xi$ be ``Weyl-invariant''. This will eventually fix the weight $\tw=-1/d$ of BMS supertranslations under celestial diffeomorphisms.

Note that one could also define a ``Lifschitz'' scaling where fundamental vector fields are identified as $\xi\sim\Omega^{-z}\,\xi$ for some exponent $z\in\RR$ instead of \eqref{confequiv} \cite{Duval:2014lpa}. This leads to a different weight of ``supertranslations'' under diffeomorphisms of the base space and no longer corresponds to the structure normally encountered in asymptotically flat gravity, but it is conceivable that it would find applications elsewhere, similarly to Lifschitz symmetries in generalisations of AdS/CFT \cite{Baggio:2011ha,Kachru:2008yh}. We will not study this situation in depth, but we shall from time to time mention the generalisation of conformal (BMS) results to the Lifschitz case since our geometrical approach makes the extension straightforward.

A bundle automorphism of $(\sM,\xi,\gamma)$ that preserves a given Carrollian conformal structure, so that $\xi'=\Omega^{-1}\,\xi$ (rescaling of Carroll time) and	$\gamma'=\Omega^2\,\gamma$ (local rescaling of the metric) is a \itt{Carrollian conformal map}.\footnote{To be precise, if $F:\sM\to\sM$ is the automorphism, we write the transformed fundamental vector field and metric as $\xi':=F_*\xi$ and $\gamma':=(F^{-1})^*\gamma$.\label{football}} Its projection on the base is a conformal transformation of $(\bar\sM,\bbgamma)$, with $\bbgamma'=\overline{\Omega}^2\,\bbgamma$. (One could similarly define a Carrollian Lifschitz map by replacing the first condition with $\xi'=\Omega^{-z}\,\xi$.)

\paragraph{Carrollian conformal vector fields.} Having chosen a Carrollian conformal structure $(\sM,[\xi,\gamma])$, pick a representative in the equivalence class \eqref{confequiv}, \ie a Carrollian spacetime $(\sM,\xi,\gamma)$. Then any projectable vector field $X\in\mX(\sM)$ such that
\be
\label{s20}
\cL_X\xi=-f\,\xi
\qquad\text{and}\qquad
\cL_X\gamma=2f\,\gamma\,,
\ee
for some function $f$ on $\sM$, is a \itt{Carrollian conformal vector field} that generates Carrollian conformal maps of $(\sM,\xi,\gamma)$. One can check that $X$ is also a Carrollian conformal vector field for any representative $(\xi',\gamma')$ in the equivalence class \eqref{confequiv}, \ie $\cL_X\xi'=-f'\,\xi'$ and $\cL_X\gamma'=2f'\,\gamma'$ with $f'=f+\cL_X(\ln\Omega)$. Note that the conditions \eqref{s20} imply $\cL_\xi f=0$ since $\gamma$ is basic by definition, so  $f=\pi^*\bar{f}$ is invariant and $X$ is automatically \sproj. The projection $\bar{X}=\pi_*(X)$ is a conformal Killing vector field of $(\bar{\sM},g)$, since the condition $\cL_X\gamma=2f\,\gamma$ projects to the conformal Killing equation $\cL_{\bar X}\bbgamma=2\bar{f}\bbgamma$ where $\bar{f}=\tfrac1{d}\nabla\cdot\bar X$. Accordingly, the conditions \eqref{s20} are equivalent to
\be
\label{confCK}
\cL_X\xi=-f\,\xi\,,
\qquad
\cL_{\bar X}\bbgamma=2\bar{f}\,\bbgamma\,,
\ee
which will be called the \itt{Carrollian conformal Killing equations} of $(\sM,[\xi,\gamma])$.

The structure of Carrollian conformal vector fields is most easily revealed in adapted coordinates $(u,x^a)$, where \sproj\ vector fields take the form \eqref{sprovec}. The second conformal condition in \eqref{confCK} then leaves the function $\alpha(x)$ in \eqref{sprovec} arbitrary, while fixing $f(x)=\tfrac{1}{d}\nabla_aX^a(x)$, where $d$ is the dimension of the base manifold and $\nabla$ is its Levi-Civita connection. Any Carrollian conformal vector field can thus be written as
\be
\label{vecc}
X
=
X^a(x)\der_a
+\big(\alpha(x)+\tfrac{u}{d}\nabla_aX^a(x)\big)\der_u
\ee
for some conformal Killing vector $\bar X=X^a(x)\der_a$ of the base manifold. The coincidence with the components of BMS vector fields \eqref{xiiu}--\eqref{xiia} is manifest. It readily follows that the space of Carrollian conformal vector fields is a Lie algebra with bracket \eqref{liba} whose abstract structure is that of a semi-direct sum
\be
\label{dimsum}
\mathfrak{conf}(\sM)
\cong
\mathfrak{conf}(\bar\sM)
\inplus
C^{\infty}_{\tw}(\bar\sM)
\qquad\text{with}\qquad\tw=-1/d,
\ee
where the subscript $\tw$ stresses that local fibrewise translations $u\to u+\alpha(x)$ are densities with weight $\tw=-1/d$ under diffeomorphisms of $\bar\sM$, in the sense of eq.\ \eqref{deltabi}.

Note for completeness that the ``Lifschitz'' generalisation of the above definitions is immediate: since the identification is now declared to be $\xi\sim\Omega^{-z}\xi$, the Lifschitz version of eqs.\ \eqref{confCK} is obtained by replacing the first condition by $\cL_X\xi=-zf\,\xi$ while leaving the rest unchanged. The coordinate expression of a ``Lifschitz Carrollian conformal vector field'' then takes the form \eqref{vecc} with the factor $u/d$ replaced by $z\,u/d$, which in turn implies that fibrewise translations $\alpha(x)$ are densities with weight $-z/d$ instead of $-1/d$. This is so similar to the standard structure of Carrollian conformal maps that we will no longer return to the Lifschitz example in what follows.

\paragraph{Example: BMS algebras.} The definition of Carrollian conformal structures is motivated by the example of null infinity $\sI$ at the boundary of asymptotically flat spacetimes, whose Carrollian conformal maps are really BMS transformations \cite{Duval:2014uva,Duval:2014lpa,Penrose:1965am}. Thus, by definition, Carrollian conformal vector fields on null infinity $\sI\cong\RR\times S^d$ span the (extended) \itt{BMS algebra}, \ie the semi-direct sum \eqref{dimsum} with a base manifold given by the celestial sphere:
\be
\label{ebms}
\mathfrak{bms}_{d+2}
\cong
\mathfrak{conf}(S^d)\inplus C^{\infty}_{\tw}(S^d)
\qquad\text{with}\qquad\tw=-1/d.
\ee
Fibrewise translations $u\to u+\alpha(x)$ spanning the Abelian ideal $C^{\infty}_{\tw}(S^d)$ are called \itt{supertranslations}, where the function $\alpha$ transforms as a density of weight $\tw=-1/d$ under diffeomorphisms of $S^d$; consequently, it is a scalar conformal primary of scaling dimension $\Delta=+1$ under the conformal algebra of $S^d$. The latter always contains the Lorentz algebra $\mathfrak{so}(d+1,1)$, whose elements are interpreted as bulk Lorentz transformations; the two actually coincide for $d\geqslant3$, but an infinite-dimensional enhancement to \itt{superrotations} occurs for $d=1,2$ \cite{Barnich:2009se,Barnich:2010eb,Barnich:2011ct}, so that
\be
\label{eebms}
\mathfrak{conf}(S^d)\cong\left\{
\begin{array}{cc}
\mathfrak{so}(d+1,1) & \text{for}\quad d\geqslant 3, \\
\mX(S^1)\oplus\mX(S^1) & \text{for}\quad d=2, \\
\mX(S^1) & \text{for}\quad d=1.
\end{array}
\right.
\ee
The terminology is justified by the fact that supertranslations (respectively superrotations) extend finite-dimensional bulk spacetime translations (respectively rotations and boosts). A minor subtlety is that the embedding $\mathfrak{so}(d+1,1)\subset\mathfrak{bms}_{d+2}$ is not canonical---it is equivalent to picking a ``good cut'' \cite{Newman:1976gc,Geroch1977,Adamo:2010ey,Herfray:2020rvq} at null infinity, or equivalently a bulk point in Minkowski spacetime---but we will not dwell on that.

BMS algebras as defined in \eqref{ebms} will be the starting point of our higher-spin symmetry construction in section \ref{symsrac}. Before moving to that topic, however, a generalisation of Carrollian conformal maps is required.

\subsection{Carrollian volumes and generalised BMS}
\label{se34}

In the context of scattering amplitudes and asymptotic symmetries, a natural generalisation of the BMS groups defined above consists in extending conformal superrotations to arbitrary (generally non-conformal) diffeomorphisms of celestial spheres \cite{Campiglia:2014yka,Campiglia:2015yka,Colferai:2020rte}. We now review the Carrollian formulation of this enhancement, generalising to arbitrary dimensions the $d=2$ case treated in \cite[sec.\ 4]{Campiglia:2014yka}. The key point is to require ``generalised BMS transformations'' (\itt{gBMS} for short\footnote{We stress that the shorthand ``gBMS'' does \itt{not} refer to the global BMS group (where superrotations are restricted to Lorentz transformations).}) to preserve a \itt{volumic} structure rather than a metric structure, but this requires a few geometric preliminaries. Accordingly, we now review notions of Carrollian volume forms and their interplay with Carrollian clocks, then introduce generalised conformal maps. As an application, the Hermitian form \eqref{sachspo} is recast in a geometric (\ie coordinate-\-in\-de\-pen\-dent) way, generalising the Sachs inner product \cite{Sachs2} to any Carrollian measured space. This will make generalised BMS symmetry manifest and provides a key prerequisite for the higher-spin considerations of section \ref{symsSachs}, where the Sachs Hermitian form will be used to define higher-spin algebras admitting a unitary representation.

\paragraph{Carrollian measured spaces.} Let $(\sM,\xi)$ be a principal $\RR$-bundle and assume $\sM$ is orientable. Then an \itt{invariant volume form} on $(\sM,\xi)$ is a nowhere-vanishing top form $\eps\in\Omega^{d+1}(\sM)$ such that $\cL_\xi\eps=0$. In adapted coordinates $(u,x^a)$, any invariant volume form reads $\eps=\pm \eps(x)\,\dd u\wedge\dd^dx$, where the function $\eps(x)$ is strictly positive on $\bar\sM$. A triplet $(\sM,\xi,\eps)$ made of a principal $\RR$-bundle with orientable total space $\sM$ and  fundamental vector field $\xi$\,, together with an invariant volume form $\varepsilon$, will be called a \itt{Carrollian measured space} since the volume form is really a measure (recall section \ref{senote}).

It is intuitively clear that a Carrollian measured space $(\sM,\xi,\eps)$ induces a volume form on the base manifold $\bar\sM$. This is not as trivial as for Carrollian metrics: the contraction $\cV:=i_\xi\eps$ does not vanish, so $\eps$ is not basic and cannot be written as the pullback of a form on $\bar\sM$. Note, however, that $\cV\in\Omega^d(\sM)$ is closed ($\dd\cV=0$) thanks to the Cartan  formula along with the invariance condition $\cL_\xi\eps=0$; also note that it is horizontal ($i_\xi\cV=0$) since $i_\xi$ is nilpotent. It follows that $\cV$ is basic and uniquely determined by a volume form $\bar\eps\in\Omega^d(\bar\sM)$ on the base manifold: 
\be
\label{calV}
\cV\,
:=
\,\pi^*\bar\eps\,\in\,\Omega^d(\sM)\,.
\ee
In adapted coordinates where $\eps=\pm \eps(x)\,\dd u \wedge\dd^dx$, one simply has $\cV=\pm \eps(x)\,\dd^dx$. Combined with the notion of clocks introduced in section \ref{se32}, this elementary observation will now allow us to define volume forms on Carrollian manifolds starting from a volume form on the base manifold.

\paragraph{Inducing volume forms from the base.} Let $\bar\eps$ be a volume form on $\bar\sM$ and define $\cV$ by \eqref{calV}. If in addition $\sM$ is endowed with an Ehresmann connection $A$, one can define an invariant volume form $\eps:=A\wedge\cV$ on $\sM$ (indeed, $\eps$ is clearly a nowhere vanishing top form on $\sM$ and $\cL_{\xi}\eps=\cL_{\xi}A\wedge\cV+A\wedge\cL_{\xi}\cV=0$ since $A$ and $\cV$ are both invariant). The volume form thus obtained is independent of the choice of connection, since any two Ehresmann connections differ by a horizontal form, and any horizontal $(d+1)$-form automatically vanishes. In this sense, any Ehresmann connection does the job equally well; in what follows we will systematically use this to induce a Carrollian measured space $(\sM,\xi,\eps)$ from the (equivalent) data $(\sM,\xi,\bar\eps)$ made of a principal $\RR$-bundle $(\sM,\xi)$ and a volume form $\bar\eps$ on the orientable base space $\bar\sM$.

For example, consider a Carrollian spacetime $(\sM,\xi,\gamma)$ that projects down to a Riemannian space $(\bar{\sM},\bbgamma)$. The metric $\bbgamma$ on $\bar\sM$ determines its canonical volume form $\bar\eps:=*1\in\Omega^d(\bar{\sM})$, so an invariant volume form $\eps=A\wedge\cV$ with $\cV=\pi^*\bar\eps$ can be defined once any Carrollian clock $A$ has been chosen. This provides another definition (valid in any dimension) of the volume form used \eg in  \cite[sec.\ II.B.4]{Ashtekar:1987tt} or \cite[sec.\ III.C]{Ashtekar:2014zsa}. We shall refer to it as the \itt{Carrollian volume form} of the canonical Carrollian measured space $(\sM,\xi,\eps)$ induced by $(\sM,\xi,\gamma)$, in the same way that any Riemannian metric  determines a canonical volume form. In adapted coordinates, the construction is nearly trivial since the Carrollian volume form is simply $\dd u\wedge\dd^dx\,\sqrt{\bbgamma(x)}$\,.

It is useful, for future reference, to know how bundle automorphisms affect invariant volume forms. Namely, let $F:\sM\to\sM$ be an automorphism, so that it admits a projection $\bar F:\bar\sM\to\bar\sM$ satisfying \eqref{fipi}. Then note that
\be
F^*\cV
=
\pi^*(\rho_{\bar F})\,\cV
\label{peps}
\ee
in terms of the Radon-Nikodym derivative \eqref{RNN} of the base volume form under $\bar F$. This elementary fact will soon allow us to define the Carrollian analogue of the Radon densities defined in section \ref{senote}.

\paragraph{Campiglia-Laddha structures.} We are finally ready to define the equivalence class of Carrollian measured spaces that generalises the conformal class \eqref{confequiv} and will eventually lead to generalised BMS transformations. Thus, we call \itt{Campiglia-Laddha structure} an equivalence class $(\sM,[\xi,\eps])$ of Carrollian measured spaces, with respect to the relation
\be
\label{ASequiv}
(\xi,\eps)\sim(\Omega^{-1}\,\xi\,,\Omega^{d+1}\,\eps)
\ee
where $\Omega=\pi^*\bar\Omega$ is any positive invariant function. These conditions can equivalently be expressed in terms of the volume form $\bar\eps$ on the base, namely
\be
\label{ASequivbar}
(\xi,\bar\eps)\sim(\Omega^{-1}\,\xi\,,\bar\Omega^d\bar\eps)\,.
\ee
To the best of our knowledge, the first discussion of such equivalence classes in their own right, characterising generalised BMS transformations, appeared in \cite[sec.\ 4.1]{Campiglia:2014yka} for $d=2$; hence the terminology.
In particular, any Carrollian conformal structure $(\sM,[\xi,\gamma])$ yields a canonical Campiglia-Laddha structure $(\sM,[\xi,\eps])$ by virtue of the above construction of the Carrollian volume form, since $A\sim\Omega A$ (due to the condition $i_\xi A=1$) and $\cV\sim\Omega^d\cV$ (due to $\bar\eps\sim\bar\Omega^d\bar\eps$). 

In contrast to conformal structures, the extension of the equivalence \eqref{ASequiv} to a Lifschitz-like relation $\xi\sim\Omega^{-z}\,\xi$ also requires a modification of the relation imposed on Carrollian volumes. This is because the normalisation condition $i_{\xi}A=1$ for Carrollian clocks requires $A\sim\Omega^zA$ in the Lifschitz case, which in turn implies that the volume form must now be identified as $\eps\sim\Omega^{d+z}\,\eps$ instead of the second condition in \eqref{ASequiv}. (By contrast, the Lifschitz generalisation of the conformal equivalence \eqref{confequiv} only affected the fundamental vector field, not the metric.) Aside from this modification, the discussion of ``Lifschitz-generalised'' conformal maps and vector fields is entirely analogous to that presented in section \ref{se33} for the conformal case, so we will not dwell on this extension of the formalism.

Similarly to section \ref{se33}, any diffeomorphism of $\sM$ that preserves a given Campiglia-Laddha structure $(\sM,[\xi,\eps])$, so that $\xi'=\Omega^{-1}\,\xi$ and $\eps'=\Omega^{d+1}\,\eps$ for any representative $(\xi,\eps)$, will be called a \itt{\CL\ map}.\footnote{With the notation of footnote \ref{football}, one has $\xi':=F_*\xi$ and $\eps':=(F^{-1})^*\eps$.} It is necessarily a bundle automorphism (since $\xi'=\Omega^{-1}\,\xi$) and its projection on the base can be any orientation-preserving diffeomorphism of $\bar\sM$ as per eq.\ \eqref{ASequivbar}. In adapted coordinates $(u,x^a)$, any generalised conformal map $F:\sM\to\sM$ reads
\be
\label{s31b}
(u,x)\to(u',x')
=
\big(\rho_{\bar F}(x)^{1/d}u+\alpha(x),\bar F(x)\big)
\ee
for some diffeomorphism $\bar F:\bar\sM\to\bar\sM$, with $\rho_{\bar F}=\Omega^d$ the corresponding Radon-Nikodym derivative \eqref{RNN} on the base and $\alpha(x)$ any ``supertranslation'' function. The similarity with the BMS transformation \eqref{coortaf} is striking indeed. We stress that such maps are much less constrained than conformal maps, thanks to the fact that the identifications \eqref{ASequiv} are much broader than those defining conformal structures in \eqref{confequiv}. Indeed, \itt{any} volume form on the base (with fixed orientation) is solely determined by a single positive function, so the second identification in \eqref{ASequiv} simply says that any two volume forms belong to the same equivalence class (provided they define the same orientation). Put differently, the orbit of a volume form is the set of all possible (oriented) volume forms,\footnote{One should still keep in mind that the equivalence class applies to pairs $(\xi,\varepsilon)$ so, strictly speaking, the various volume forms $\varepsilon'$ in the orbit of $\varepsilon$ are distinguished by their partner $\xi'$ inside the pair  $(\xi',\varepsilon')\sim (\xi,\varepsilon)$.} and a Campiglia-Laddha structure as defined by \eqref{ASequiv} generally contains many distinct conformal structures as defined by \eqref{confequiv}. The only exception occurs for one-dimensional base manifolds ($d=1$), where conformal maps and their generalisations coincide.

\paragraph{Generalised conformal vector fields.} Again mimicking section \ref{se33}, it is natural to call \itt{\CL\ vector field} any projectable vector field $X\in\mX(\sM)$ such that
\be
\label{s200}
\cL_X\xi=-f\,\xi
\qquad\text{and}\qquad
\cL_X\eps=(d+1)f\,\eps
\ee
for some function $f$ on $\sM$. Equivalently,
\be
\label{s200'}
\cL_X\xi=-f\,\xi
\qquad\text{and}\qquad
\cL_{\bar X}\bar\eps=d\,\bar f\,\bar\eps
\ee
in terms of the base volume form. This is the volumic generalisation of the conformal conditions \eqref{s20}--\eqref{confCK}. In particular, any $X\in\mX(\sM)$ that satisfies eqs.\ \eqref{s200} generates \CL\ maps on $\sM$. As in section \ref{se33}, eqs.\ \eqref{s200} imply $\cL_\xi f=0$ since $\eps$ is invariant by definition, so $X$ is necessarily \sproj. However, in contrast to section \ref{se33}, the projection $\bar{X}=\pi_*(X)$ is now an arbitrary vector field on $\bar\sM$. Note that the last condition in \eqref{s200'} implies $\bar f=\tfrac1{d}\,\text{div}\bar X$ with the divergence defined by \eqref{didef}, in agreement with the special case $\bar{f}=\tfrac1{d}\nabla\cdot\bar X$ obtained above \eqref{vecc} for Carrollian conformal vector fields.

In adapted coordinates $(u,x^a)$, any \CL\ vector field takes the form \eqref{vecc} with an \itt{arbitrary} vector field $X^a\der_a$ on the base manifold. It readily follows that the space of \CL\ vector fields is a Lie algebra, with a bracket \eqref{liba} that generalises that of Poincar\'e and BMS vector fields. Abstractly, the algebra is a semi-direct sum
\be
\label{dimsumbis}
\mathfrak{gconf}(\sM)
\cong
\mX(\bar\sM)
\inplus C^{\infty}_{\tw}(\bar\sM)
\qquad\text{with}\qquad\tw=-1/d,
\ee
where the subscript stresses once more that local fibrewise translations $u\to u+\alpha(x)$ are densities with weight $\tw=-1/d$ under diffeomorphisms of $\bar\sM$, in the sense of eq.\ \eqref{deltabi}.

\paragraph{Example: Generalised BMS.} It is immediate to apply the definition of \CL\ maps and vector fields to null infinity $\sI\cong S^d\times\RR$. The algebra \eqref{dimsumbis} with $\bar\sM=S^d$ is then referred to as a \itt{generalised BMS algebra} \cite{Campiglia:2014yka,Campiglia:2015yka} (see also \cite{Grumiller:2019fmp})
\be
\label{semidirect}
\mathfrak{gbms}_{d+2}
\cong
\mX(S^d)
\inplus C^{\infty}_{\tw}(S^d)
\qquad\text{with}\qquad\tw=-1/d,
\ee
where the Abelian ideal $C^\infty_{\tw}(S^d)$ is still spanned by supertranslations, while the subalgebra $\mX(S^d)$ generates arbitrary diffeomorphisms of celestial spheres that generalise the superrotations introduced around eq.\ \eqref{ebms}. Conversely, the conformal BMS algebras \eqref{eebms} are Lie subalgebras of generalised BMS algebras, leading to the hierarchy
\be
\label{hierarchy}
\mathfrak{iso}(d+1,1)
\subset
\mathfrak{bms}_{d+2}
\subseteq
\mathfrak{gbms}_{d+2}
\subset
\mX_{\text{spro}}(\sI_{d+1})
\subset
\mX_{\text{pro}}(\sI_{d+1})
\,,
\ee
where the second inclusion is an equality only for $\mathfrak{bms}_3=\mathfrak{gbms}_3$. The higher-spin extensions of the symmetry groups involved in this sequence will be studied in detail in section \ref{symsSachs}. However, this first requires that we return to the Sachs form \eqref{sachspo} and rephrase it in a manifestly Carrollian and coordinate-independent way after defining suitable Carrollian densities.

\paragraph{Carrollian Hermitian form.} Consider a Carrollian measured space $(\sM,\xi,\varepsilon)$ and define the basic form $\cV=i_\xi\varepsilon=\pi^*\bar\eps$ induced from the volume form $\bar\eps$ on $\bar\sM$, as in \eqref{calV}. Then, for any two functions $\phi$ and $\psi$ on $\sM$, the wedge product of $\phi^*\dd\psi$ and $\cV$ is a top form given by
\be
\label{identityLxi}
\phi^*\dd\psi\wedge\cV=(\phi^*\,\cL_\xi\psi)\,\eps\,,
\ee
as follows from the relation $\cL_\xi\psi=i_\xi(\dd\psi)$.\footnote{Indeed, $\cL_\xi\psi\,\eps=i_\xi(\dd\psi)\wedge\eps=-\dd\psi\wedge i_\xi\eps=-\dd\psi\wedge\cV$ where the second equality holds because the $(d+2)-$form $\dd\psi\wedge\eps=0$ vanishes identically (its degree exceeds $\text{dim}(\sM)=d+1$).} This top form allows us to define the integral
\be
\label{phipsi}
\langle\phi|\psi\rangle
\,:=\,
i\int_{\sM}\phi^*\,\dd\psi\wedge\cV\,
=
\,i\int_{\sM}(\phi^*\,\cL_\xi\psi)\,\eps\,,   
\ee
which is a Hermitian form when $\phi,\psi$ satisfy suitable boundary conditions since the property $\langle\phi|\psi\rangle=\langle\psi|\phi\rangle^*$ follows from integration by parts of the middle integral, along with $\dd\cV=0$. We shall refer to it in full generality as the \itt{Carrollian Hermitian form} of $(\sM,\xi,\eps)$. In adapted coordinates $(u,x^a)$, it reproduces the earlier Sachs inner product \eqref{sachspo}, generalising the result in \cite{Sachs2} to any dimension and any Carrollian measured space. It may roughly be seen as a ``matrix element'' of the Carroll Hamiltonian $i\xi$ between the ``asymptotic data'' $\phi$ and $\psi$.

Note in passing that the Hermitian form \eqref{phipsi} is intimately related to the symplectic form \eqref{symplectic} applied to radiative solutions ($\Phi=\cO(r^{-d/2})$) of the d'Alembert equation \eqref{daleq}. Indeed, the symplectic form \eqref{symplectic} may be recast in Carrollian terms as
\be
\label{phi1phi2}  
\Omega[\delta\phi]
=i\int_{\sM}\big(\delta\phi^*\wedge\dd(\delta\phi)\big)\wedge\cV
=i\int_{\sM}\big(\delta\phi^*\wedge\cL_\xi(\delta\phi)\big)\wedge\eps\,,
\ee
where $\phi$ is understood as the leading term of the expansion \eqref{expa} with $\Delta=d/2$. The notation in \eqref{phi1phi2} may be confusing: the wedge product that follows $\delta\phi^*$ involves differential forms in field space, while that in $\eps=A\wedge\cV$ involves forms on $\sM$. (Fortunately, this is the only time this issue will affect us.)

\paragraph{Carrollian densities.} Having defined Carrollian measured spaces, it is straightforward to define notions of densities analogous to those of section \ref{senote}. Let therefore $(\sM,\xi,\eps)$ be a Carrollian measured space with base volume form $\bar\eps$ and let $F:\sM\to\sM$ be a bundle automorphism so that the identity \eqref{peps} holds. This allows us to define a (scalar) \itt{Carrollian density} with weight $\tw$, as a function $\phi$ on $\sM$ that transforms under any bundle automorphism $F$ according to $\phi\to\phi':=F\cdot\phi$ with
\be
\label{e90}
F\cdot\phi
:=
(\pi^*\rho_{\bar F^{-1}})^{\tw}\,\,(F^{-1})^*\phi\,,
\ee
where $\rho_{\bar F^{-1}}$ is the Radon-Nikodym derivative \eqref{RNN} of the base volume form $\bar\eps$ under $\bar F^{-1}$. This applies in particular to any generalised conformal map, in which case $\eps'=(F^{-1})^*\eps=\Omega^{d+1}\,\eps$ and $\xi'=F_*\xi=\Omega^{-1}\xi$ with $\Omega=(\pi^*\rho_{\bar F^{-1}})^{1/d}$. We stress that eq.\ \eqref{e90} is a straightforward analogue of the transformation of densities and primary fields encountered \eg in CFT. This is manifest in adapted coordinates $(u,x^a)$, where any bundle automorphism takes the form \eqref{e68} so that the transformation law \eqref{e90} becomes
\be
\label{s34b}
\phi'\big(u'(u,x),x'(x)\big)
=
\rho_{\bar F}(x)^{-\tw}\,
\phi(u,x)\,,
\ee
which is nothing but the generalisation of the BMS transformation law \eqref{bact} to any bundle automorphism, with $\tw=\Delta/d$. Indeed, eqs.\ \eqref{bact} and \eqref{s34b} coincide in the special case where $F$ is a generalised conformal map \eqref{s31b}.

The infinitesimal transformation corresponding to eq.\ \eqref{e90} can be found in the same way as for standard densities in section \ref{senote}. Thus, for any projectable vector field $X$, eq.\ \eqref{e90} yields
\be
\label{ss34b}
\delta\phi
=
\cL_X\phi+\tw\,\text{div}(\bar X)\,\phi
\ee
where the divergence is defined by \eqref{didef}. When $\bar\sM$ carries a metric, $\text{div}(\bar X)=\nabla_a\bar X^a$ as before. It is again manifest that eq.\ \eqref{ss34b} reduces to the infinitesimal BMS transformation law \eqref{deph} when expressed in adapted coordinates and applied to a generalised conformal vector field.

\paragraph{Manifest generalised BMS symmetry of the Sachs form.} We showed in section \ref{sebemin} that BMS transformations acting on densities with weight $\tw=1/2$ leave the Sachs form \eqref{sachspo} invariant. This remains true even when the conformal assumption is relaxed so that BMS is enhanced to \itt{generalised} BMS, allowing for any diffeomorphism of the base manifold. Indeed, if the wavefunctions $\phi,\psi$ appearing in the Sachs form \eqref{phipsi} transform as Carrollian densities \eqref{e90} with weight $\tw=1/2$, then any \CL\ map of $\sM$ sends $\eps\to\eps'=\Omega^{d+1}\eps$,\, $\xi\to\xi'=\Omega^{-1}\xi$,\, $\phi\to\phi'=\Omega^{-d/2}\phi$ and $\psi\to\psi'=\Omega^{-d/2}\psi$ with $\Omega=(\pi^*\rho_{\bar F^{-1}})^{1/d}$, which manifestly implies that the top form $(\phi^*\,\cL_\xi\psi)\,\eps$ and the Carrollian Hermitian form \eqref{phipsi} are invariant under \CL\ transformations. One can actually be even more general: \itt{any} bundle automorphism leaves the Sachs form invariant provided $\phi,\psi$ are Carrollian densities with weight $\tw=1/2$. The proof is immediate in terms of the middle expression in \eqref{phipsi} upon using the transformation law \eqref{e90} with $\tw=1/2$. We will also confirm this independently in section \ref{symsSachs} in terms of vector fields.

Following section \ref{sebemin}, a corollary of these observations is a neat sequence of symmetry enhancements summarised by the inclusions \eqref{hierarchy}. Indeed, we saw below eq.\ \eqref{e52} that any scalar massless \UIR\ of the Poincar\'e group lifts to a massless \UIR\ of the BMS group, but we now know that this even lifts to a massless \UIR\ of the generalised BMS group, and even further to the group of all bundle automorphisms. As counter-intuitive as it may sound, the Hilbert space of radiative modes of a massless particle on Minkowski spacetime carries a \UIR\ of all Carrollian symmetries of null infinity. In section \ref{symsSachs}, this property will be extended to the corresponding higher-spin algebras.

\section{Higher-spin BMS algebras}
\label{seHS}

Having established our playground, we now turn to the construction of higher-spin extensions of (generalised) BMS algebras. Our strategy will be the same in all cases and follows a standard higher-spin pattern now applied to Carrollian structures:
\begin{itemize}
\item[1.] Start by considering suitable Lie algebras of vector fields on a Carrollian manifold---typically the Carrollian conformal Killing vector fields (standard BMS) of section \ref{se33}, or the \CL\ vector fields (generalised BMS) of section \ref{se34}.
\item[2.] Build the corresponding, much larger, universal enveloping algebras (``$\sU$'') whose elements are suitable families of differential operators of arbitrarily high order.
\item[3.] In each case, $\sU$ admits an ideal (an annihilator ``$\sU_0$'') related to the fact that some of its elements are trivial in the corresponding realisation (\eg all powers of the conformal Laplacian in the case of the WRac). Mod out this annihilator to obtain higher-spin algebras of the general form $\sU/\sU_0$. This quotient is actually automatic for the Sachs module, but in that case one further restricts attention to operators that are Hermitian with respect to the form \eqref{sachspo}.
\end{itemize}
As announced in the \hyperref[semott]{introduction}, this investigation is motivated by the desire to define a Minkowskian analogue of the Rac in AdS. We shall focus on the two structures encountered in section \ref{seSCAL}: the WRac and the Sachs module, respectively studied in sections \ref{symsrac} and \ref{symsSachs}. Both will provide candidate higher-spin extensions of BMS, and both have overlaps with the asymptotic higher-spin symmetries of \cite{Campoleoni:2017mbt,Campoleoni:2017qot,Campoleoni:2018uib,Campoleoni:2020ejn}, investigated in our language in section \ref{sehispin}. However, none of them will reproduce the ``flat Rac'' obtained in \cite{Campoleoni:2021blr} from a genuine Minkowskian limit of the Rac in AdS space. This suggests that more work is required in order to understand the relation between scalar kinematics at infinity, Carrollian geometry, and flat limits---an issue that we will not address here.

Note that we will rely heavily on the conventions and Carrollian language of section \ref{seCAR}, as this will allow us to avoid local coordinates when studying differential operators on Carrollian spacetimes. We will also use various generic properties of differential operators that will not be reviewed at length. (We refer \eg to \cite{saunders_1989,olver_1995} for detailed introductions.) In particular, all differential operators are henceforth understood to act on scalar fields $f\in C^{\infty}(\sM)$.\footnote{Note that these fields are allowed to be densities with non-zero weight, which will manifest itself in the presence of zeroth-order factors $\propto\text{div}\,X$ in first-order differential operators specified by a vector field $X$, as in the transformation law \eqref{deltabi}.} Homogeneous differential operators of orders zero and one in the partial derivatives are, respectively, functions (acting by local multiplication) and vector fields (acting by local derivation), while operators of order $k$ are defined recursively as those linear operators $\DD$ on $C^{\infty}(\sM)$ such that the commutator $[\DD,C^{\infty}(\sM)]$ be a space of operators of order $k-1$. Thus, in local coordinates, an operator of order $k$ reads as $\DD=\sum_{n=0}^kf^{a_1...a_n}(x)\,\der_{a_1}...\der_{a_n}$, where the components $f^{a_1...a_n}(x)$ are totally symmetric in their indices (note that they do \itt{not} transform as tensor fields, except for $n=k$). In particular, the \itt{leading symbol} of a differential operator of order $k$ is the totally-symmetric contravariant tensor field of rank $k$ encoded by the operator's leading piece in the number of derivatives (\ie the one involving exactly $k$ derivatives); in the example just given, this leading symbol is $f^{a_1...a_k}(x)\,\der_{a_1}...\der_{a_k}$. We shall write as $\cD^k(\sM)$ the vector space of all operators of order $k$. Note that the composition of operators is an associative binary operation, so the space $\cD(\sM)$ of all differential operators on a manifold $\sM$ is an associative algebra; it is also a Lie algebra with respect to the standard commutator. 

\subsection{Higher symmetries of the WRac}
\label{symsrac}

Recall from section \ref{seSCAL} that a WRac is an ``overleading'' solution of the d'Alembert equation \eqref{daleq}. In the simplest case \eqref{e55}, it behaves at infinity as $\Phi(r,u,x)\sim r^{1-d/2}\phi(x)$ and satisfies the Yamabe equation \eqref{yamyam} owing to bulk equations of motion. Since the Yamabe condition has no smooth solutions, the field $\phi(x)$ is generally singular; it is nevertheless natural to study it from a higher-spin perspective. Accordingly, we now build the higher-spin algebra obtained by (i) extending the conformal BMS vector fields of section \ref{se33} to higher-order differential operators, (ii) quotienting them by all time derivatives (owing to $\dot\phi=0$ in eqs.\ \eqref{timeindep}) and all powers of the conformal Laplacian (owing to the Yamabe equation \eqref{yamyam}). The end result will be an algebra very similar to that of a standard Rac in AdS, with BMS supertranslations essentially set to zero. This oversimplification ultimately rules out the WRac as a well-behaved analogue of the usual singleton, and justifies the later study of radiative scalars (section \ref{symsSachs}).

The plan is as follows. We start by rephrasing the Carrollian conformal Killing equations \eqref{s20}--\eqref{confCK} in terms of commutators involving differential operators and the Yamabe operator of eq.\ \eqref{yamyam}, which provides a geometric definition of the WRac for any Carrollian spacetime. The WRac's conformal invariance is then extended to a symmetry under higher-order differential operators, obtained by adapting to Carrollian manifolds the definition of higher symmetries of conformal Laplacians \cite{Eastwood:2002su,Nikitin,shapovalov1992symmetry}. Finally, the condition $\dot\phi=0$ leads to the definition of vertical operators and their modding out of the higher-spin symmetry algebra; the latter eventually reduces (after a further quotient by trivial Laplacian symmetries) to the Eastwood-Vasiliev algebra \cite{Eastwood:2002su,Vasiliev:2003ev} that would be relevant in bosonic higher-spin theories on de Sitter space \cite{Anninos:2011ui} for which the present Minkowski spacetime would play the role of ambient space, dS$_{d+1}\subset\RR^{d+1,1}$.

\paragraph{Proposition: Carrollian conformal vector fields as operators.}\label{propconf} Let $(\sM,\xi,\gamma)$ be a Carrollian spacetime as defined at the end of section \ref{sebundle}. Then any Carrollian conformal vector field $X$ that satisfies eqs.\ \eqref{s20} is the leading symbol of a first-order differential operator $\DD$ that satisfies the commutation relations
\be
\label{cCK'}
[\hxi,\DD\,]=f\,\hxi\,,
\qquad
[\Delta,\bar\DD\,]=2\bar{f}\,\Delta\,,
\ee
where $\xi$ is understood as a homogeneous first-order differential operator, $f=\pi^*\bar f$, and $\Delta:=\nabla^2-\tfrac{d-2}{4(d-1)}\cR$ is the conformal Laplacian of the Riemannian metric $\bbgamma$ (\ie the Yamabe operator in eq.\ \eqref{yamyam}). This completion $X\to\DD$ is unique up to an additive constant in the zeroth-order part of $\DD$. In this sense, the Carrollian conformal Killing equations \eqref{confCK} are equivalent to the commutation relations \eqref{cCK'}.

\proof{
Let $X\in\mX(\sM)$ be a vector field. Any one of its completions to a first-order differential operator reads $\DD=X+h$ for some function $h\in C^{\infty}(\sM)$. Our goal is to show that $h$ can be chosen such that the commutators \eqref{cCK'} hold, assuming that $X$ satisfies eqs.\ \eqref{s20}--\eqref{confCK}. To begin, the first condition in \eqref{cCK'} can explicitly be written as $[\xi,X+h]\stackrel{\text{\eqref{s20}}}{=}f\xi+\cL_{\xi}h\stackrel{!}{=}f\xi$. This holds iff $\cL_{\xi}h=0$, \ie whenever $h=\pi^*\bar h$ is an invariant function. Then $\bar\DD=\bar X+\bar h$ is well-defined and the second condition in \eqref{cCK'} reads
\be
\label{s33t}
\tfrac{2}{d}\nabla_a\bar X^a\,\nabla^2
+(2\nabla^a\bar h+\nabla^2\bar X^a)\der_a
+(\nabla^2\bar h+\tfrac{d-2}{4(d-1)}\bar X^a\der_a\cR)
\stackrel{!}{=}
2\bar f(\nabla^2-\tfrac{d-2}{4(d-1)}\cR)\,.
\ee
Here the leading symbols of both sides automatically coincide for $\bar f=\nabla_a\bar X^a/d$, which is the standard factor on the right-hand side of the conformal Killing equation in \eqref{confCK}. It only remains to solve the first- and zeroth-order terms of \eqref{s33t}, namely
\be
\label{saxo}
2\nabla^a\bar h+\nabla^2\bar X^a=0
\qquad\text{and}\qquad
\tfrac{4(d-1)}{d-2}\nabla^2\bar h
+\bar X^a\der_a\cR
+\tfrac{2}{d}\nabla_a\bar X^a\cR=0\,.
\ee
The first of these conditions sets $\bar h=(\tfrac{1}{2}-\tfrac{1}{d})\nabla_a\bar X^a+\text{cst}$, whereupon the second condition holds automatically. Note that the factor $\tfrac{1}{2}-\tfrac{1}{d}=\frac{d-2}{2d}$ in front of the divergence encodes the proper weight of the conformal scalar field, \cf section \ref{serac}. Since $h=\pi^*\bar h$, this proves as announced that the completion $X\to\DD$ is unique up to an additive constant.
\qed}

Note that the explicit computations \eqref{s33t}--\eqref{saxo} are unnecessary if one uses well-known facts on conformal geometry and differential operators. Indeed, the leading symbol of the first equation in \eqref{cCK'} is $[X,\xi]=-f\,\xi$, which is a mere rewriting of the first equation in \eqref{confCK}; and the leading symbol of the second equation in \eqref{cCK'} is equivalent to the conformal Killing equation $\cL_{\bar X}\bbgamma=2\bar{f}\,\bbgamma$. But the conformal Laplacian is designed so that any conformal map can be lifted to a symmetry of the Yamabe equation upon assigning a suitable weight to the scalar field, so any conformal Killing vector field $\bar X$ on $\bar\sM$ can be completed into a first-order operator $\bar\DD=\bar X+\bar h$ that satisfies the second equation in \eqref{cCK'}.

\paragraph{Corollary: Invariance of the WRac.} Any solution $\bar\phi$ of the Yamabe equation $\Delta\bar{\phi}=0$ can be lifted to an invariant function $\phi=\pi^*\bar\phi$ on $\sM$. The above proposition then trivially implies that the space of solutions of the \itt{WRac equations}
\be
\label{wreq}
\cL_\xi\phi=0\,,
\qquad
\Delta\bar{\phi}=0     
\ee
is preserved by all Carrollian conformal vector fields defined in section \ref{se33}. In the case of null infinity, this provides a coordinate-free description of the WRac introduced in sections \ref{sescad}--\ref{serac}: it is the $\mathfrak{bms}_{d+2}\,$-module spanned by functions $\phi\in C^\infty(\sI_{d+1})$ that solve \eqref{wreq}, where the first condition expresses the equation $\dot\phi=0$ encountered in \eqref{timeindep} while the second is the Yamabe equation \eqref{yamyam}.

In the remainder of this subsection, our goal will be to generalise the first-order conformal symmetries of the WRac to higher-order differential operators. This requires that we first introduce notions of projectable and invariant operators, after which we shall adapt the notion of higher symmetries of the Laplacian to the WRac equations \eqref{wreq}. We will then finally introduce vertical operators and ``trivial'' symmetries, and mod them both out.

\paragraph{Projectable and invariant differential operators.} We have seen around eq.\ \eqref{procond} how the bundle structure of $(\sM,\xi)$ motivates the definition of projectable and invariant vector fields; we now extend this to differential operators. Namely, a differential operator $\DD\in\cD(\sM)$ is \itt{projectable} if
\be
\label{projop}
\hxi\circ\DD=\EE\circ\hxi\qquad\text{for some}\qquad\EE\in\cD(\sM)\,.
\ee
Note that this immediately implies that the leading symbols of $\EE$ and $\DD$ coincide. Furthermore, a projectable operator is \itt{invariant} if $\EE=\DD$ in \eqref{projop}, \ie if $[\hxi,\DD] =0$. (The intermediate class of \sproj\ operators will be introduced in section \ref{symsSachs}.) The terminology here is consistent with that of section \ref{sebundle}: a first-order operator is projectable/invariant iff its leading symbol is a projectable/invariant vector field. 

These definitions ensure that projectable operators preserve the space of invariant functions, so any such operator $\DD$ admits a well-defined projection $\bar\DD$ on $\bar\sM$; it is obtained by restricting $\DD$ to the subspace $C_{\text{inv}}^\infty(\sM)\subset C^\infty(\sM)$ and using the isomorphism $C_{\text{inv}}^\infty(\sM)\cong C^\infty(\bar{\sM})$. Note that projectable differential operators span an associative algebra $\cD_{\text{pro}}(\sM)$, and that the projection $\pi_*:\cD_{\text{pro}}(\sM)\twoheadrightarrow\cD(\bar\sM):\DD\mapsto\bar\DD$ is a surjective morphism (\ie $\pi_*(\DD\circ\EE)=\pi_*(\DD)\circ\pi_*(\EE)$). Invariant operators similarly span an associative algebra $\cD_{\text{inv}}(\sM)$, isomorphic to the tensor product of the algebra $\cD(\bar\sM)$ of differential operators on $\bar\sM$ with the commutative algebra $\RR[\xi]$ of polynomials in the fundamental vector field:
\be
\label{Dinv}
\cD_{\text{inv}}(\sM)\cong\cD(\bar\sM)\otimes\RR[\xi]\,.
\ee
As in the expression of Carrollian symmetry groups as semi-direct products in section \ref{seCAR}, the isomorphism \eqref{Dinv} requires a choice of (synchronisable) Carrollian clock $A$. Indeed, this makes it possible to define a horizontal lift of differential operators generalising the lift \eqref{s22t} of vector fields:
\be
\label{lift}
\ell
\,:\,
\cD(\bar\sM)\hookrightarrow\cD_{\text{inv}}(\sM)
\,:\,
\bar\DD\mapsto\ell(\bar\DD),
\ee
where $\ell(\bar\DD)$ is the unique invariant horizontal differential operator whose projection on the base is $\bar D$. The isomorphism \eqref{Dinv} then states that any invariant differential operator $\DD\in\cD^k_{\text{inv}}(\sM)$ of order $k$ takes the form
\be
\label{invgenform}
\DD
=
\sum\limits_{n=0}^k \ell(\bar\DD_{n})\circ \xi^n
\qquad\text{with}\qquad
\bar\DD_n\in\cD^{k-n}(\bar\sM)\,,
\ee
and is thus entirely specified by differential operators living on the base $\bar\sM$ only. (This will be used in section \ref{symsSachs} when characterising \sproj\ differential operators.)

\paragraph{Higher Carrollian conformal symmetries.} In \cite{Eastwood:2002su} Eastwood defined a \itt{higher symmetry} of the conformal Laplacian $\Delta$ as a differential operator $\bar\DD\in\cD(\bar{\sM})$ on a conformal manifold $\bar{\sM}$ such that $\Delta\circ\bar\DD=\bar\FF\circ\Delta$ for some $\bar\FF\in\cD(\bar{\sM})$. Such higher symmetries span an associative algebra $\cD_{\text{sym}}(\bar{\sM})$, to which we shall return below (see eq.\ \eqref{projsyms}). By analogy, we call \itt{higher Carrollian conformal symmetry} any projectable differential operator $\DD\in\cD_{\text{pro}}(\sM)$ such that 
\be
\label{hisywrac}
\hxi^2\circ\DD=\FF\circ\hxi^2
\qquad\text{and}\qquad
\Delta\circ\bar\DD=\bar\FF\circ\Delta
\qquad\text{for some}\qquad
\FF\in\cD_{\text{pro}}(\sM)\,,
\ee 
where $\bar\FF$ is the projection of $\FF$ on the base.\footnote{In contrast to sections \ref{seSCAL}--\ref{seCAR}, $\FF$ and $\bar\FF$ now denote differential operators rather than diffeomorphisms.} In the BMS context, such symmetries should be seen as higher-spin generalisations of infinitesimal supertranslations and superrotations, understood here in the sense of conformal maps on celestial spheres \cite{Barnich:2009se,Barnich:2010eb}. Non-conformal generators are excluded by design, since generalised conformal maps do not preserve the kernel of the Yamabe operator. Note that the square of $\xi$ in \eqref{hisywrac} is required to ensure that the first and second conditions are mutually compatible. In fact, the operators $\xi^2$ and $\Delta$ have the same conformal weight ($w=-2$), as they should in order for the conditions \eqref{hisywrac} to reduce to \eqref{cCK'} for first-order symmetries (see below).\footnote{By analogy, one expects the GJMS generalisation of \eqref{hisywrac} to be $\hxi^{2n}\circ\DD=\FF\circ\hxi^{2n}$ and $P_{2n}\circ\bar\DD=\bar\FF\circ P_{2n}$, where $P_{2n}$ is the GJMS operator mentioned in eq.\ \eqref{s5t} for $\bar\sM=S^d$. It is natural to anticipate that the corresponding algebras of non-trivial higher symmetries are (Wick-rotated versions of) the higher-spin algebras discussed in \cite{Bekaert:2013zya} and references therein.} Also note that the second condition in \eqref{hisywrac} merely says that the projected operator $\bar\DD$ is a higher symmetry of the Yamabe operator on the base, while the projectability condition ensures that $\DD$ preserves the time-independence of the WRac in \eqref{wreq}. This implies the following result:

\paragraph{Proposition: Higher symmetries of the WRac.} If a first-order operator $\DD$ is a higher Carrollian conformal symmetry, then $D$ is the completion of a Carrollian conformal Killing vector field $X$ (in the sense of the proposition on page \pageref{propconf}) and thus preserves the WRac equations \eqref{wreq}. More generally, \itt{all} higher symmetries of the Carrollian conformal Laplacian preserve the space of solutions of \eqref{wreq}; they span an associative algebra, denoted $\cD_{\text{sym}}(\sM)$.

\proof{%
Let $\DD\in\cD^1_{\text{pro}}(\sM)$ be a projectable first-order operator that is also a higher Carrollian conformal symmetry. Then $\DD$ satisfies the conditions \eqref{hisywrac}, that is,
\be
\label{HScK}
[\hxi^2,\DD\,]
=
\GG\circ\hxi^2\,,
\qquad
[\Delta,\bar\DD\,]
=
\bar\GG\circ\Delta
\ee
for $\GG=\FF-\DD\in\cD^0_{\text{pro}}(\sM)$. We need to show that these conditions imply the commutators \eqref{cCK'}. To see this, use the fact that $\DD$ is projectable along with eq.\ \eqref{projop}, which is equivalent to the condition $[\xi,\DD]=f\,\xi$ for some $f\in C^{\infty}(\sM)$. But this implies $[\xi^2,\DD]=(\cL_{\xi}f+2f\xi)\circ\xi$, which reproduces the first equation of \eqref{HScK} by assumption, so $\cL_{\xi}f=0$ and $\GG=2f$. Hence $f=\pi^*\bar f$ and $\bar\GG=2\bar f$, which allows us to identify the second equation in \eqref{HScK} with the Laplacian commutator in \eqref{cCK'}, proving that $\DD$ is indeed the completion of a Carrollian conformal vector field.

We now turn to higher-order differential operators. Accordingly, let $\DD\in\cD_{\text{pro}}(\sM)$ satisfy eqs.\ \eqref{hisywrac} and let $\phi$ solve the WRac equations \eqref{wreq}. It is then obvious that $\DD\phi$ solves the same equations, since $\cL_{\xi}(\DD\phi)=\EE(\cL_{\xi}\phi)=0$ and $\Delta(\bar\DD\bar\phi)=\bar\FF(\Delta\bar\phi)=0$. Furthermore, eqs.\ \eqref{hisywrac} and the projectability condition \eqref{projop} are linear in $\DD$, so the set of higher symmetries of the WRac is a vector space. Finally, it is an associative algebra because the composition of
two symmetries $\DD_1,\DD_2\in\cD_{\text{pro}}(\sM)$ with associated operators $\FF_1,\FF_2\in\cD_{\text{pro}}(\sM)$ is a symmetry $\DD_1\circ\DD_2$ with associated operator $\FF_1\circ\FF_2$.
\qed}

\paragraph{Vertical symmetries.} It is now time to turn to the ideal that will eventually be modded out of the WRac higher-spin symmetry algebra. This ideal will consist of two pieces: vertical symmetries and ``trivial'' Laplacian symmetries. We begin with the former. Accordingly, a differential operator $\DD$ on $(\sM,\xi)$ will be called \itt{vertical} if it is the composition of the fundamental vector field with another differential operator, \ie if it reads $\DD=\EE\circ\hxi$ for some $\EE\in\cD(\sM)$. Any vertical operator is automatically projectable; in fact it projects to zero on the base manifold (\ie $\bar\DD=0$). 

Vertical higher Carrollian conformal symmetries are higher-spin generalisations of infinitesimal supertranslations. They span an associative algebra $\cD_{\text{vsym}}(\sM)$ which is a bilateral ideal of the algebra $\cD_{\text{sym}}(\sM)$ of all higher symmetries.

\paragraph{Proposition: Projected WRac symmetries.} The quotient of the algebra $\cD_{\text{sym}}(\sM)$ of higher Carrollian conformal symmetries by the ideal $\cD_{\text{vsym}}(\sM)$ of vertical higher symmetries is isomorphic to the algebra of higher symmetries of the conformal Laplacian on the base,
\be
\label{projsyms}
\cD_{\text{sym}}(\sM)/\cD_{\text{vsym}}(\sM)\cong\cD_{\text{sym}}(\bar\sM)\,.
\ee

\proof{The higher symmetries of the Carrollian conformal Killing-Laplacian are projectable differential operators, so the pushforward morphism $\pi_*:\cD_{\text{sym}}(\sM)\to\cD(\bar\sM)$ is well-defined. Its kernel is $\cD_{\text{vsym}}(\sM)$ and its image is $\cD_{\text{sym}}(\bar\sM)$, so \eqref{projsyms} follows as an isomorphism of algebras.
\qed}

\paragraph{Higher-spin algebra of the WRac.} In \cite{Eastwood:2002su} Eastwood defined a \itt{trivial symmetry} of the conformal Laplacian $\Delta$ on $\bar\sM$ as being any differential operator of the form $\bar\DD=\bar\EE\circ\Delta$ for some $\bar\EE\in\cD(\bar\sM)$. Such trivial symmetries span a bilateral ideal of the algebra of higher symmetries of the conformal Laplacian, whose quotient by this ideal will be called the \itt{Eastwood-Vasiliev higher-spin algebra} of $(\bar\sM,\Delta)$ \cite{Eastwood:2002su,Vasiliev:2003ev,Nikitin,shapovalov1992symmetry}, and denoted $\hs(\bar\sM)$. The Carrollian generalisation is straightforward: the algebra of non-trivial higher Carrollian conformal symmetries is the quotient of \eqref{projsyms} by the ideal of higher symmetries $\DD$ that project to trivial symmetries $\bar\DD=\bar\EE\circ\Delta$ of the conformal Laplacian on the base. In the case of null infinity, this quotient is the \itt{algebra of non-trivial higher symmetries of the WRac} \eqref{wreq}.

\paragraph{Proposition: Non-trivial symmetries of the WRac.} The algebra of non-trivial higher symmetries of the WRac on null infinity $\sI\cong\RR\times S^d$ is isomorphic to the Eastwood-Vasiliev higher-spin algebra $\hs(S^d)$ of non-trivial higher symmetries of the conformal Laplacian on the celestial sphere\,.

\proof{%
The proof is immediate since the system \eqref{wreq} is merely an equivalent description of the conformal scalar field on $S^d$ as a scalar field on $\sI\cong\RR\times S^d$. Specifically, the isomorphism \eqref{projsyms} states that the quotient $\cD_{\text{sym}}(\sI)/\cD_{\text{vsym}}(\sI)\cong\cD_{\text{sym}}(S^d)$ is isomorphic to the algebra of higher symmetries of the conformal Laplacian on $S^d$, which reduces the problem to the results in \cite{Eastwood:2002su,Nikitin,shapovalov1992symmetry}. \qed}

The Eastwood-Vasiliev algebra found here is manifestly too small to be a proper candidate higher-spin symmetry of asymptotically flat spacetimes, since all supertranslations (and their powers) are modded out of \eqref{projsyms}.\footnote{However, note in relation with footnote \ref{winfty} that $\hs(S^2)\cong\hs(S^1)\oplus\hs(S^1)$, where $\hs(S^1)\cong{\cal D}(S^1)$ may be thought of as a %sort of
$w_{1+\infty}$ algebra %with vanishing central charge 
\cite[sec.\ 4]{Bekaert:2006zoe}, enhancing to ${\cal W}[\lambda]$ as in standard higher-spin AdS$_3$/CFT$_2$ \cite{Henneaux:2010xg,Gaberdiel:2010ar,Gaberdiel:2010pz,Campoleoni:2011hg}. It is conceivable that the WRac fits in an interesting higher-spin structure in that sense, but we will not investigate this possibility.} This confirms that the WRac is \itt{not} a promising Minkowskian analogue of the singleton in AdS, and leads us to consider the other candidate encountered in section \ref{seSCAL}: the Sachs module.

\subsection{Higher symmetries of the Sachs module}
\label{symsSachs}

Similarly to section \ref{symsrac}, our goal here is to build a higher-spin extension of the BMS algebra that preserves a suitable Carrollian structure. We shall achieve this by considering all differential operators obtained as powers of \sproj\ vector fields. Indeed, in contrast to section \ref{symsrac}, there will be no need to restrict attention to conformal generators: we consider generalised BMS transformations \cite{Campiglia:2014yka,Campiglia:2015yka,Colferai:2020rte} throughout. Furthermore, the condition that symmetry generators preserve the Sachs form \eqref{sachspo}--\eqref{phipsi} will force them to be Hermitian. The resulting symmetries will span a Lie algebra, but \itt{not} an associative one as was the case in section \ref{symsrac}.

The plan is as follows. We start by defining \sproj\ differential operators and working out their (associative) algebra. Then we introduce higher symmetries of the Sachs module---namely differential operators that satisfy a suitable Hermiticity condition---and show that they span a Lie algebra to be thought of as a higher-spin extension of BMS. Finally, we relate this extension to the symmetries of partially massless higher-spin theory. The link with asymptotic symmetries \cite{Campoleoni:2017mbt,Campoleoni:2017qot,Campoleoni:2018uib,Campoleoni:2020ejn,Campoleoni:2021blr} is relegated to section \ref{sehispin}.

\paragraph{Super-projectable differential operators.} Let $(\sM,\xi)$ be the principal $\RR$-bundle of section \ref{seCAR} and let $\DD\in\cD^k_{\text{pro}}(\sM)$ be a projectable differential operator of order $k$. Then $\DD$ is \itt{\sproj} if $[\hxi,\DD]=\EE\circ\hxi$ for some \sproj\ operator $\EE\in\cD^{k-1}_{\text{pro}}(\sM)$ of order $k-1$. This definition may seem circular, but it is not: rather, it provides an inductive characterisation of super-projectability. For instance, a \sproj\ operator $f$ of order zero is an invariant function ($\cL_\xi f=0$); and any \sproj\ operator of order one is the sum of a \sproj\ vector field (recall section \ref{sebundle}) and an invariant function. Note the ensuing inclusions of algebras generalising \eqref{s19b}:
\be
\begin{array}{ccccccc}
\cD_{\text{inv}}(\sM) & \subset & \cD_{\text{spro}}(\sM) & \subset 
& \cD_{\text{pro}}(\sM) & \subset & \cD(\sM)
 \\
\cup && \cup 
&& \cup&& \cup\\
\mX_{\text{inv}}(\sM) & \subset & \mX_{\text{spro}}(\sM) & \subset 
& \mX_{\text{pro}}(\sM) & \subset & \mX(\sM)
\end{array}
\ee
where the first line involves associative algebras while the second one involve Lie algebras. We will soon encounter a similar hierarchy involving Poincar\'e and (generalised) BMS algebras along with their higher-spin extensions.

Recall from section \ref{se32} that \sproj\ vector fields span a semi-direct sum \eqref{s22q}, where the fibrewise $\mathfrak{igl}(1)$ subalgebra is generated by the fundamental vector field $\xi$ and the canonical Euler vector field $\eta$ determined by a choice of global section. This structure now carries over to \sproj\ differential operators:

\paragraph{Proposition: Super-projectable operators.} The associative algebra $\cD_{\text{spro}}(\sM)$ of \sproj\ differential operators is a tensor product
\be
\label{e95}
\cD_{\text{spro}}(\sM)\cong 
\cD(\bar{\sM})\otimes U\big(\mathfrak{igl}(1)\big)\,
\ee
where $U\big(\mathfrak{igl}(1)\big)$ is the associative algebra generated by the fundamental vector field $\xi$ and the canonical Euler vector field $\eta$. Thus, as a vector space, $\cD_{\text{spro}}(\sM)$ is isomorphic to the tensor product $\Gamma(\odot \,T\bar\sM)\otimes\RR[\xi,\eta]$, where $\Gamma(\odot\, T\bar\sM)$ is the space of symmetric multivector fields on the base while $\RR[\xi,\eta]$ is the space of polynomials in the fundamental vector field and the Euler vector field. Moreover, any \sproj\ differential operator $\DD\in\cD^k_{\text{spro}}(\sM)$ of order $k$ takes the form
\be
\label{sform}
\DD
=
\sum\limits_{m=0}^k \DD_{m}\circ \eta^m\,,
\qquad\text{where}\qquad
\DD_{m}\in\cD^{k-m}_{\text{inv}}(\sM)\,.
\ee
Note from the vector space isomorphism $\cD_{\text{spro}}(\sM)\cong\cD(\bar\sM)\otimes\RR[\xi,\eta]$ that \eqref{Dinv} is obviously a subalgebra of \eqref{e95}.

\proof{%
The structure of the algebra \eqref{e95} is an immediate consequence of the expression \eqref{s22q} for the algebra of \sproj\ vector fields. Thus it only remains to prove the decomposition \eqref{sform}; we do this by induction and in adapted coordinates where $\xi=\der_u$ and $\eta=u\,\der_u$, since computations involving differential operators are always local. Accordingly, suppose $\DD\in\cD^0_{\text{spro}}(\sM)$ is of order zero; then $\DD=f(x)$ is an invariant function, which trivially satisfies \eqref{sform} with $k=0$. If $\DD\in\cD^1_{\text{spro}}(\sM)$ has order one, then it is the sum of an invariant function and a \sproj\ vector field; the coordinate expression \eqref{sprovec} then ensures that \eqref{sform} is still valid. Now suppose that \eqref{sform} holds for all $k=0,1,...,K$. Then a \sproj\ operator $\DD\in\cD^{K+1}_{\text{spro}}(\sM)$ of order $K+1$ is such that $[\hxi,\DD]=\EE\circ\hxi$ with $\EE\in\cD^K_{\text{spro}}(\sM)$, of order $K$, of the form \eqref{sform}. The inductive definition thus becomes a first-order differential equation in $u$, whose solution $\DD$ is readily found to take the form \eqref{sform} with $K$ replaced by $K+1$. Explicitly,
\be
\EE
=
\sum\limits_{m=0}^K\sum\limits_{n=0}^{K-m} G_{m,n}\,u^m\der_u^{m+n}
~~\Rightarrow~~
\DD
=
\sum\limits_{m=1}^{K+1}\sum\limits_{n=0}^{K+1-m}\frac{G_{m-1,n}}{m} \,u^m\der_u^{m+n}
+
\sum_{n=0}^{K+1}F_{0,n}\,\der_u^n
\ee
where the $G_{m,n}$'s and the $F_{m,n}$'s are horizontal invariant differential operators and the $u$-independent operator on the far right-hand side is an ``integration constant''. This proves \eqref{sform} upon recalling \eqref{invgenform}. \qed}

We are now ready to investigate the algebra of \sproj\ higher symmetries of the Sachs form. This will involve a notion of Hermiticity, eventually producing a Lie subalgebra of \eqref{e95} that greatly extends the generalised BMS symmetries of section \ref{se34}.

\paragraph{Higher symmetries of the Sachs module.} Consider a Campiglia-Laddha structure $(\sM,[\xi,\eps])$. Pick a representative $(\xi,\eps)$ and introduce the non-degenerate Hermitian form 
\be
\label{phipsi'}
(\phi|\psi)
\,:=\,
\int_{\sM}\phi^*\,\psi\,\eps
\qquad\forall\phi,\psi\in C^\infty(\sM,\CC)\,,
\ee
which is nothing but the standard inner product of complex-valued ``wavefunctions'' on $\sM$ with volume form $\eps$. The Carrollian inner product \eqref{phipsi} can then be written as
\be
\label{Hprod}
\langle\phi|\psi\rangle
\,=\,
i\,(\phi|\hxi|\psi)\,.
\ee
Hermitian conjugation with respect to \eqref{phipsi'} will be denoted by a dagger, \ie $(\DD^{\dagger}\phi,\psi):=(\phi,\DD\psi)$ for any differential operator $\DD\in\cD(\sM)$. Thus, the Hermitian conjugate of a function is just its complex conjugate, while that of a vector field $X$ is 
\be
X^{\dagger}=-X^*-\text{div}(X^*)\,,
\ee
with the divergence defined by \eqref{didef} in terms of the invariant volume form. In the particular case of a Carrollian spacetime, this divergence simply reads $\text{div}(X)=\der_uX^u+\nabla_aX^a$ in adapted coordinates $(u,x^a)$.

We now define a \itt{higher symmetry of the Carrollian Hermitian form} \eqref{phipsi}--\eqref{Hprod} (or a \textit{Hermitian symmetry} for short) as being a differential operator $\DD\in\cD(\sM)$ such that the infinitesimal transformations $\delta\phi=i\DD\phi$ and $\delta\psi=i\DD\psi$ preserve $\langle\phi|\psi\rangle$ for all $\phi,\psi$. The relation \eqref{Hprod} shows that this requirement is equivalent to the \itt{Hermiticity condition}
\be
\label{Hform}
\hxi\circ\DD
=
\DD^\dagger\circ\hxi\,.
\ee
For instance, a zeroth order operator is Hermitian in the above sense if it is a real invariant function. And a first-order operator $\hat X=X+h$ is Hermitian if $X$ is a purely imaginary projectable vector field while $\text{Im}(h)=\tfrac{1}{2}\,\text{div}\,X$ and $\text{Re}(h)$ is an invariant function. Thus we recover, as expected, the statement of the end of section \ref{se34} that \itt{all} bundle automorphisms preserve the Carrollian Hermitian form provided wavefunctions are Carrollian densities with weight $\tw=1/2$. This includes all \sproj\ vector fields, which in turn includes all \CL\ vector fields and of course all Carrollian conformal vector fields. Also note that, given a Carrollian clock, there exists a large collection of Euler vector fields $\eta$ such that $\xi\circ\eta=-\eta^\dagger\circ\xi$: in adapted coordinates, they take the form $\eta=u\,\der_u +X^a(x)\der_a$ with $\bar X=X^a(x)\der_a$ divergenceless with respect to $\bar\eps$. In the case of null infinity, the higher symmetries of the Carrollian Hermitian form \eqref{sachspo} will be called \itt{higher symmetries of the Sachs module}; they include, in particular, higher-order extensions of generalised BMS generators.

Hermitian higher symmetries span a real Lie algebra with Lie bracket $i\times$commutator. In contrast to section \ref{symsrac}, the algebra is \itt{not} associative because the composition of two Hermitian operators is generally not Hermitian. Given a representative Carrollian measured space $(\sM,\xi,\eps)$ endowed with a synchronised clock, this algebra is completely characterised as follows:

\paragraph{Proposition: All Hermitian symmetries.}  The Lie algebra $\cH_{\text{sym}}(\sM)$ of all higher symmetries of the Carrollian Hermitian form \eqref{phipsi}--\eqref{Hprod} is isomorphic to the semi-direct sum
\be
\label{Dsusy}
\cH_{\text{sym}}(\sM)
\cong 
\cH(\bar{\sM})\,\inplus\,\cH_{\text{vsym}}(\sM)\,
\ee
of the Lie algebra $\cH(\bar{\sM})$ of differential operators on the base that are Hermitian with respect to
\be
\label{barphipsi}
(\bar\phi|\bar\psi)
\,:=\,
\int_{\bar{\sM}}\bar\phi^*\,\bar\psi\,\bar\eps\qquad\forall\bar\phi,\bar\psi\in C^\infty(\bar{\sM}),
\ee
and the Lie ideal $\cH_{\text{vsym}}(\sM)$ of all vertical higher symmetries. More precisely, any higher symmetry $\DD$ decomposes as $\DD=\EE\circ\xi+\ell(\bar H)$ where $\EE$ is antihermitian with respect to \eqref{phipsi'} while $\bar H$ is hermitian with respect to \eqref{barphipsi}, and $\ell$ is the horizontal lift \eqref{lift}.

\proof{To start, consider a vertical differential operator $\DD=\EE\circ\hxi$. In order for it to satisfy \eqref{Hform}, one needs $\hxi\circ \EE\circ\hxi=-\hxi\circ\EE^\dagger\circ\hxi$, hence $\EE^\dagger=-\EE$ since $\hxi$ is nowhere vanishing. Now check that vertical symmetries span an ideal by computing $D_3:=i[D_1,D_2]$ for $D_1=A\circ\hxi$ a vertical symmetry and $D_2$ any symmetry, which shows that $D_3=i(A\circ D_2^\dagger-D_2\circ A)\circ\hxi$ is also a vertical symmetry. Thus the quotient $\cH_{\text{sym}}(\sM)/\cH_{\text{vsym}}(\sM)$ is a Lie algebra. Given a horizontal lift, one can consider representatives that are invariant horizontal differential operators. Moreover, the Hermitian conjugations with respect to  \eqref{phipsi'} and \eqref{barphipsi} commute with the horizontal lift:\footnote{For simplicity, we slightly abuse notation by using the same symbol for both conjugations.} ${}^\dagger\circ\ell=\ell\circ {}^\dagger$, as is obvious in coordinates. Thus the  condition \eqref{Hform} for the invariant horizontal lift $D=\ell(\bar D)$ of a differential operator $\bar D$ on the base is equivalent to the hermiticity condition $\bar D^\dagger=\bar D$ on the base, and the isomorphism \eqref{Hform} follows.
\qed}

It is clear from this result that the algebra of all higher symmetries is huge, so some criterion is needed to restrict attention to those symmetries that are actually relevant. Here we adopt the point of view that the ``interesting'' generators are \sproj\ (which includes \CL\ transformations), for which the following holds:

\paragraph{Proposition: Super-projectable Hermitian symmetries.} Let $(\sM,\xi,\eps)$ be a Carrollian measured space endowed with a global section defining the canonical Euler vector field $\eta$. Then, the Lie algebra $\cH_{\text{spro}}(\sM)$ of \sproj\ higher symmetries of the Carrollian Hermitian form \eqref{phipsi}--\eqref{Hprod}
 is a tensor product
\be
\label{Dspro}
\cH_{\text{spro}}(\sM)
\cong 
\cH(\bar{\sM})\otimes U_+\big(\mathfrak{igl}(1)\big)\,
\ee
where $\cH(\bar{\sM})$ is the Lie algebra of differential operators on the base that are Hermitian with respect to \eqref{barphipsi}, while $U_+\big(\mathfrak{igl}(1)\big)$ is the real form of the universal enveloping algebra which is spanned by Weyl-ordered polynomials in $i\,\xi$ and $i\,\eta$. Thus, as a vector space, $\cH_{\text{spro}}(\sM)$ is isomorphic to the tensor product $\Gamma_{\RR}(\odot T\bar\sM)\otimes\RR[i\xi,i\eta]$ where the subscript in $\Gamma_{\RR}$ stresses that one only considers real symmetric multivector fields.

\proof{Let $\ell$ be the horizontal lift \eqref{lift}. Using the decompositions \eqref{invgenform}--\eqref{sform}, any \sproj\ differential operator $\DD$ can be written as
\be
\label{supergenform'}
\DD=\sum\limits_{m=0}^k\sum\limits_{n=0}^{k-m} \ell\big(\bar\DD_{m,n}\big)\circ\big[\,(i\eta)^m\circ(i\xi)^n+\cdots\big]\,,\qquad\bar\DD_{m,n}\in\cD^{k-m-n}(\bar\sM)\,,
\ee
where the dots stand for terms that enforce Weyl-ordering of the polynomial in $i\eta$ and $i\xi$. What is the Hermitian conjugate of \eqref{supergenform'}? To answer this, note first that $\eta$ and $\xi$ commute with any invariant horizontal differential operator by definition. Also recall that the Hermitian conjugations with respect to  \eqref{phipsi'} and \eqref{barphipsi} commute with the horizontal lift: ${}^\dagger\circ\ell=\ell\circ {}^\dagger$. The Hermitian conjugate of \eqref{supergenform'} therefore reads
\be
\label{supergenform''}
\DD^\dagger
=
\sum\limits_{m=0}^k\sum\limits_{n=0}^{k-m} \ell\big(\bar\DD^\dagger_{m,n}\big)\circ\Big[\,\big((i\eta)^\dagger\big)^m\circ\big((i\xi)^\dagger\big)^n+\cdots\big]\,,
\ee
where the dots still stand for Weyl-ordering terms. Now using the fact that $i\eta$ and $i\hxi$ are higher symmetries of \eqref{phipsi}, the Hermiticity condition \eqref{Hform} becomes
\be
\label{supergenformsym}
\sum\limits_{m=0}^k\sum\limits_{n=0}^{k-m} \ell\big(\bar\DD^\dagger_{m,n}-\bar\DD_{m,n}\big)\circ\xi\circ\big[\,(i\eta)^m\circ(i\xi)^n+\cdots\big]
=
0\,.
\ee
This implies that $\ell\big(\bar\DD^\dagger_{m,n}-\bar\DD_{m,n}\big)=0$ for all $m,n$, which requires $\bar\DD^\dagger_{m,n}=\bar\DD_{m,n}$ since the horizontal lift $\ell$ is injective, proving the isomorphism \eqref{Dsusy}. \qed}

This concludes the characterisation of Carrollian higher-spin symmetries in full generality. From now on, and until the end of this work, we focus on the application of these ideas to BMS symmetry at null infinity.

\paragraph{Proposition: Higher-spin extensions.} Any real Lie algebra $\mathfrak{g}$ in the hierarchy \eqref{hierarchy} admits a higher-spin extension $\mathfrak{hg}$, built as the real Lie subalgebra of higher symmetries of the Sachs module spanned by Weyl-ordered products of the generators of $\mathfrak{g}$. The hierarchy \eqref{hierarchy} then extends to the following inclusions:
\be
\label{columns}
\begin{array}{ccccccc}
\mathfrak{hiso}(d+1,1) & \subset & \mathfrak{hbms}_{d+2} & \subseteq
& \mathfrak{hgbms}_{d+2} & \subset &\cD_{\text{spro}}(\sI_{d+1}) \\
\cup && \cup 
&& \cup&& \cup\\
\mathfrak{iso}(d+1,1) & \subset & \mathfrak{bms}_{d+2} & \subseteq
& \mathfrak{gbms}_{d+2} & \subset & \mX_{\text{spro}}(\sI_{d+1})
\end{array}
\ee

\proof{%
The proof is constructive and applies to any Lie algebra, so we start by spelling it out in full generality before applying it to the case of interest here. Let $\mathfrak{g}_{\CC}$ be the complexification of a real Lie algebra $\mathfrak{g}_{\RR}$ and denote the corresponding universal enveloping algebras by $U(\mathfrak{g}_{\RR})$ and $U(\mathfrak{g}_{\CC})$ (the former is a real form of the latter). Let also $\iota$ be an involution (\ie an involutive antilinear antiautomorphism) of $U(\mathfrak{g}_{\CC})$, and call $\mathfrak{g}_\pm\subset\mathfrak{g}_{\CC}$ the eigenspace of $\iota$ with eigenvalue $\pm1$, respectively, so that $\mathfrak{g}_{\CC}=\mathfrak{g}_+\oplus\mathfrak{g}_-$ as real vector spaces. Now pick a basis $\{T_i\}$ of $\mathfrak{g}_{\CC}$ whose elements are eigenvectors of $\iota$ with eigenvalue one, yielding $\mathfrak{g}_+=\text{span}_{\RR}\{T_i\}\cong\mathfrak{g}_{\RR}$ and $\mathfrak{g}_-=\text{span}_{\RR}\{\text{i}\,T_i\}$. Since $\mathfrak{g}_+$ is a real Lie algebra with $i\times$commutator as Lie bracket, we shall assume that $\mathfrak{g}_+$ is isomorphic to $\mathfrak{g}_{\RR}$. It only remains to extend this to enveloping algebras: let $U_\pm(\mathfrak{g}_{\CC})\subset U(\mathfrak{g}_{\CC})$ denote the eigenspace of $\iota$ with eigenvalue $\pm1$ (respectively), so that $U_+(\mathfrak{g}_{\CC})$ endowed with the bracket $i\times$commutator is a real Lie algebra $\mathfrak{hg}_{\RR}$ such that $\mathfrak{g}_{\RR}\subset\mathfrak{hg}_{\RR}$ is a Lie subalgebra. The Poincar\'e-Birkhoff-Witt theorem then ensures that $U(\mathfrak{g}_{\CC})=\text{span}_{\CC}\{T_{i_1}\cdots T_{i_r}\}$ is spanned by Weyl-ordered products of generators with complex coefficients. Thus $U_+(\mathfrak{g}_{\CC})=\text{span}_{\RR}\{T_{i_1}\cdots T_{i_r}\}$, which proves that $\mathfrak{hg}_{\RR}$ is the real Lie algebra spanned by Weyl-ordered products of generators with real coefficients.

In the specific examples above, the involution $\iota$ is Hermitian conjugation with respect to the Carrollian inner product \eqref{Hprod}: $\langle\phi,\DD\psi\rangle=\langle\,\iota(\DD)\phi,\psi\rangle$, and the proof applies in exactly the same way from that point on. A minor subtlety is that the higher-spin extensions considered in the proposition are \itt{not} actually universal enveloping algebras, but quotients thereof (by the annihilator on the Sachs module). The proof applies nevertheless since, by definition, the higher-spin extensions in the proposition are spanned by Weyl-ordered products of generators.
\qed}

\paragraph{Generalised BMS higher-spin algebra.} Recall that any \CL\ vector field $X$ is \sproj\ and admits an extension to a first-order symmetry of the Sachs module. An analogous property remains true for Weyl-ordered polynomials of such first-order symmetries, so $\mathfrak{gbms}_{d+2}$ is a Lie subalgebra of the Lie algebra $\cH_{\text{spro}}(\sI_{d+1})$ of all \sproj\ symmetries of the Sachs module.

The $\mathfrak{gbms}_{d+2}$ subalgebra can be characterised in detail by focussing on the underlying vector space. Indeed, as a vector space, the Lie algebra \eqref{dimsumbis}-\eqref{semidirect} of \CL\ vector fields is isomorphic to the space \eqref{al3} of invariant vector fields because the time-dependence of generalised conformal vector fields is entirely fixed by data on $\bar\sM$. In particular, $\mathfrak{gbms}_{d+2}\cong\mX_{\text{inv}}(\sI_{d+1})$ as vector spaces. This remains true for Weyl-ordered products of \CL\ vector fields: as vector spaces, the higher-spin extension of the generalised BMS algebra is isomorphic to the algebra \eqref{Dinv} of invariant differential operators on null infinity, that is, $\mathfrak{hgbms}_{d+2}\cong\cD_{\text{inv}}(\sI_{d+1})$. One can also see this in local coordinates, where any element $\DD$ of $\mathfrak{hgbms}_{d+2}$ is uniquely determined by an invariant differential operator
\be
\label{D_0}
\DD_0\,
=
\,\sum_{r,q\geqslant 0} K^{a_1\cdots a_r}_{(q)}(x)\nabla_{a_1}\cdots\nabla_{a_r}\der^q_u\,,
\ee
in the sense that $\DD_0$ admits a unique completion of the form $\DD=\DD_0+{\cO}(u)$ such that $\DD\in\mathfrak{hgbms}_{d+2}$. We shall return to this type of expression in section \ref{sehispin}, where $K$'s will be interpreted as residual gauge parameters in a higher-spin gravity theory. Note for later purposes that $K^{a_1\cdots a_r}_{(q)}(x)$ has scaling dimension $-(r+q)$, as follows from elementary dimensional analysis.

\paragraph{Minkowski higher-spin algebras.} The real Lie algebra spanned by Hermitian differential operators on Minkowski spacetime that commute with the d'Alembertian operator was first discussed in \cite{Vasiliev:2005zu}, and dubbed \itt{off-shell Minkowski higher-spin algebra} in \cite{Bekaert:2008sa} because it is the ``flat limit'' (really an In\" on\" u-Wigner contraction) of the off-shell AdS higher-spin algebra introduced in \cite{Vasiliev:2003ev}. (See \eg Proposition 2 and Section 3.2.3 in \cite{Bekaert:2008sa} for a proof.) It is spanned by all Weyl-ordered products of Killing vector fields spanning the Poincar\'e algebra (see \cite[Cor.\ 6]{Bekaert:2008sa}). By construction, the d'Alembertian operator is a central element in this algebra, so all products of the d'Alembertian with elements of the off-shell Minkowski higher-spin algebra span an ideal. The quotient of the off-shell Minkowski higher-spin algebra by this ideal will be called the \itt{partially-massless Minkowski higher-spin algebra} since it is a flat limit \cite[sec.\ 6.1.2]{Campoleoni:2021blr} of the partially-massless AdS higher-spin algebra introduced in \cite{Joung:2015jza}. It can be seen as the realisation of off-shell Minkowski higher-spin algebra on the space of solutions of the d'Alembert equation, since the ideal that is quotiented out corresponds to the operators annihilating the solutions of the d'Alembert equation. We can now identify this structure in the first column of \eqref{columns}:

\paragraph{Proposition: Higher-spin algebra on Minkowski.} The partially-massless Min\-kow\-ski higher-spin algebra on $\mathbb{R}^{d+1,1}$ is isomorphic to the Lie subalgebra $\mathfrak{hiso}(d+1,1)$ of higher symmetries of the Sachs module on $\sI_{d+1}$.

\proof{%
The Poincar\'e algebra is a subalgebra $\mathfrak{iso}(d+1,1)\subset\mathfrak{bms}_{d+2}$ of the Lie algebra of Carrollian conformal vector fields of null infinity $\sI_{d+1}$. It consists of Carrollian conformal vector fields that can be extended to Killing vector fields in the interior of Minkowski spacetime. Accordingly, one can extend Weyl-ordered products of Carrollian conformal vector fields generating $\mathfrak{iso}(d+1,1)$ to Weyl-ordered products of Killing vector fields in Minkowski spacetime. This defines an injective linear map sending higher symmetries of the Sachs module to differential operators on compactified Minkowski spacetime. (To prove injectivity, note that the kernel vanishes since the restriction of a vanishing differential operator on compactified Minkowski spacetime obviously vanishes on the conformal boundary.) Furthermore, the map is invertible since Weyl-ordered products of Killing vector fields on Minkowski spacetime acting on the space of solutions of the d'Alembert equation span the partially-massless Minkowski higher-spin algebra; in particular, their action on radiative solutions (\ie on the Sachs module \eqref{e52} seen as on-shell fields in Minkowski spacetime) induces their action on the corresponding boundary data (\ie the Sachs module seen as fields at null infinity). It follows that Weyl-ordered products of Carrollian conformal vector fields on $\sI_{d+1}$ that span the Lie subalgebra $\mathfrak{iso}(d+1,1)\subset\mathfrak{bms}_{d+2}$ form a Lie algebra isomorphic to the partially-massless Minkowski higher-spin algebra.
\qed}

\subsection{Asymptotic higher-spin symmetries}
\label{sehispin}

To conclude this work, we now relate the structures found above to the asymptotic higher-spin symmetries studied in \cite{Campoleoni:2017mbt,Campoleoni:2017qot,Campoleoni:2018uib,Campoleoni:2020ejn}. We start by recalling elementary aspects of higher-spin gravity theories and their linearised gauge transformations, and briefly review three classes of fall-off conditions for Fronsdal fields. We then compare the resulting asymptotic symmetries to the WRac and Sachs higher symmetries investigated in sections \ref{symsrac}--\ref{symsSachs}.

\paragraph{Asymptotic Killing tensors.} The starting point of \cite{Campoleoni:2017mbt,Campoleoni:2017qot} was to write the Fronsdal equations of motion for a totally-symmetric, doubly-traceless tensor gauge field $\phii$ of rank $s$ on Minkowski spacetime $\RR^{d+1,1}$ in retarded Eddington-Finkelstein coordinates $(r,u,x^a)$, as we did for a scalar in section \ref{sescad}. The Bondi-like gauge $\phii_{r\,\cdots}=0$ and $\bbgamma^{ab}\phii_{ab\,\cdots}=0$ then implies that the non-vanishing components of $\phii$ are of the form $\phii_{u(s-k)\,a_1\cdots a_k}$, where $u(t)$ is a shorthand for $t$ indices $u$. The question raised in \cite{Campoleoni:2017qot} was to find the corresponding asymptotic Killing tensors, \ie higher-spin gauge transformations that preserve this gauge in addition to suitable fall-off conditions at null infinity. This led to the conditions $\delta\phii_{u(s-k)\,a_1\cdots a_k}=\cO(1/r^{d/2-k})$, which for $s=k$ yield \cite[Eq.\ (172)]{Campoleoni:2017qot}
\be
\label{residualsyms}
\delta\phii_{a_1\cdots a_s}(r,u,x)=\sum\limits_{t=1}^{s}\frac{1}{r^{t+1-2s}}\,\nabla^{}_{(a_1}\cdots\nabla^{}_{a_t}K^{(t-1)}_{a_{t+1}\cdots a_s)_0}(x)+\cO(u)=\cO(1/r^{d/2-s})\,,
\ee
where $K^{(t-1)}_{a_1\cdots a_{s-t}}(x)$ is a residual gauge parameter. The round bracket in $T_{(a_1\cdots a_s)_0}$ stands for the traceless (with respect to the metric $g$ on $S^d$) and totally symmetric part of any rank-$s$ tensor $T_{a_1\cdots a_s}$. Aside from tracelessness (to which we shall return), the tensor $K$ in \eqref{residualsyms} is the same as the multivector $K$ appearing in the differential operator \eqref{D_0}.

Now, the global symmetries of a higher-spin field configuration $\phii$ are gauge transformations such that $\delta\phii(r,u,x)=0$. Thus the asymptotic symmetry \eqref{residualsyms} is a global symmetry if 
\be
\label{HSconfKeq}
\nabla^{}_{(a_1} \cdots\nabla^{}_{a_t}K^{(t-1)}_{a_{t+1}\ldots\,a_s)_0}(x)=0\,.
\ee
This equation is well-known in conformal geometry: for $s\geqslant t$, a \itt{conformal Killing tensor 
of rank $s-t$ and depth $t$} on a conformally flat manifold is a symmetric traceless tensor field $K^{(t-1)}_{b_1\ldots \,b_{s-t}}(x)$ which is a primary field of scaling dimension $1-s$ and which satisfies, for any metric in the equivalence class, the generalised conformal Killing equation \cite{Nikitin}
\be
\label{genKeq}
\nabla^{}_{(a_1} \cdots\nabla^{}_{a_t}K^{(t-1)}_{a_{t+1}\ldots\,a_s)}(x)
=
g^{}_{(a_1a_2}(x)\chi^{}_{a_3\ldots\,a_s)}(x)
\ee
where $\chi$ can be any symmetric tensor field. This is indeed equivalent to the requirement \eqref{HSconfKeq}. The usual conformal Killing tensors are the particular case of depth one. In this sense, eq.\ \eqref{residualsyms} readily displays the Carrollian conformal structure that pervades flat space holography. However, the structure is generally quite rigid: the space of conformal Killing tensors of rank $s-t$ and depth $t$ on $S^d$ is a finite-dimensional irreducible $\mathfrak{so}(d+1,1)$-module $\cD(1-s,s-t)$, labelled by a Young diagram $Y=(s-1,s-t)$ with a first row of length $s-1$ and a second row of length $s-t$ (see \eg  \cite[sec.\ 3.2]{Bekaert:2013zya} for details). Note that this module satisfies the inclusion $\cD(1-s,s-t)\subset \cD(-1,1)^{\otimes s-t}\otimes\cD(-1,0)^{\otimes t-1}$, in accordance with the fact that the space of conformal Killing tensors of rank $s-t$ and depth $t$ on $S^d$ is spanned by symmetrised products of $s-t$ conformal Killing vectors (rank-one, depth-one) and $t-1$ conformal Killing scalars of depth two (rank-zero, depth-two).

The issue, then, is to understand to what extent fall-off conditions force $K$ in \eqref{residualsyms} to be a heavily constrained conformal Killing tensor or, to the contrary, allow it to be pretty much any tensor field. This whole spectrum of possibilities is actually available in the literature:
\begin{itemize}
\item[1.] The strong fall-offs $\phii_{u(s-k)\,a_1\cdots a_k}=\cO(1/r^{d/2-k})$ of \cite{Campoleoni:2017qot} were originally proposed as a higher-spin analogue of the usual ones \eqref{e52} for scalars. They turn out to force $K^{(t-1)}_{a_1\cdots a_{s-t}}(x)$ to be a conformal Killing tensor of depth $t$ whenever $d>2$. An exception only occurs for $d=2$, in which case the residual gauge parameters $K^{(s-1)}(x)$ are left free and correspond to the higher-spin generalisation of supertranslations.
\item[2.] The slightly weaker fall-offs $\phii_{u(s-k)\,a_1\cdots a_k}={\cO}(1/r^{1-k})$ advocated in \cite{Campoleoni:2020ejn} allow higher-spin supertranslations to occur in any spacetime dimension.
\item[3.] Finally, the much weaker fall-off conditions $\phii_{u(s-k)\,a_1\cdots a_k}={\cO}(r^{2-s-k})$ were proposed in \cite{Campoleoni:2020ejn} as higher-spin extensions of the gravitational ones leading to generalised BMS in \cite{Campiglia:2014yka,Campiglia:2015yka,Colferai:2020rte}. In that case, the symmetry enhancement is maximal, as \eqref{residualsyms} leaves all residual gauge parameters $K^{(t-1)}_{a_1\cdots a_{s-t}}(x)$ completely unconstrained.
\end{itemize}
In what follows, we restrict ourselves to the two extreme classes of fall-offs---the ``strong'' and the ``much weaker'' ones---and attempt to match their asymptotic Killing tensors with the two candidate algebras of asymptotic symmetries obtained at the end of section \ref{symsSachs}. (The WRac is discarded from the get-go since it fails to capture supertranslations.) In short, the matching would be spot on if it weren't for trace conditions.

\paragraph{Higher-spin algebras of asymptotic symmetries.} Our two candidate algebras of asymptotic symmetries are, respectively, the partially-massless Minkowski algebra and the generalised BMS higher-spin algebra. They are, respectively, higher-spin extensions of Poincar\'e and generalised BMS algebras. As vector spaces, they are spanned by some symmetric tensor fields $K^{a_1\cdots a_r}_{(q)}(x)$ on null infinity, as in \eqref{D_0}. The only discrepancy with the residual gauge parameters in \eqref{residualsyms} is that the latter are traceless, while they are traceful for the  higher-spin generalised BMS algebra, \cf \eqref{D_0}.

Trace conditions are a recurring problem of tentative higher-spin algebras in Min\-kow\-ski spacetime (see \eg \cite{Bekaert:2008sa,Campoleoni:2021blr}), because they preclude the interpretation of traceful tensors as algebras of global symmetries of massless gauge fields.  This suggests to look for ``exotic'' higher-spin gravity theories whose spectra contain extra propagating fields beyond the usual tower of Fronsdal massless fields of all spins. A tantalising option would be an ``unconstrained'' higher-spin theory with higher-derivative equations (see \eg \cite{Joung:2012qy,Francia:2012rg} and references therein) since, na\"ively, its global symmetries are traceful Killing tensors on Minkowski spacetime, suggesting that asymptotic symmetries are similarly traceful tensors at null infinity. However, determining rigorously the spectrum of such symmetries and the underlying degrees of freedom is a subtle issue for higher-derivative gauge theories, and it remains to be achieved in complete generality.

Another way out was suggested in  \cite[sec.\ 6.2]{Campoleoni:2021blr}:
the collection of generators of the partially-massless Minkowski algebra precisely matches the spectrum of global symmetries for a tower of higher-spin gauge fields of all spin $s=1,2,3,\ldots$ and all odd depths $t=1,3,5,\ldots$ around Minkowski spacetime; see \eg \cite{Bekaert:2013zya,Joung:2015jza} for the (A)dS counterparts of such a matching. To be precise, in Minkowski spacetime such gauge fields are only ``partially-massless-like'' (in the sense of \cite{Campoleoni:2021blr}). Unfortunately, exotic theories on Minkowski spacetime including such partially-massless-like fields do not seem to be unitary; in fact, partially-massless fields are only unitary on dS, not on AdS \cite{Deser:2001xr}. This second option is thus problematic in its own way.

The important questions raised by trace conditions will be left for future investigations. Nevertheless, it is tempting to conjecture that an exotic higher-spin gravity around Minkowski space, whose spectrum is a tower of partially-massless-like fields of all spins and all odd depths, admits 
as algebra of asymptotic symmetries the partially-massless Minkowski (respectively, generalised BMS) higher-spin algebra for suitable fall-off conditions that generalise to higher depths
the strong fall-offs in \cite{Campoleoni:2017mbt} (respectively, the weak ones in \cite{Campoleoni:2020ejn}) of the massless case. It would be fascinating to see such an asymptotic symmetry computation carried out explicitly.

\section*{Acknowledgements}

The authors thank G.\ Barnich, A.\ Campoleoni, B.\ Estienne, D.\ Francia, M.\ Grigoriev, Y.\ Herfray, K.\ Morand, S.\ Pekar and R.\ Ruzziconi for useful discussions. X.B.\ thanks the Erwin Schr\"{o}dinger International Institute for Mathematics and Physics (Vienna) for hospitality during the thematic programme ``Geometry for Higher Spin Gravity: Conformal Structures, PDEs, and Q-manifolds'' (August 23rd--September 17th, 2021) where part of this work was pursued. The work of B.O.\ is supported by the ANR grant \itt{TopO} No.\ ANR-17-CE30-0013-01 and the European Union’s Horizon 2020 research and innovation programme under the Marie Sk{\l}odowska-Curie grant agreement No.\ 846244.

\addcontentsline{toc}{section}{References}
% \bibliographystyle{utphys}
% \bibliography{HiBMS.bib}

\begin{thebibliography}{100}
\setlength{\itemsep}{0.37em}

\bibitem{Arnowitt}
R.~L. Arnowitt, S.~Deser, and C.~W. Misner, ``{Dynamical Structure and
  Definition of Energy in General Relativity},'' {\em Phys. Rev.} \textbf{116}
  (1959) 1322--1330.

\bibitem{Bondi}
H.~Bondi, M.~G.~J. van~der Burg, and A.~W.~K. Metzner, ``{Gravitational waves
  in general relativity. 7. Waves from axisymmetric isolated systems},'' {\em
  {Proceedings of the Royal Society of London}} \textbf{A269} (1962)
21--52.
%%CITATION = PRSLA,A269,21;%%.

\bibitem{Sachs1}
R.~K. Sachs, ``{Gravitational waves in general relativity. 8. Waves in
  asymptotically flat space-times},'' {\em {Proceedings of the Royal Society of
  London}} \textbf{A270} (1962)
103--126.
%%CITATION = PRSLA,A270,103;%%.

\bibitem{Sachs2}
R.~Sachs, ``{Asymptotic symmetries in gravitational theory},'' {\em {Physical
  Review}} \textbf{128} (1962)
2851--2864.
%%CITATION = PHRVA,128,2851;%%.

\bibitem{Brown}
J.~D. Brown and M.~Henneaux, ``{Central Charges in the Canonical Realization of
  Asymptotic Symmetries: An Example from Three-Dimensional Gravity},'' {\em
  {Communications in Mathematical Physics}} \textbf{104} (1986)
207--226.
%%CITATION = CMPHA,104,207;%%.

\bibitem{Barnich:2009se}
G.~Barnich and C.~Troessaert, ``{Symmetries of asymptotically flat
  4-dimensional spacetimes at null infinity revisited},'' {\em Phys. Rev.
  Lett.} \textbf{105} (2010) 111103,
\href{http://www.arXiv.org/abs/0909.2617}{\texttt{0909.2617}}.
%%CITATION = ARXIV:0909.2617;%%.

\bibitem{Barnich:2010eb}
G.~Barnich and C.~Troessaert, ``{Aspects of the BMS/CFT correspondence},'' {\em
  JHEP} \textbf{05} (2010) 062,
  \href{http://www.arXiv.org/abs/1001.1541}{\texttt{1001.1541}}.

\bibitem{Bagchi:2009my}
A.~Bagchi and R.~Gopakumar, ``{Galilean Conformal Algebras and AdS/CFT},'' {\em
  JHEP} \textbf{07} (2009) 037,
  \href{http://www.arXiv.org/abs/0902.1385}{\texttt{0902.1385}}.

\bibitem{Bagchi:2012yk}
A.~Bagchi, S.~Detournay, and D.~Grumiller, ``{Flat-Space Chiral Gravity},''
  {\em Phys. Rev. Lett.} \textbf{109} (2012) 151301,
\href{http://www.arXiv.org/abs/1208.1658}{\texttt{1208.1658}}.
%%CITATION = ARXIV:1208.1658;%%.

\bibitem{He:2014laa}
T.~He, V.~Lysov, P.~Mitra, and A.~Strominger, ``{BMS supertranslations and
  Weinberg's soft graviton theorem},'' {\em JHEP} \textbf{05} (2015) 151,
\href{http://www.arXiv.org/abs/1401.7026}{\texttt{1401.7026}}.
%%CITATION = ARXIV:1401.7026;%%.

\bibitem{Strominger:2017zoo}
A.~Strominger, ``{Lectures on the Infrared Structure of Gravity and Gauge
  Theory},'' \href{http://www.arXiv.org/abs/1703.05448}{\texttt{1703.05448}}.

\bibitem{Raclariu:2021zjz}
A.-M. Raclariu, ``{Lectures on Celestial Holography},''
  \href{http://www.arXiv.org/abs/2107.02075}{\texttt{2107.02075}}.

\bibitem{Prema:2021sjp}
A.~B. Prema, G.~Comp\`ere, L.~P. de~Gioia, I.~Mol, and B.~Swidler, ``{Celestial
  holography: Lectures on asymptotic symmetries},'' {\em SciPost Phys. Lect.
  Notes} \textbf{47} (2022) 1,
  \href{http://www.arXiv.org/abs/2109.00997}{\texttt{2109.00997}}.

\bibitem{Pasterski:2021raf}
S.~Pasterski, M.~Pate, and A.-M. Raclariu, ``{Celestial Holography},'' in {\em
  {2022 Snowmass Summer Study}}.
\newblock 11, 2021.
\newblock \href{http://www.arXiv.org/abs/2111.11392}{\texttt{2111.11392}}.

\bibitem{Campiglia:2014yka}
M.~Campiglia and A.~Laddha, ``{Asymptotic symmetries and subleading soft
  graviton theorem},'' {\em Phys. Rev. D} \textbf{90} (2014), no.~12, 124028,
  \href{http://www.arXiv.org/abs/1408.2228}{\texttt{1408.2228}}.

\bibitem{Campiglia:2015yka}
M.~Campiglia and A.~Laddha, ``{New symmetries for the Gravitational
  S-matrix},'' {\em JHEP} \textbf{04} (2015) 076,
  \href{http://www.arXiv.org/abs/1502.02318}{\texttt{1502.02318}}.

\bibitem{Bekaert:2010hw}
X.~Bekaert, N.~Boulanger, and P.~Sundell, ``{How higher-spin gravity surpasses
  the spin two barrier: no-go theorems versus yes-go examples},'' {\em Rev.
  Mod. Phys.} \textbf{84} (2012) 987--1009,
  \href{http://www.arXiv.org/abs/1007.0435}{\texttt{1007.0435}}.

\bibitem{Rahman:2015pzl}
R.~Rahman and M.~Taronna, ``{From Higher Spins to Strings: A Primer},''
  \href{http://www.arXiv.org/abs/1512.07932}{\texttt{1512.07932}}.

\bibitem{Ponomarev:2022vjb}
D.~Ponomarev, ``{Basic introduction to higher-spin theories},''
  \href{http://www.arXiv.org/abs/2206.15385}{\texttt{2206.15385}}.

\bibitem{Gross:1988ue}
D.~J. Gross, ``{High-Energy Symmetries of String Theory},'' {\em Phys. Rev.
  Lett.} \textbf{60} (1988) 1229.

\bibitem{Sagnotti:2003qa}
A.~Sagnotti and M.~Tsulaia, ``{On higher spins and the tensionless limit of
  string theory},'' {\em Nucl. Phys. B} \textbf{682} (2004) 83--116,
  \href{http://www.arXiv.org/abs/hep-th/0311257}{\texttt{hep-th/0311257}}.

\bibitem{Bekaert:2012ux}
X.~Bekaert, E.~Joung, and J.~Mourad, ``{Comments on higher-spin holography},''
  {\em Fortsch. Phys.} \textbf{60} (2012) 882--888,
  \href{http://www.arXiv.org/abs/1202.0543}{\texttt{1202.0543}}.

\bibitem{Giombi:2016ejx}
S.~Giombi, ``{Higher Spin \textemdash{} CFT Duality},'' in {\em {Theoretical
  Advanced Study Institute in Elementary Particle Physics}: {New Frontiers in
  Fields and Strings}}, pp.~137--214.
\newblock 2017.
\newblock \href{http://www.arXiv.org/abs/1607.02967}{\texttt{1607.02967}}.

\bibitem{Sleight:2017krf}
C.~Sleight, ``{Metric-like Methods in Higher Spin Holography},'' {\em PoS}
  \textbf{Modave2016} (2017) 003,
  \href{http://www.arXiv.org/abs/1701.08360}{\texttt{1701.08360}}.

\bibitem{Dirac:1963ta}
P.~A.~M. Dirac, ``{A Remarkable representation of the 3 + 2 de Sitter group},''
  {\em J. Math. Phys.} \textbf{4} (1963) 901--909.

\bibitem{Flato:1978qz}
M.~Flato and C.~Fronsdal, ``{One Massless Particle Equals Two Dirac Singletons:
  Elementary Particles in a Curved Space. 6.},'' {\em Lett. Math. Phys.}
  \textbf{2} (1978) 421--426.

\bibitem{Henneaux:2010xg}
M.~Henneaux and S.-J. Rey, ``{Nonlinear $W_{infinity}$ as Asymptotic Symmetry
  of Three-Dimensional Higher Spin Anti-de Sitter Gravity},'' {\em JHEP}
  \textbf{12} (2010) 007,
  \href{http://www.arXiv.org/abs/1008.4579}{\texttt{1008.4579}}.

\bibitem{Campoleoni:2010zq}
A.~Campoleoni, S.~Fredenhagen, S.~Pfenninger, and S.~Theisen, ``{Asymptotic
  symmetries of three-dimensional gravity coupled to higher-spin fields},''
  {\em JHEP} \textbf{11} (2010) 007,
  \href{http://www.arXiv.org/abs/1008.4744}{\texttt{1008.4744}}.

\bibitem{Gaberdiel:2010ar}
M.~R. Gaberdiel, R.~Gopakumar, and A.~Saha, ``{Quantum $W$-symmetry in
  $AdS_3$},'' {\em JHEP} \textbf{02} (2011) 004,
  \href{http://www.arXiv.org/abs/1009.6087}{\texttt{1009.6087}}.

\bibitem{Gaberdiel:2010pz}
M.~R. Gaberdiel and R.~Gopakumar, ``{An AdS$_{3}$ Dual for Minimal Model
  CFTs},'' {\em Phys. Rev. D} \textbf{83} (2011) 066007,
  \href{http://www.arXiv.org/abs/1011.2986}{\texttt{1011.2986}}.

\bibitem{Campoleoni:2011hg}
A.~Campoleoni, S.~Fredenhagen, and S.~Pfenninger, ``{Asymptotic W-symmetries in
  three-dimensional higher-spin gauge theories},'' {\em JHEP} \textbf{09}
  (2011) 113, \href{http://www.arXiv.org/abs/1107.0290}{\texttt{1107.0290}}.

\bibitem{Afshar:2013vka}
H.~Afshar, A.~Bagchi, R.~Fareghbal, D.~Grumiller, and J.~Rosseel, ``{Spin-3
  Gravity in Three-Dimensional Flat Space},'' {\em Phys. Rev. Lett.}
  \textbf{111} (2013), no.~12, 121603,
  \href{http://www.arXiv.org/abs/1307.4768}{\texttt{1307.4768}}.

\bibitem{Gonzalez:2013oaa}
H.~A. Gonzalez, J.~Matulich, M.~Pino, and R.~Troncoso, ``{Asymptotically flat
  spacetimes in three-dimensional higher spin gravity},'' {\em JHEP}
  \textbf{09} (2013) 016,
  \href{http://www.arXiv.org/abs/1307.5651}{\texttt{1307.5651}}.

\bibitem{Campoleoni:2015qrh}
A.~Campoleoni, H.~A. Gonzalez, B.~Oblak, and M.~Riegler, ``{Rotating Higher
  Spin Partition Functions and Extended BMS Symmetries},'' {\em JHEP}
  \textbf{04} (2016) 034,
  \href{http://www.arXiv.org/abs/1512.03353}{\texttt{1512.03353}}.

\bibitem{Campoleoni:2016vsh}
A.~Campoleoni, H.~A. Gonzalez, B.~Oblak, and M.~Riegler, ``{BMS Modules in
  Three Dimensions},'' {\em Int. J. Mod. Phys. A} \textbf{31} (2016), no.~12,
  1650068, \href{http://www.arXiv.org/abs/1603.03812}{\texttt{1603.03812}}.
  
\bibitem{Ammon:2020fxs}
M.~Ammon, M.~Pannier, and M.~Riegler, ``{Scalar Fields in 3D Asymptotically
  Flat Higher-Spin Gravity},'' {\em J. Phys. A} \textbf{54} (2021), no.~10,
  105401, \href{http://www.arXiv.org/abs/2009.14210}{\texttt{2009.14210}}.

\bibitem{Campoleoni:2017mbt}
A.~Campoleoni, D.~Francia, and C.~Heissenberg, ``{On higher-spin
  supertranslations and superrotations},'' {\em JHEP} \textbf{05} (2017) 120,
  \href{http://www.arXiv.org/abs/1703.01351}{\texttt{1703.01351}}.

\bibitem{Campoleoni:2017qot}
A.~Campoleoni, D.~Francia, and C.~Heissenberg, ``{Asymptotic Charges at Null
  Infinity in Any Dimension},'' {\em Universe} \textbf{4} (2018), no.~3, 47,
  \href{http://www.arXiv.org/abs/1712.09591}{\texttt{1712.09591}}.

\bibitem{Campoleoni:2018uib}
A.~Campoleoni, D.~Francia, and C.~Heissenberg, ``{Asymptotic symmetries and
  charges at null infinity: from low to high spins},'' {\em EPJ Web Conf.}
  \textbf{191} (2018) 06011,
  \href{http://www.arXiv.org/abs/1808.01542}{\texttt{1808.01542}}.

\bibitem{Campoleoni:2020ejn}
A.~Campoleoni, D.~Francia, and C.~Heissenberg, ``{On asymptotic symmetries in
  higher dimensions for any spin},'' {\em JHEP} \textbf{12} (2020) 129,
  \href{http://www.arXiv.org/abs/2011.04420}{\texttt{2011.04420}}.

\bibitem{Campoleoni:2021blr}
A.~Campoleoni and S.~Pekar, ``{Carrollian and Galilean conformal higher-spin
  algebras in any dimensions},'' {\em JHEP} \textbf{02} (2022) 150,
  \href{http://www.arXiv.org/abs/2110.07794}{\texttt{2110.07794}}.

\bibitem{Duval:2014uoa}
C.~Duval, G.~W. Gibbons, P.~A. Horvathy, and P.~M. Zhang, ``{Carroll versus
  Newton and Galilei: two dual non-Einsteinian concepts of time},'' {\em Class.
  Quant. Grav.} \textbf{31} (2014) 085016,
  \href{http://www.arXiv.org/abs/1402.0657}{\texttt{1402.0657}}.

\bibitem{Levy1965}
J.-M. Lévy-Leblond, ``Une nouvelle limite non-relativiste du groupe de
  poincaré,'' {\em Annales de l'I.H.P. Physique théorique} \textbf{3} (1965),
  no.~1, 1--12.

\bibitem{Gupta1966}
N.~D. {Sen Gupta}, ``{On an analogue of the Galilei group},'' {\em Nuovo
  Cimento A Serie} \textbf{44} (July, 1966) 512--517.

\bibitem{Henneaux:1979vn}
M.~Henneaux, ``{Geometry of Zero Signature Space-times},'' {\em Bull. Soc.
  Math. Belg.} \textbf{31} (1979) 47--63.

\bibitem{Duval:2014uva}
C.~Duval, G.~W. Gibbons, and P.~A. Horvathy, ``{Conformal Carroll groups and
  BMS symmetry},'' {\em Class. Quant. Grav.} \textbf{31} (2014) 092001,
  \href{http://www.arXiv.org/abs/1402.5894}{\texttt{1402.5894}}.

\bibitem{Duval:2014lpa}
C.~Duval, G.~W. Gibbons, and P.~A. Horvathy, ``{Conformal Carroll groups},''
  {\em J. Phys. A} \textbf{47} (2014), no.~33, 335204,
  \href{http://www.arXiv.org/abs/1403.4213}{\texttt{1403.4213}}.

\bibitem{Penrose:1965am}
R.~Penrose, ``{Zero rest mass fields including gravitation: Asymptotic
  behavior},'' {\em Proc. Roy. Soc. Lond. A} \textbf{284} (1965) 159.

\bibitem{Penrose:1972ea}
R.~Penrose, ``{Relativistic Symmetry Groups},'' in {\em {NATO Advanced Study
  Institute -- Group theory in non-linear problems}}.
\newblock 1972.

\bibitem{Geroch1977}
R.~Geroch, {\em Asymptotic Structure of Space-Time}, pp.~1--105.
\newblock Springer US, Boston, MA, 1977.

\bibitem{Ashtekar:1987tt}
A.~Ashtekar, {\em {Asymptotic Quantization: Based on 1984 Naples Lectures}}.
\newblock Bibliopolis, 1987.

\bibitem{Ciambelli:2018xat}
L.~Ciambelli, C.~Marteau, A.~C. Petkou, P.~M. Petropoulos, and K.~Siampos,
  ``{Covariant Galilean versus Carrollian hydrodynamics from relativistic
  fluids},'' {\em Class. Quant. Grav.} \textbf{35} (2018), no.~16, 165001,
  \href{http://www.arXiv.org/abs/1802.05286}{\texttt{1802.05286}}.

\bibitem{Ciambelli:2018wre}
L.~Ciambelli, C.~Marteau, A.~C. Petkou, P.~M. Petropoulos, and K.~Siampos,
  ``{Flat holography and Carrollian fluids},'' {\em JHEP} \textbf{07} (2018)
  165, \href{http://www.arXiv.org/abs/1802.06809}{\texttt{1802.06809}}.

\bibitem{Campoleoni:2018ltl}
A.~Campoleoni, L.~Ciambelli, C.~Marteau, P.~M. Petropoulos, and K.~Siampos,
  ``{Two-dimensional fluids and their holographic duals},'' {\em Nucl. Phys. B}
  \textbf{946} (2019) 114692,
  \href{http://www.arXiv.org/abs/1812.04019}{\texttt{1812.04019}}.

\bibitem{Petkou:2022bmz}
A.~C. Petkou, P.~M. Petropoulos, D.~R. Betancour, and K.~Siampos,
  ``{Relativistic Fluids, Hydrodynamic Frames and their Galilean versus
  Carrollian Avatars},''
  \href{http://www.arXiv.org/abs/2205.09142}{\texttt{2205.09142}}.

\bibitem{Ciambelli:2019lap}
L.~Ciambelli, R.~G. Leigh, C.~Marteau, and P.~M. Petropoulos, ``{Carroll
  Structures, Null Geometry and Conformal Isometries},'' {\em Phys. Rev. D}
  \textbf{100} (2019), no.~4, 046010,
  \href{http://www.arXiv.org/abs/1905.02221}{\texttt{1905.02221}}.

\bibitem{Donnay:2022aba}
L.~Donnay, A.~Fiorucci, Y.~Herfray, and R.~Ruzziconi, ``{Carrollian Perspective
  on Celestial Holography},'' {\em Phys. Rev. Lett.} \textbf{129} (2022),
  no.~7, 071602,
  \href{http://www.arXiv.org/abs/2202.04702}{\texttt{2202.04702}}.

\bibitem{Campiglia:2017dpg}
M.~Campiglia, L.~Coito, and S.~Mizera, ``{Can scalars have asymptotic
  symmetries?},'' {\em Phys. Rev. D} \textbf{97} (2018), no.~4, 046002,
  \href{http://www.arXiv.org/abs/1703.07885}{\texttt{1703.07885}}.

\bibitem{Campiglia:2017xkp}
M.~Campiglia and L.~Coito, ``{Asymptotic charges from soft scalars in even
  dimensions},'' {\em Phys. Rev. D} \textbf{97} (2018), no.~6, 066009,
  \href{http://www.arXiv.org/abs/1711.05773}{\texttt{1711.05773}}.

\bibitem{Satishchandran:2019pyc}
G.~Satishchandran and R.~M. Wald, ``{Asymptotic behavior of massless fields and
  the memory effect},'' {\em Phys. Rev. D} \textbf{99} (2019), no.~8, 084007,
  \href{http://www.arXiv.org/abs/1901.05942}{\texttt{1901.05942}}.

\bibitem{Mccarthy:1972ry}
P.~J. McCarthy, ``{Asymptotically flat space-times and elementary particles},''
  {\em Phys. Rev. Lett.} \textbf{29} (1972)
817--819.
%%CITATION = PRLTA,29,817;%%.

\bibitem{McCarthy01}
P.~J. McCarthy, ``{Representations of the Bondi-Metzner-Sachs Group. I.
  Determination of the Representations},'' {\em Proc. Roy. Soc. Lond.}
  \textbf{A330} (1972), no.~1583, 517--535.

\bibitem{McCarthy00}
P.~J. McCarthy, ``{Structure of the Bondi-Metzner-Sachs Group},'' {\em J. Math.
  Phys.} \textbf{13} (1972), no.~11, 1837--1842.

\bibitem{McCarthy317}
P.~J. McCarthy, ``{Representations of the Bondi-Metzner-Sachs Group. II.
  Properties and Classification of the Representations},'' {\em Proceedings of
  the Royal Society of London A: Mathematical, Physical and Engineering
  Sciences} \textbf{333} (1973), no.~1594, 317--336.

\bibitem{McCarthy301}
P.~J. McCarthy and M.~Crampin, ``{Representations of the Bondi-Metzner-Sachs
  Group. III. Poincar\'e Spin Multiplicities and Irreducibility},'' {\em
  Proceedings of the Royal Society of London A: Mathematical, Physical and
  Engineering Sciences} \textbf{335} (1973), no.~1602, 301--311.

\bibitem{McCarthy489}
P.~J. McCarthy, ``{The Bondi-Metzner-Sachs Group in the Nuclear Topology},''
  {\em Proceedings of the Royal Society of London A: Mathematical, Physical and
  Engineering Sciences} \textbf{343} (1975), no.~1635, 489--523.

\bibitem{Eastwood:2002su}
M.~G. Eastwood, ``{Higher symmetries of the Laplacian},'' {\em Annals Math.}
  \textbf{161} (2005) 1645--1665,
  \href{http://www.arXiv.org/abs/hep-th/0206233}{\texttt{hep-th/0206233}}.

\bibitem{Vasiliev:2003ev}
M.~A. Vasiliev, ``{Nonlinear equations for symmetric massless higher spin
  fields in (A)dS(d)},'' {\em Phys. Lett. B} \textbf{567} (2003) 139--151,
  \href{http://www.arXiv.org/abs/hep-th/0304049}{\texttt{hep-th/0304049}}.

\bibitem{Nguyen:2022nnx}
K.~Nguyen and P.~West, ``{Conserved asymptotic charges for any massless
  particle},'' \href{http://www.arXiv.org/abs/2208.08234}{\texttt{2208.08234}}.

\bibitem{Ashtekar:1981bq}
A.~Ashtekar and M.~Streubel, ``{Symplectic Geometry of Radiative Modes and
  Conserved Quantities at Null Infinity},'' {\em Proc. Roy. Soc. Lond. A}
  \textbf{376} (1981) 585--607.

\bibitem{Skenderis:2002wp}
K.~Skenderis, ``{Lecture notes on holographic renormalization},'' {\em Class.
  Quant. Grav.} \textbf{19} (2002) 5849--5876,
  \href{http://www.arXiv.org/abs/hep-th/0209067}{\texttt{hep-th/0209067}}.

\bibitem{GJMS}
C.~R. Graham, R.~Jenne, L.~J. Mason, and G.~A.~J. Sparling, ``{Conformally
  Invariant Powers of the Laplacian, I: Existence},'' {\em Journal of the
  London Mathematical Society} \textbf{s2-46} (1992), no.~3, 557--565.

\bibitem{Branson}
T.~P. Branson, ``Sharp inequalities, the functional determinant, and the
  complementary series,'' {\em Transactions of the American Mathematical
  Society} \textbf{347} (1995) 3671--3742.

\bibitem{Angelopoulos:1980wg}
E.~Angelopoulos, M.~Flato, C.~Fronsdal, and D.~Sternheimer, ``{Massless
  Particles, Conformal Group and De Sitter Universe},'' {\em Phys. Rev. D}
  \textbf{23} (1981) 1278.

\bibitem{Royden}
H.~L. Royden and P.~Fitzpatrick, {\em Real analysis}, vol.~32.
\newblock Macmillan New York, 1988.

\bibitem{Barut}
A.~Barut and R.~Raczka, {\em {Theory of Group Representations and
  Applications}}.
\newblock World Scientific, 1986.

\bibitem{Oblak:2016eij}
B.~Oblak, {\em {BMS Particles in Three Dimensions}}.
\newblock PhD thesis, U. Brussels, Brussels U., 2016.
\newblock \href{http://www.arXiv.org/abs/1610.08526}{\texttt{1610.08526}}.

\bibitem{Dirac:1936fq}
P.~A.~M. Dirac, ``{Wave equations in conformal space},'' {\em Annals Math.}
  \textbf{37} (1936) 429--442.

\bibitem{Bekaert:2011js}
X.~Bekaert, ``{Singletons and their maximal symmetry algebras},'' in {\em {6th
  Summer School in Modern Mathematical Physics}}, pp.~71--89.
\newblock 11, 2011.
\newblock \href{http://www.arXiv.org/abs/1111.4554}{\texttt{1111.4554}}.

\bibitem{Kobayashi:2001nq}
T.~Kobayashi and B.~Orsted, ``{Analysis on the minimal representation of O(p,
  q). 1. Realization via conformal geometry},'' {\em Advances in Mathematics}
  \textbf{180} (2003), no.~2, 486--512.

\bibitem{Anninos:2011ui}
D.~Anninos, T.~Hartman, and A.~Strominger, ``{Higher Spin Realization of the
  dS/CFT Correspondence},'' {\em Class. Quant. Grav.} \textbf{34} (2017),
  no.~1, 015009, \href{http://www.arXiv.org/abs/1108.5735}{\texttt{1108.5735}}.

\bibitem{Bekaert:2013zya}
X.~Bekaert and M.~Grigoriev, ``{Higher order singletons, partially massless
  fields and their boundary values in the ambient approach},'' {\em Nucl. Phys.
  B} \textbf{876} (2013) 667--714,
  \href{http://www.arXiv.org/abs/1305.0162}{\texttt{1305.0162}}.

\bibitem{Oblak:2015qia}
B.~Oblak, ``{From the Lorentz Group to the Celestial Sphere},'' in {\em
  Proceedings of the Seventh Brussels Summer School of Mathematics}.
\newblock 8, 2015.
\newblock \href{http://www.arXiv.org/abs/1508.00920}{\texttt{1508.00920}}.

\bibitem{Strominger:2021mtt}
A.~Strominger, ``{$w_{1+\infty}$ Algebra and the Celestial Sphere: Infinite
  Towers of Soft Graviton, Photon, and Gluon Symmetries},'' {\em Phys. Rev.
  Lett.} \textbf{127} (2021), no.~22, 221601,
  \href{http://www.arXiv.org/abs/2105.14346}{\texttt{2105.14346}}.

\bibitem{Himwich:2021dau}
E.~Himwich, M.~Pate, and K.~Singh, ``{Celestial operator product expansions and
  w$_{1+\infty}$ symmetry for all spins},'' {\em JHEP} \textbf{01} (2022) 080,
  \href{http://www.arXiv.org/abs/2108.07763}{\texttt{2108.07763}}.

\bibitem{Freidel:2021ytz}
L.~Freidel, D.~Pranzetti, and A.-M. Raclariu, ``{Higher spin dynamics in
  gravity and $w_{1 + \infty}$ celestial symmetries},''
  \href{http://www.arXiv.org/abs/2112.15573}{\texttt{2112.15573}}.

\bibitem{Longhi:1997zt}
G.~Longhi and M.~Materassi, ``{A Canonical realization of the BMS algebra},''
  {\em J. Math. Phys.} \textbf{40} (1999) 480--500,
  \href{http://www.arXiv.org/abs/hep-th/9803128}{\texttt{hep-th/9803128}}.

\bibitem{Gomis:2015ata}
J.~Gomis and G.~Longhi, ``{Canonical realization of Bondi-Metzner-Sachs
  symmetry: Quadratic Casimir},'' {\em Phys. Rev. D} \textbf{93} (2016), no.~2,
  025030, \href{http://www.arXiv.org/abs/1508.00544}{\texttt{1508.00544}}.

\bibitem{Batlle:2017llu}
C.~Batlle, V.~Campello, and J.~Gomis, ``{Canonical realization of
  (2+1)-dimensional Bondi-Metzner-Sachs symmetry},'' {\em Phys. Rev. D}
  \textbf{96} (2017), no.~2, 025004,
  \href{http://www.arXiv.org/abs/1703.01833}{\texttt{1703.01833}}.

\bibitem{Batlle:2022hwf}
C.~Batlle, V.~Campello, and J.~Gomis, ``{Polyharmonic Green Functions and
  Nonlocal BMS Transformations of a Free Scalar Field},''
  \href{http://www.arXiv.org/abs/2207.12299}{\texttt{2207.12299}}.

\bibitem{Carmeli:2000af}
M.~Carmeli, {\em {Group theory and general relativity: Representations of the
  Lorentz group and their applications to the gravitational field}}.
\newblock Imperial College Press, 2000.

\bibitem{Colferai:2020rte}
D.~Colferai and S.~Lionetti, ``{Asymptotic symmetries and the subleading soft
  graviton theorem in higher dimensions},'' {\em Phys. Rev. D} \textbf{104}
  (2021), no.~6, 064010,
  \href{http://www.arXiv.org/abs/2005.03439}{\texttt{2005.03439}}.

\bibitem{Steenrod}
N.~Steenrod, {\em The topology of fibre bundles}, vol.~44.
\newblock Princeton university press, 1999.

\bibitem{Lee}
J.~M. Lee, {\em Manifolds and differential geometry}, vol.~107 of {\em Graduate
  studies in mathematics}.
\newblock American Mathematical Society, 2009.

\bibitem{Nakahara}
M.~Nakahara, {\em {Geometry, Topology and Physics}}.
\newblock Graduate student series in physics. Taylor \& Francis, 2003.

\bibitem{Herfray:2021qmp}
Y.~Herfray, ``{Carrollian manifolds and null infinity: A view from Cartan
  geometry},'' \href{http://www.arXiv.org/abs/2112.09048}{\texttt{2112.09048}}.

\bibitem{Bekaert:2015xua}
X.~Bekaert and K.~Morand, ``{Connections and dynamical trajectories in
  generalised Newton-Cartan gravity II. An ambient perspective},'' {\em J.
  Math. Phys.} \textbf{59} (2018), no.~7, 072503,
  \href{http://www.arXiv.org/abs/1505.03739}{\texttt{1505.03739}}.

\bibitem{Detournay:2012pc}
S.~Detournay, T.~Hartman, and D.~M. Hofman, ``{Warped Conformal Field
  Theory},'' {\em Phys. Rev. D} \textbf{86} (2012) 124018,
  \href{http://www.arXiv.org/abs/1210.0539}{\texttt{1210.0539}}.

\bibitem{Afshar:2015wjm}
H.~Afshar, S.~Detournay, D.~Grumiller, and B.~Oblak, ``{Near-Horizon Geometry
  and Warped Conformal Symmetry},'' {\em JHEP} \textbf{03} (2016) 187,
  \href{http://www.arXiv.org/abs/1512.08233}{\texttt{1512.08233}}.

\bibitem{Ashtekar:2014zsa}
A.~Ashtekar, ``{Geometry and Physics of Null Infinity},''
  \href{http://www.arXiv.org/abs/1409.1800}{\texttt{1409.1800}}.

\bibitem{Baggio:2011ha}
M.~Baggio, J.~de~Boer, and K.~Holsheimer, ``{Anomalous Breaking of Anisotropic
  Scaling Symmetry in the Quantum Lifshitz Model},'' {\em JHEP} \textbf{07}
  (2012) 099, \href{http://www.arXiv.org/abs/1112.6416}{\texttt{1112.6416}}.

\bibitem{Kachru:2008yh}
S.~Kachru, X.~Liu, and M.~Mulligan, ``{Gravity duals of Lifshitz-like fixed
  points},'' {\em Phys. Rev. D} \textbf{78} (2008) 106005,
  \href{http://www.arXiv.org/abs/0808.1725}{\texttt{0808.1725}}.

\bibitem{Barnich:2011ct}
G.~Barnich and C.~Troessaert, ``{Supertranslations call for superrotations},''
  {\em PoS} (2010) 010,
  \href{http://www.arXiv.org/abs/1102.4632}{\texttt{1102.4632}}.
[Ann. U. Craiova Phys.21,S11(2011)].
%%CITATION = ARXIV:1102.4632;%%.

\bibitem{Newman:1976gc}
E.~T. Newman, ``{Heaven and Its Properties},'' {\em Gen. Rel. Grav.} \textbf{7}
  (1976) 107--111.

\bibitem{Adamo:2010ey}
T.~M. Adamo and E.~T. Newman, ``{The Generalized Good Cut Equation},'' {\em
  Class. Quant. Grav.} \textbf{27} (2010) 245004,
  \href{http://www.arXiv.org/abs/1007.4215}{\texttt{1007.4215}}.

\bibitem{Herfray:2020rvq}
Y.~Herfray, ``{Asymptotic shear and the intrinsic conformal geometry of
  null-infinity},'' {\em J. Math. Phys.} \textbf{61} (2020), no.~7, 072502,
  \href{http://www.arXiv.org/abs/2001.01281}{\texttt{2001.01281}}.
  
\bibitem{Grumiller:2019fmp}
D.~Grumiller, A.~P\'erez, M.~M. Sheikh-Jabbari, R.~Troncoso, and C.~Zwikel,
  ``{Spacetime structure near generic horizons and soft hair},'' {\em Phys.
  Rev. Lett.} \textbf{124} (2020), no.~4, 041601,
  \href{http://www.arXiv.org/abs/1908.09833}{\texttt{1908.09833}}.

\bibitem{saunders_1989}
D.~J. Saunders, {\em The Geometry of Jet Bundles}.
\newblock London Mathematical Society Lecture Note Series. Cambridge University
  Press, 1989.

\bibitem{olver_1995}
P.~J. Olver, {\em Equivalence, Invariants and Symmetry}.
\newblock Cambridge University Press, 1995.

\bibitem{Nikitin}
A.~G.~Nikitin, ``Generalized Killing tensors of arbitrary rank and order,'' {\em
  Ukrainian Mathematical Journal} \textbf{43} (1991), no.~6, 734--743.

\bibitem{shapovalov1992symmetry}
A.~V. Shapovalov and I.~Shirokov, ``Symmetry algebras of linear differential
  equations,'' {\em Theoretical and Mathematical Physics} \textbf{92} (1992),
  no.~1, 697--703.

\bibitem{Bekaert:2006zoe}
X.~Bekaert, ``{Higher spin algebras as higher symmetries},'' {\em Ann. U.
  Craiova Phys.} \textbf{16} (2006), no.~II, 58--65,
  \href{http://www.arXiv.org/abs/0704.0898}{\texttt{0704.0898}}.

\bibitem{Vasiliev:2005zu}
M.~A. Vasiliev, ``{Actions, charges and off-shell fields in the unfolded
  dynamics approach},'' {\em Int. J. Geom. Meth. Mod. Phys.} \textbf{3} (2006)
  37--80,
  \href{http://www.arXiv.org/abs/hep-th/0504090}{\texttt{hep-th/0504090}}.

\bibitem{Bekaert:2008sa}
X.~Bekaert, ``{Comments on higher-spin symmetries},'' {\em Int. J. Geom. Meth.
  Mod. Phys.} \textbf{6} (2009) 285--342,
  \href{http://www.arXiv.org/abs/0807.4223}{\texttt{0807.4223}}.

\bibitem{Joung:2015jza}
E.~Joung and K.~Mkrtchyan, ``{Partially-massless higher-spin algebras and their
  finite-dimensional truncations},'' {\em JHEP} \textbf{01} (2016) 003,
  \href{http://www.arXiv.org/abs/1508.07332}{\texttt{1508.07332}}.

\bibitem{Joung:2012qy}
E.~Joung and K.~Mkrtchyan, ``{A note on higher-derivative actions for free
  higher-spin fields},'' {\em JHEP} \textbf{11} (2012) 153,
  \href{http://www.arXiv.org/abs/1209.4864}{\texttt{1209.4864}}.

\bibitem{Francia:2012rg}
D.~Francia, ``{Generalised connections and higher-spin equations},'' {\em
  Class. Quant. Grav.} \textbf{29} (2012) 245003,
  \href{http://www.arXiv.org/abs/1209.4885}{\texttt{1209.4885}}.

\bibitem{Deser:2001xr}
S.~Deser and A.~Waldron, ``{Null propagation of partially massless higher spins
  in (A)dS and cosmological constant speculations},'' {\em Phys. Lett. B}
  \textbf{513} (2001) 137--141,
  \href{http://www.arXiv.org/abs/hep-th/0105181}{\texttt{hep-th/0105181}}.

\end{thebibliography}

\providecommand{\href}[2]{#2}

\end{document}